\documentclass[gmd, manuscript]{copernicus} 
\AtBeginDocument{\nolinenumbers} 

\usepackage{multirow}
\usepackage{amsthm}
\usepackage{float}
\usepackage{subfig}
\usepackage{soul}
\usepackage[dvipsnames]{xcolor}

\begin{document}

\title{PRecover 1.0: Process Rate Recovery with Machine Learning}

\Author[1][miriam.simm@kit.edu]{Miriam}{Simm} 
\Author[2,3]{Tom}{Beucler}
\Author[1]{Corinna}{Hoose}

\affil[1]{Institute of Meteorology and Climate Research, Karlsruhe Institute of Technology, Karlsruhe, Germany}
\affil[2]{Faculty of Geosciences and Environment, University of Lausanne, Lausanne, Switzerland}
\affil[3]{Expertise Center for Climate Extremes, University of Lausanne, Lausanne, Switzerland}

\runningtitle{Process Rate Recovery with Machine Learning}
\runningauthor{M. Simm et al.}

\received{}
\pubdiscuss{} 
\revised{}
\accepted{}
\published{}

\firstpage{1}

\maketitle

\begin{abstract}
Comprehensive information on cloud microphysical process rates from numerical simulations allows for better understanding of precipitation formation pathways and aerosol-cloud interactions. However, resource limitations often make it impractical to include all microphysical process rates in the model output, limiting in-depth analyses. To address this shortcoming, we introduce PRecover, a data-driven post-processing approach to recover microphysical process rates that are not stored during runtime from standard output of a numerical weather prediction model. In particular, we train random forests, gradient boosting models, and feed-forward neural networks to recover microphysical process rates from a two-moment bulk microphysics scheme in the ICOsahedral Nonhydrostatic (ICON) model. We use cloud variables as input, obtained from high-resolution simulations in a limited-area setup over Europe. Warm-rain and ice microphysical process rates are recovered with a two-step classification-regression approach for both instantaneous process rates and process rates accumulated over output time steps ranging from one to 60 minutes. As a physics-based baseline, we assess whether process rates can be directly recalculated from stored ICON output variables. Accurate recalculation is possible for process rates such as accretion and self-collection but not for the autoconversion, rain melting or heterogeneous ice nucleation rate. Using PRecover, we successfully recover most of the process rates that are accumulated over output time steps of 10 minutes or less, but the values are increasingly difficult to recover for rates accumulated over longer accumulation intervals. For a model output time step of 10 minutes, the final combined classification-regression models achieve a deterministic performance of ${R^2} = 0.66$ for instantaneous process rates and ${R^2} = 0.40$ for accumulated process rates, respectively. The recovery fails for the heterogeneous ice nucleation rate, likely due to the unavailability of the number of activated ice nuclei in the model output. Excluding QI\_HET, the mean scores increase to $R^2 = 0.69$ for instantaneous rates and $R^2 = 0.41$ for accumulated rates. To quantify predictive uncertainty, we provide calibrated prediction intervals through conformalized quantile regression, achieving a prediction interval coverage probability of $88.09\%$ and $86.95\%$ for instantaneous and accumulated process rates, respectively. We demonstrate spatial transferability of the models with two case studies over different regional domains and simulation settings unseen during training. PRecover opens the possibility of obtaining information about microphysical process rates in a more resource-efficient and flexible way, allowing for in-depth studies of cloud microphysics even when the microphysical process rates were not saved initially.
\end{abstract}

\introduction[Introduction: Data-driven recovery of microphysical process rates]
Any computational model used in practice is constrained to a finite set of possible output variables, posing an inherent bound on the information that can be obtained from the model output. In Earth system modeling, this is particularly pronounced as a consequence of the complexity of the models and the spatio-temporal extent of simulations. Current research drives progress towards finer model resolutions, i.e. smaller time steps and grid spacing, which are able to better resolve small-scale processes and provide more accurate forecasts and more reliable climate projections \citep{hoeflerEarthVirtualizationEnginesTechnical2023}. This evolution is supported by computational advances \citep{bauerQuietRevolutionNumerical2015} and the deployment of machine learning (ML) methods \citep{deburgh-dayMachineLearningNumerical2023,beuclerMachineLearningClouds2023,molinaReviewRecentEmerging2023,eyringPushingFrontiersClimate2024a}. However, finer resolutions are accompanied by a growing number of grid cells, prognostic variables and time steps, leading to a drastic increase in the amount of data that needs to be stored. This issue has been addressed in recent studies on compression methods for Earth system model (ESM) and climate data \citep{bakerAMethodologyEvaluating2014,bakerEvaluatingLossyData2016,huang2025errorboundedcompressionweather}, identifying data output and storage as a major bottleneck for high-resolution climate modeling. This underscores the need to move beyond fixed archives toward on-demand access to user-relevant Earth observation and ESM information from compressed \citep{gomesLossyNeuralCompression2025} or streamed \citep{grayson2025ESM_data_streaming} data; in that context, reconstructing missing variables from routinely saved or streamed ones, which remains challenging, is likely to become increasingly important.

Within numerical weather prediction and climate models, processes at scales smaller than the model's grid spacing need to be represented empirically using parameterizations. The representation of cloud microphysics poses a distinct challenge, which stems from incomplete understanding, especially of ice- and mixed-phase clouds \citep{boucherClimateChange2014,beuclerMachineLearningClouds2023}. Cloud microphysics describes the interactions of cloud and precipitation particles and the phase transformations of condensed water in the atmosphere, with numerous cloud particles of varying shapes and sizes interacting through multiple nonlinear processes over a wide range of spatio-temporal scales. In the ICOsahedral Nonhydrostatic (ICON) model, microphysical processes are typically parameterized with a bulk moment parameterization. Microphysical process rates are computed at interim steps to update the prognostic cloud variables. The one-moment microphysics scheme \citep{baldaufOperationalConvectiveScaleNumerical2011} predicts the mass mixing ratios of cloud water, rain water, cloud ice and snow and is used operationally \citep{prillWorkingWithTheICONModel2024}, whereas the two-moment microphysics scheme \citep{seifertbehengTwomomentCloudMicrophysics2006} includes graupel and hail and solves the prognostic equations for the mass mixing ratios and number concentrations. Thus, the two-moment scheme allows for a more detailed representation of cloud microphysics, while maintaining the computational efficiency of a bulk parameterization scheme. Currently, this scheme is used operationally for nowcasting within the Rapid Update Cycle ICON-D2-RUC \citep{dwd2025ICONRUC,reinert2025DWDDatabase}.

ML has shown great potential to improve our ability to model and study sub-grid scale processes \citep{gentineDeepLearningParametrization2021}, including the parameterization of cloud microphysics \citep{lamb2025perspective_cloud_microphysics}. By replacing an existing parameterization with a data-driven counterpart, trained on data from simulations with a bin microphysics scheme \citep{gettelmanMachineLearningWarmRainProcess2021}, Lagrangian super-droplet simulations \citep{seifertPotentialLimitationsMachine2020,seifertMLBasedP3LikeMultimodal2024,sharmaSuperdropnetStableAccurate2025} or high-resolution simulations with a one-moment microphysics scheme \citep{sarauerAphysicsinformedMachineLearning2025}, these approaches aim to improve accuracy and speed up simulations. However, their potential to improve microphysical process understanding remains underexplored.

Knowledge of microphysical processes is needed for advancing our understanding of cloud microphysics and to inform parameterization improvements. Motivated by the lack of archived microphysical process rate data, \citet{kiszlerMicrophysicalProcessesInvolving2024} developed a microphysics parameterization wrapper for the study of Arctic low-level clouds, a diagnostic tool that runs the two-moment microphysics scheme offline from previously saved model output of the ICON-LEM model (ICON model in the large-eddy version). This wrapper was used to analyze several months of cloud-resolving ICON simulations over Svalbard and study the phase partitioning in Arctic low-level mixed-phase clouds. While this approach is an important step toward process-oriented diagnostics, it relies on the availability of several days of the required model output and access to the underlying microphysics code. Here, we address the problem differently: rather than replaying the microphysics scheme offline, we ask whether process rates can be directly recovered from routinely saved variables using ML. This would broaden access to process-oriented diagnostics \citep{maloneyProcessOrientedEvaluationClimate2019}, which are currently limited to small-scale and often idealized simulations \citep{barthlottAerosolEffectsClouds2018,heikenfeldAerosolEffectsDeep2019,barthlottImpactAerosolsModel2024,barrettHooseMicrophysicalPathwaysActive2023}. 

To facilitate the study of cloud microphysics, we introduce a data-driven approach to recover microphysical process rates from high-resolution simulations of weather and climate. The purpose of our study is to assess the feasibility of this approach and to build the foundation for more detailed studies of cloud microphysics with regional or global storm-resolving model simulations, such as those conducted within the DYnamics of the Atmospheric general circulation Modeled On Non-hydrostatic Domains (DYAMOND) \citep{stevensDYAMONDDynamicsAtmospheric2019} or nextGEMS \citep{seguraNextgemsEnteringEra2025,nextGEMSCatalogsFull2024} project \citep[e.g., ][]{klocke2025computingearth1km} or in \citet{preinSingleStormsGlobal2026}. Here, we investigate microphysical process rates corresponding to the two-moment microphysics parameterization scheme \citep{seifertbehengTwomomentCloudMicrophysics2006}. We aim to recover process rates describing warm-rain and ice microphysical processes with a two-step ML-based classification-regression framework. To generate datasets for training, validation and testing, we conduct high-resolution ICON model simulations in a limited-area configuration over Europe with output time steps ranging from one to 60 minutes. Additionally, we evaluate whether the models are transferable to other spatial domains and model setups.

To quantify predictive uncertainty, we employ conformal prediction (CP) \citep{vovkLeaningByTransduction1998,vovkAlgorithmicLearningRandom2022,angelopoulosGentleIntroductionConformal2022} which allows us to complement the deterministic predictions with prediction intervals. Conformal prediction is a framework for constructing statistically valid prediction intervals for any underlying classification or regression model, under the assumption of data exchangeability. Through calibration with a separate calibration dataset, we obtain prediction intervals that are designed to contain the true value of the target variable with a pre-defined probability, e.g. 90\%. While uncertainty quantification methods, such as ensemble approaches \citep[e.g.,][]{behrensSimulatingAtmosphericProcesses2025, shinParameterizationStochasticallyEntraining2022, gagneMachineLearningStochastic2020} are often complex and computationally demanding, CP methods are computationally inexpensive and require no model modifications. \citet{simm2026calibratedconformalpredictionintervals} compared two popular CP methods, split conformal prediction and conformalized quantile regression, with respect to their capability to provide prediction intervals for microphysical process rates. Conformalized quantile regression (CQR) \citep{angelopoulosGentleIntroductionConformal2022,romanoConformalizedQuantileRegression2019} is a variant of CP that better adapts to heteroscedasticity but requires training a quantile regression model. While both methods produced well-calibrated prediction intervals on average, it was demonstrated that CQR is the preferable method for process rates that span many orders of magnitude. In this work, we therefore implement CQR. Furthermore, we round the interval bounds to nearest order-of-magnitude bounds in order to improve interpretability and empirical coverage.

This paper is organized as follows. Section~\ref{sec:data} introduces the ICON model simulations and the data pre-processing, including a review of the two-moment microphysics scheme. In Sect.~\ref{sec:methods}, we outline the methods and the training procedure, present the ML setup and describe how conformal prediction is used to obtain prediction intervals. In Sect.~\ref{sec:results}, we present our results, before discussing them in Sect.~\ref{sec:conclusions} and concluding.
\section{Data}\label{sec:data}
We first review the parameterization of microphysical processes within the two-moment microphysics scheme in the ICON model in Sect.~\ref{sec:two_moment_scheme}, followed by a brief description of the relevant microphysical processes. In Sect.~\ref{sec:model_simulations}, we provide details of the ICON model simulations performed to obtain the data used for training, validation and testing. We outline the sampling and pre-processing procedure in Sect.~\ref{sec:sampling} and Sect.~\ref{sec:preprocessing}, respectively. In Sect.~\ref{sec:transferability_other_regions}, we introduce the case studies and the corresponding simulation setups.

\subsection{The two-moment bulk microphysics scheme in ICON}\label{sec:two_moment_scheme}
The two-moment microphysics scheme \citep{seifertbehengTwomomentCloudMicrophysics2006} as implemented in the ICON model \citep{zanglICONICOsahedralNonhydrostatic2015} parameterizes cloud microphysical processes with six hydrometeor categories: cloud water, rain water, cloud ice, snow, graupel and hail. 
Cloud droplets and raindrops are both liquid hydrometeors, which are distinguished by their mass to account for their different collisional behavior. Cloud droplets and raindrops are defined to have masses $x_\text{cloud droplet} < x^\ast = \unit{2.5 \times 10^{-10}\,kg}$ and $x_\text{raindrop} \geq x^\ast$.
The formulation of the bulk microphysics scheme is based on the quasi-stochastic collection equation \citep{zanglICONICOsahedralNonhydrostatic2015, pruppacherklettMicrophysicsCloudPrecipitation2010chapter11}. 
In its original form, it describes the time evolution of the dimensionless size distribution function $f(x)$ for particles of (dimensionless) masses $x$ and $y$ undergoing binary collisions
\begin{equation}
\label{eq:sce}
    \frac{\partial f (x, t)}{\partial t} = \frac{1}{2} \int_0^x f(x-y, t) f(y, t) K(x-y, y) \, \mathrm{d} y - f(x,t)\int_0^\infty f(y,t) K(x, y) \, \mathrm{d}y \, , 
\end{equation}
where $f(x)$ denotes the dimensionless concentration of droplets in the mass range $x \in [x, x+\mathrm{d}x]$ and 
the kernel function $K(x,y)$ is a non-negative symmetric function.

The two-moment microphysics scheme aims to describe the particle size distribution (PSD) in terms of two of its partial moments $M$
\begin{equation}\label{eq:partial_moments}
\begin{split}
    &n_k = M^{(0)}_k = \int_0^\infty f_k(x) \, \mathrm{d} x \, , \\
    &q_k \propto M^{(1)}_k = \int_0^\infty x f_k(x) \, \mathrm{d} x \, ,
\end{split}
\end{equation}
which correspond to the number concentration $n_k$ and mass mixing ratio $q_k$ for each hydrometeor category $k$. A fundamental assumption of any bulk microphysics scheme is that the PSD of each hydrometeor category can be described by an analytic functional form which is completely determined by only a few parameters. A common choice is a gamma distribution
\begin{equation*}
    f_k(x)= N_0 x^\mu \mathrm{e}^{- \lambda x^\nu} \, ,
\end{equation*}
where $x$ is the particle mass, $N_0$ is the intercept, $\lambda$ the slope and $\mu$ the shape parameter. Here, we use $\mu = 1$ and $\nu = 1$. From Eq.~\eqref{eq:sce} and Eq.~\eqref{eq:partial_moments}, we can derive equations for the microphysical process rates. For a comprehensive description, we refer the reader to \citet{seifertbehengDoublemomentParameterization2001, seifertbehengTwomomentCloudMicrophysics2006}.

In ICON, the two-moment microphysics scheme is implemented as a set of subroutines that are called sequentially. Within each subroutine, the microphysical process rates $\partial_t q_k^X$ and $\partial_t n_k^X$ for each microphysical process $X$ are computed. The hydrometeor mass mixing ratios $q_k$ and number concentrations $n_k$ are prognostic variables, which are updated accordingly at each fast-physics time step $t^\text{fast}$. Here, the index $k \in \{c, \,r, \, i, \,s, \,g, \,h\}$ refers to the cloud droplet, raindrop, ice, snow, graupel and hail category, respectively. Consequently, the microphysics parameterization has the same time step as the fast-physics time step. All $n_k$ and $q_k$ are set to be non-negative before and after the microphysical process rates are computed. Accordingly, minimum and maximum values for the mean particle mass are specified for each hydrometeor category which translate to limits on the number density. If not indicated otherwise, the process rates are strictly non-negative. In the following, we briefly describe the microphysical processes studied in this work\footnote{Our naming convention for the abbreviations of the process rates is as follows: Each name consists of two parts, separated by an underscore. The first part refers to the hydrometeor category that is affected by the process, e.g. ``QC'' for cloud droplets. Here, ``QX'' indicates the mass mixing ratio and ``QNX'' the number concentration. The second part refers to the process itself, e.g. ``AC'' for autoconversion or ``DEP'' for deposition.}. All process rates (targets) are listed in Table~\ref{table:microphysical_process_rates}.

\subsubsection*{Condensation and evaporation}
In conditions with low supersaturation with respect to liquid water, cloud droplets grow by condensation of water vapor onto the droplet's surface; evaporation describes the reverse process. In the ICON model, the \textit{saturation adjustment} establishes thermodynamic equilibrium between water vapor and liquid water by condensation and evaporation of cloud droplets.  Temperature and water vapor mixing ratio are adjusted to establish saturation with respect to liquid water. The saturation adjustment is called twice, before and after the microphysical processes, to ensure that the subsequent slow physics processes are started with an equilibrated state.
Evaporation of rain (QR\_EVAP) occurs in sub-saturated conditions\footnote{Note that process rates describing condensation and evaporation of cloud droplets are not included here.}. 

\subsubsection*{Autoconversion, accretion and self-collection}
Autoconversion, accretion and self-collection constitute the warm-rain processes. 
Autoconversion (QR\_AC) describes the formation of raindrops from coagulating cloud droplets, whereas accretion (QR\_ACC) describes the increase in rain water mass due to the collection of cloud droplets by raindrops. Self-collection of cloud droplets (QNC\_SC) and raindrops (QNR\_SC) refers to the coagulation of the respective hydrometeors and describes a change in the corresponding number concentration. The cloud droplet self-collection rate is strictly non-positive as the number of cloud droplets can only decrease in this process. However, the raindrop self-collection rate can take both positive and negative values as the number of raindrops can increase due to breakup, which is not implemented as a separate process in our version of the microphysics parameterization. Note that the process rates for self-collection are the only two process rates that describe a change in the number concentration of hydrometeors that are discussed here.

\subsubsection*{Homogeneous and heterogeneous ice nucleation}
At temperatures $T \lesssim -38^\circ\,\unit{C}$ liquid droplets freeze homogeneously (QI\_HOM) \citep{hoosemoehlerHeterogeneousIceNucleation2012}. At higher temperatures, freezing depends on the availability of insoluble aerosols which act as ice nucleating particles (INPs). This process is called heterogeneous ice nucleation (QI\_HET) and describes immersion freezing (the freezing of a supercooled liquid droplet with an INP immersed within) and deposition nucleation (ice is formed directly by deposition of water vapor onto the surface of an aerosol).

\subsubsection*{Vapor deposition and sublimation}
In non-equilibrium conditions with sometimes large supersaturation, frozen hydrometeors grow by deposition of water vapor, i.e. ice particles (QI\_DEP), snow (QS\_DEP), graupel (QG\_DEP) and hail (QH\_DEP). Sublimation describes the reverse process in sub-saturated regions. In the implementation of the two-moment microphysics scheme, the deposition rates include both deposition and sublimation. Hence, the values can be positive and negative.

\subsubsection*{Riming}
Riming is the collection of supercooled liquid particles by frozen hydrometeors, a process which occurs in mixed phase clouds at $-40^\circ\,\unit{C} \lesssim T < 0^\circ \,\unit{C}$ \citep{pruppacherklettMicrophysicsCloudPrecipitation2010}. Here, we consider the riming of ice with cloud droplets (QC\_RIME\_I) and raindrops (QR\_RIME\_I), the riming of snow with cloud droplets (QC\_RIME\_S) and raindrops (QR\_RIME\_S) and the riming of graupel or hail with cloud droplets (QC\_RIME\_GH) and raindrops (QR\_RIME\_GH). Additionally, we consider the total riming (QX\_RIMING). 
\subsubsection*{Freezing of rain}
These processes describe the freezing of rain to snow, graupel and hail (QR\_RF) at temperatures
$-40^\circ\,\unit{C} \lesssim T < 0^\circ \,\unit{C}$. Additionally, we consider the freezing of rain to ice (QI\_RF), to graupel (QG\_RF) and to hail (QH\_RF) individually. Depending on their size and by partial integration of the spectrum, the frozen raindrops are categorized as cloud ice, graupel or hail.

\subsubsection*{Evaporation and melting of hydrometeors}
At temperatures $T > 0^\circ\,\unit{C}$, all frozen hydrometeors melt. Here, we consider the melting of all types of frozen hydrometeors to rain (QR\_MELT). If sufficiently small, ice particles melt to cloud droplets (QC\_MELT). Moreover, at $T > 0^\circ\,\unit{C}$, the liquid surfaces of melting snow, graupel and hail can evaporate (QSGH\_EVAP).

\subsection{ICON model simulations}\label{sec:model_simulations}
We employ the ICON modeling framework \citep{zanglICONICOsahedralNonhydrostatic2015} with the NWP physics package \citep{prillWorkingWithTheICONModel2024} in a limited-area setup. The horizontal grid is based on an icosahedral structure with triangular grid cells where the variables are located at the cell circumcenter. Here, we use the regional $\text{R}19\text{B}07$ grid\footnote{An $\text{R}n\text{B}k$ grid with  $n, k \in \mathbb{N}$ is defined by $n-1$ divisions of the edges of the base icosahedron and $k$ subsequent edge bisections. For further details we refer the reader to \citet[Sect.~2.1]{prillWorkingWithTheICONModel2024}.} with $\overline{\Delta x} \approx 2$\,\unit{km} effective horizontal grid spacing, corresponding to the operational ICON-D2 configuration used by the German Weather Service (Deutscher Wetterdienst, DWD). The domain extends from $-0.4^\circ\,$E to $17.7^\circ\,$E and from $43.7^\circ\,$N to $57.26^\circ\,$N and contains $538\,164$ grid cells. The vertical dimension is described by a height based terrain-following smooth level vertical (SLEVE) coordinate system \citep{leuenbergerAGeneralizationSLEVEVerticalCoordinate2010} with 65 height levels, which span from \unit{10\,m} to \unit{22\,km} above ground. Deep convection is explicitly resolved, while shallow convection is parameterized with a bulk mass flux convection scheme based on the Tiedtke-Bechtold convection scheme \citep{bechtoldAdvancesSimulatingAtmospheric2008, tiedtkeAComprehensiveMassFluxScheme1989}. The parameterization of homogeneous and heterogeneous ice nucleation follows that of \citet{handeSeasonalVariabilitySaharan2015} and \citet{karcherParameterizationCirrusCloud2002,karcherPhysicallyBasedParameterizationCirrus2006}. Radiation is described by the ecRad scheme \citep{bozzohoganAFlexibleEfficientRadiationScheme2018}. Beyond that, we apply the parameterization schemes that are in operational use, namely the parameterization of non-orographic gravity wave drag \citep{orrImprovedMiddleAtmosphereClimate2010}, sub-grid scale orographic drag \citep{lottmillerNewSubgridscaleOrographicDragParametrization1997}, diagnostic cloud cover, and turbulence \citep{raschendorferNewTurbulenceParameterization2001}. The internal fast-physics time step is set to $t^\text{fast} = \unit{20\,s}$. The slow physics processes are called at a reduced frequency, the radiation scheme with a time step $t^\text{slow, rad} = \unit{720\,s}$ and the convection, sub-grid and non-orographic gravity wave drag parameterization schemes all with a time step $t^\text{slow} = \unit{120\,s}$. For computational efficiency, the radiation scheme is computed on a coarser horizontal $\text{R}19\text{B}06$ grid with $\overline{\Delta x} \approx 4$\,\unit{km} effective horizontal grid spacing.
The model configuration is summarized in Table~\ref{table:icon_model_summary}.
\begin{table*}[t]
    \centering
    \caption{Model configuration of the ICON model simulations.}
    \label{table:icon_model_summary}
    \begin{tabular}{l l}
        \tophline
            {Model aspect} & {Setting} \\
        \middlehline
            Initial and boundary data & ICON-EU analyses, \unit{3\,h} update \\
            Initialization time        & 00:00 UTC \\
            Integration time          & \unit{24\,hours} \\
            Turbulence scheme & Prognostic TKE \citep{raschendorferNewTurbulenceParameterization2001} \\
            Microphysics scheme & Two-moment microphysics \citep{seifertbehengTwomomentCloudMicrophysics2006} \\ 
            Convection scheme & Explicit deep convection, parameterized shallow convection \\
            {} &  \citep{bechtoldAdvancesSimulatingAtmospheric2008, tiedtkeAComprehensiveMassFluxScheme1989} \\
            Cloud condensation nuclei activation &  Segal-Khain scheme \citep{segalDependenceDropletConcentration2006} \\ 
            Homogeneous and heterogeneous ice nucleation & \citet{karcherParameterizationCirrusCloud2002,karcherPhysicallyBasedParameterizationCirrus2006,handeSeasonalVariabilitySaharan2015} \\
            Land-surface model & Multilayer land-surface scheme TERRA \citep{schrodinheiseLandSurfaceModel2001} \\
            Radiation scheme & ecRad \citep{bozzohoganAFlexibleEfficientRadiationScheme2018}\\
        \bottomhline
    \end{tabular}
\end{table*}
We perform simulations for a total of 19 days, i.e. one day per month from January 2022 to July 2023, with an output time step of \unit{1\,min}, \unit{2\,min}, \unit{5\,min}, \unit{10\,min}, \unit{30\,min} and \unit{60\,min}. The simulations are initialized with ICON-EU analyses at 00:00 UTC and run for 24 hours. In order to obtain the process rates to be used during training from the ICON model simulations, we added these as additional output variables. The process rates are updated at the end of the corresponding ICON subroutine(s). In our setup, the full output of the ICON model simulations for each time step consists of the 25 process rates $\partial_t q_k^X$ and $\partial_t n_k^X$, $[\partial_t q_k^X] = \unit{kg\,kg^{-1}\,t^{-1}}$ and $[\partial_t n_k^X] = \unit{kg^{-1}\,t^{-1}}$, where \unit{t} is $t^\text{out}[\unit{s}]$ for the accumulated process rates and $t^\text{fast}[\unit{s}]$ for the instantaneous rates, as well as the hydrometeor mass mixing ratios $q_k$, $[q_k] = \unit{kg\,kg^{-1}}$, and the number concentrations $n_k$, $[n_k] = \unit{kg^{-1}}$. Additionally, $q_v$ is the specific humidity, $[q_v] = \unit{kg\,kg^{-1}}$, $T$ the temperature, $[T] = \unit{K}$, $p$ the pressure, $[p] = \unit{Pa}$, and $\rho$ the density, $[\rho] = \unit{kg\,m^{-3}}$, which we include in the output as well. The output variables are defined for each grid cell, height level and output time step. A list of all output variables and the process rates together with their respective dimension is given in Table~\ref{table:ICON_output_variables} and \ref{table:microphysical_process_rates}, respectively. Note that there is a temporal mismatch between the output of the process rates and the output of the remaining cloud variables from one common time step due to the processing sequence in ICON. As described in Sect.~\ref{sec:two_moment_scheme}, the subroutines for the different microphysical processes are called sequentially, followed by a second call to the saturation adjustment. This modifies the values of the output variables further down the processing sequence. Consequently, a set of output variables for one time step contains values of cloud variables that can be different from the values used for the computation of the process rates in the same set. This is also the reason why the direct recalculation of the process rates from the model output is generally not accurate.

\begin{figure*}[t]
\centering
\label{fig:fig1}
\includegraphics[width=12cm]{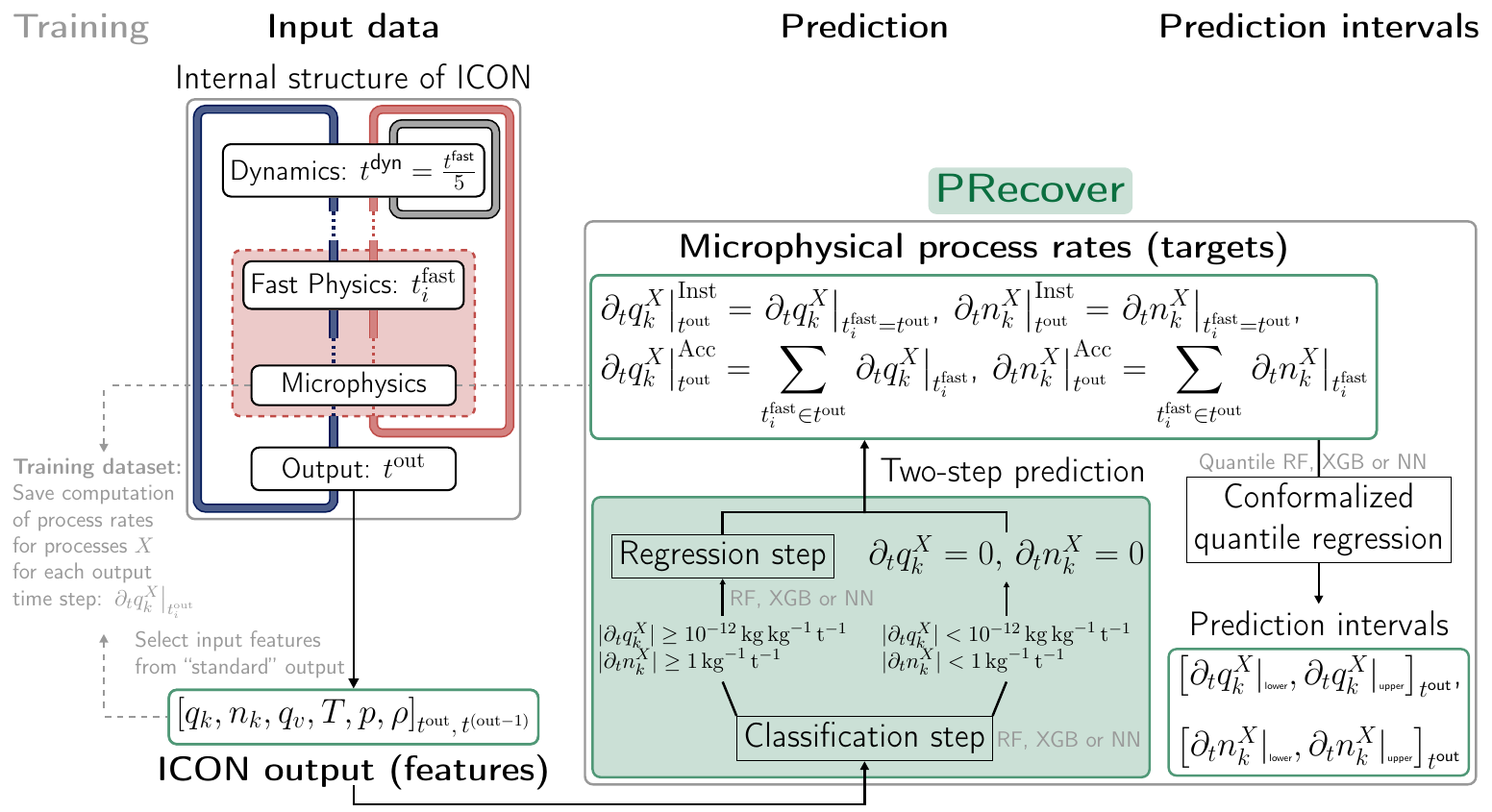}
\caption{
Overview of PRecover. Left: ICON advances the dynamical core, fast-physics package, and model output with $t^\mathrm{dyn}=t^\mathrm{fast}/5$, $t^\mathrm{fast}$, and $t^\mathrm{out}$, respectively. At each fast-physics step, it computes the microphysical process rates $\partial_t q_k^X$ and $\partial_t n_k^X$, stored alongside standard ICON variables to construct the training, validation, and test datasets. Right: PRecover predicts instantaneous and accumulated process rates from ICON output with a two-step classification--regression model based on Eq.~\eqref{eq:input_parameters} and Eq.~\eqref{eq:input_parameters_accumulated}, and augments these predictions with CQR-based prediction intervals to quantify uncertainty. The ICON schematic is adapted from \citet[Fig.~3.1]{prillWorkingWithTheICONModel2024}.}
\label{fig:data_generation_flowchart}
\end{figure*}

We distinguish two cases. In the first case, which we refer to as the \textit{instantaneous} case, the output values of the microphysical process rates are the result of the last internal computation with  the fast-physics time step, 
\begin{equation}\label{eq:process_rates_instantaneous}
\begin{split}
    \partial_t q_k^X\lvert_{t^\text{out}}^\text{Inst} &= \partial_t q_k^X\lvert_{t_i^\text{fast}=t^\text{out}} \, , \\
    \partial_t n_k^X\lvert_{t^\text{out}}^\text{Inst} &= \partial_t n_k^X\lvert_{t_i^\text{fast}=t^\text{out}} \, .
\end{split}
\end{equation}
The corresponding process rates are referred to as \textit{instantaneous process rates}. In the second, \textit{accumulated} case, the computation for each $t^\text{fast} = \unit{20\,s}$ is summed over the model output time step,
\begin{equation}\label{eq:process_rates_accumulated}
\begin{split}
    \partial_t q_k^X\lvert_{t^\text{out}}^\text{Acc} &= \sum_{t_i^\text{fast}\in t^\text{out}}\partial_t q_k^X\lvert_{t_i^\text{fast}} \, ,\\
    \partial_t n_k^X\lvert_{t^\text{out}}^\text{Acc} &= \sum_{t_i^\text{fast}\in t^\text{out}}\partial_t n_k^X\lvert_{t_i^\text{fast}}
    \, .
\end{split}
\end{equation}
Accordingly, the process rates are referred to as \textit{accumulated process rates}. For instance, the accumulated process rates for the one-minute output time step are calculated as the sum of three internal computations of the corresponding process rate. The data generation from ICON and the general framework of our proposed method is visualized in Fig.~\ref{fig:data_generation_flowchart}.
Note that all other output variables are instantaneous quantities. The simulation output from the ICON model used in this study has not been remapped to a regular grid and is defined on the original ICON Arakawa C grid \citep{zanglICONICOsahedralNonhydrostatic2015}. We find that using remapped data drastically degrades the model performance for some process rates, likely because remapping the output is an additional processing step which modulates the relation between the process rates and the state variables in the model output.

\subsection{Sampling procedure}\label{sec:sampling}
We train directly on the output of the ICON model simulations. For this purpose, we split the output data from the simulated 19 days by day, i.e. the days from February, April, June, August, October and December 2022 (even-numbered months) are used in the training dataset, the days from January, March, May, July, September and November 2022 (odd-numbered months) are used in the validation dataset and the days from January to July 2023 are used in the test dataset. In order to obtain a representative set of training, validation and test data, we select days with high amounts of precipitation in different regions. This yields a total of $n_\text{tot} \approx 7.75 \times 10^8$ available samples (number of time steps $\times$ number of grid cells) of each process rate per day for the shortest output time step of one minute. To reduce the size of the datasets we use only a subset containing two-hourly intervals of the simulation output for each day, as listed in Table~\ref{table:full_training_data_overview}. Furthermore, we perform a random sampling of the pre-selected data. The two-hourly intervals are selected such that the distribution of the values of each process rate in the training, validation and test datasets resembles the distribution of the values of the corresponding process rate in the original model output (see Fig.~\ref{fig:histograms_data_distribution_1}\nobreakdash--\ref{fig:histograms_data_distribution_3}). Additionally, we discard the output from very high altitudes where we do not expect clouds to be present, i.e. we discard level numbers 1 to 15 ($\gtrsim\unit{10\,km}$) for QC\_RIME\_\{I, S, GH\}, Q\{R, I, G, H\}\_RF, QC\_MELT and QR\_MELT,  level numbers 1 to 9 ($\gtrsim\unit{13\,km}$) for QI\_HOM and QI\_HET and level numbers 1 to 22 ($\gtrsim\unit{7.2\,km}$) for the remaining process rates. The set of input features from the ICON model output for the prediction of each process rate can be found in Table~\ref{table:features_models}. Their choice is based on the corresponding subroutine(s) in the ICON model code in which the process rates are computed. Specifically, we include all variables that either appear in the computation of the process rates or are affected by it\footnote{Note that the computation of heterogeneous ice nucleation also depends on the prognostic variable $n_\text{in,act}$ which describes the number of activated ice nuclei. $n_\text{in,act}$ is not available as model output and therefore not included in the set of input features for QI\_HET.}. For each model, the set of input features is a subset of
\begin{equation}\label{eq:input_parameters}
    \left[q_k,\,n_k, \,q_v, \,T, \,p, \,\rho \right], \quad k \in \left\{c, \,r, \,i, \,s, \,g, \,h \right\} \, .
\end{equation}

Our approach is fully local (grid-cell based), i.e. the set of input features includes data only from the current grid cell. While incorporating spatial information can hypothetically improve predictiveness, this choice of locality is based on the ICON model code, where the microphysical processes are computed for each grid cell independently. In principle, this also applies to the temporal dimension. However, for the models for the accumulated process rates, we add the values of the cloud variables $q_k$ and $n_k$ from the previous time step to the set of input features, which we denote by $q_k^{t-1}$ and $n_k^{t-1}$, so that the set of input features for the accumulated process rates is a subset of 
\begin{equation}
    \label{eq:input_parameters_accumulated}
    \left[q_k, \,q_k^{t-1}, \,n_k, \,n_k^{t-1}, \,q_v, \,T, \,p, \,\rho \right], \quad k \in \left\{c, \,r, \,i, \,s, \,g, \,h \right\} \, .
\end{equation}
Note that in this case it is not possible to predict the process rates for the first time step of the input dataset.

Based on the assumption that an absence of hydrometeors implies an absence of clouds and thus a vanishing process rate, we further reduce the size of the datasets by removing all data points where the sum of the hydrometeor mass content present in the set of input features from the current time step is smaller than a pre-defined threshold
\begin{equation}
    \label{eq:threshold_qx} 
    \sum_k \left|q_k\right|  = 10^{-12}\, \unit{kg\,kg^{-1}} \, .
\end{equation}
In the ICON model code, an equivalent threshold in the two-moment microphysics scheme is $q_\text{crit} = 10^{-12}\, \unit{kg\,kg^{-1}}$.
For the accumulated process rates, we apply the same constraint for the sum of the hydrometeor mass content from both the current and the previous time step. We find that this constraint is not satisfied for $0.7\%$ of all data points on average for the shortest output time step of one minute and for $2.7\%$ for the 30-minute output time step (computed over the values of all available process rates). This is most likely a consequence of the temporal mismatch between the process rates and the cloud variables in the model output, as described above. Despite that, we find that imposing this constraint is advantageous both for the performance of the classification and regression models as well as for keeping the size of the datasets manageable. Furthermore, with regard to the datasets used for training the regression models, we apply the constraint
\begin{equation}\label{eq:threshold_regression}
\begin{split}
    &\left|\partial_t q_k^X\right| \geq 10^{-12} \, \unit{kg\,kg^{-1}\,t^{-1}} \, , \\ &\left|\partial_t n_k^X\right| \geq 1 \, \unit{kg^{-1}\,t^{-1}} \, ,
\end{split}
\end{equation}
where $\unit{t}^{-1}$ is the inverse of either $t^\text{fast}$ or $t^\text{out}$, depending on whether instantaneous or accumulated process rates are considered.

\subsection{Pre-processing and training/validation/test split}\label{sec:preprocessing}
We obtain a total of $10^7$ samples in the training datasets and $1.25 \times 10^6$ samples in the validation and test datasets, corresponding to a $80\%/10\%/10\%$ (training/validation/test) split. The calibration dataset also contains $1.25 \times 10^6$ samples. We evaluate the performance of the final classification-regression model with a second test set which contains $10^7$ samples. 

The datasets are further preprocessed by scaling the input variables to the range $[0, 1]$ using their minimum and maximum absolute values \textit{(min-max scaling)}. Because the values of the process rates span many orders of magnitude, we take the natural logarithm of the process rates from the ICON model output when training the regression models, i.e. the models are trained to predict the logarithm of the process rates. For negative values, we apply the logarithmic transformation on absolute values\footnote{More details on the treatment of process rates which can take positive and negative values are given in Sect.~\ref{sec:methods}.}.

\subsection{Model evaluation and transferability across regional case studies}\label{sec:transferability_other_regions}
For initial training, validation, and testing, our input data is confined to a specific region, the ICON-D2 domain. To assess model performance across different regional domains, we consider three case studies. The first case, also contained in the test set (see Table~\ref{table:full_training_data_overview}), corresponds to the ICON-D2 domain and focuses on the low-pressure system ``Lambert'' on 23 June 2023. The second and third cases are based on atmospheric measurement campaigns: The International Collaborative Experiments for PyeongChang 2018 Olympic and Paralympic winter games (ICE-POP 2018) campaign on the Korean Peninsula in 2017 and 2018 and the Multidisciplinary drifting Observatory for the Study of Arctic Climate (MOSAiC) campaign in the Arctic in 2019 and 2020. For each case, we apply the models trained on the ICON-D2 data without any re-training.

\subsubsection*{ICON-D2 domain: low-pressure system ``Lambert''}
\begin{figure}[t]
    \centering
    \includegraphics[width=8.0cm]{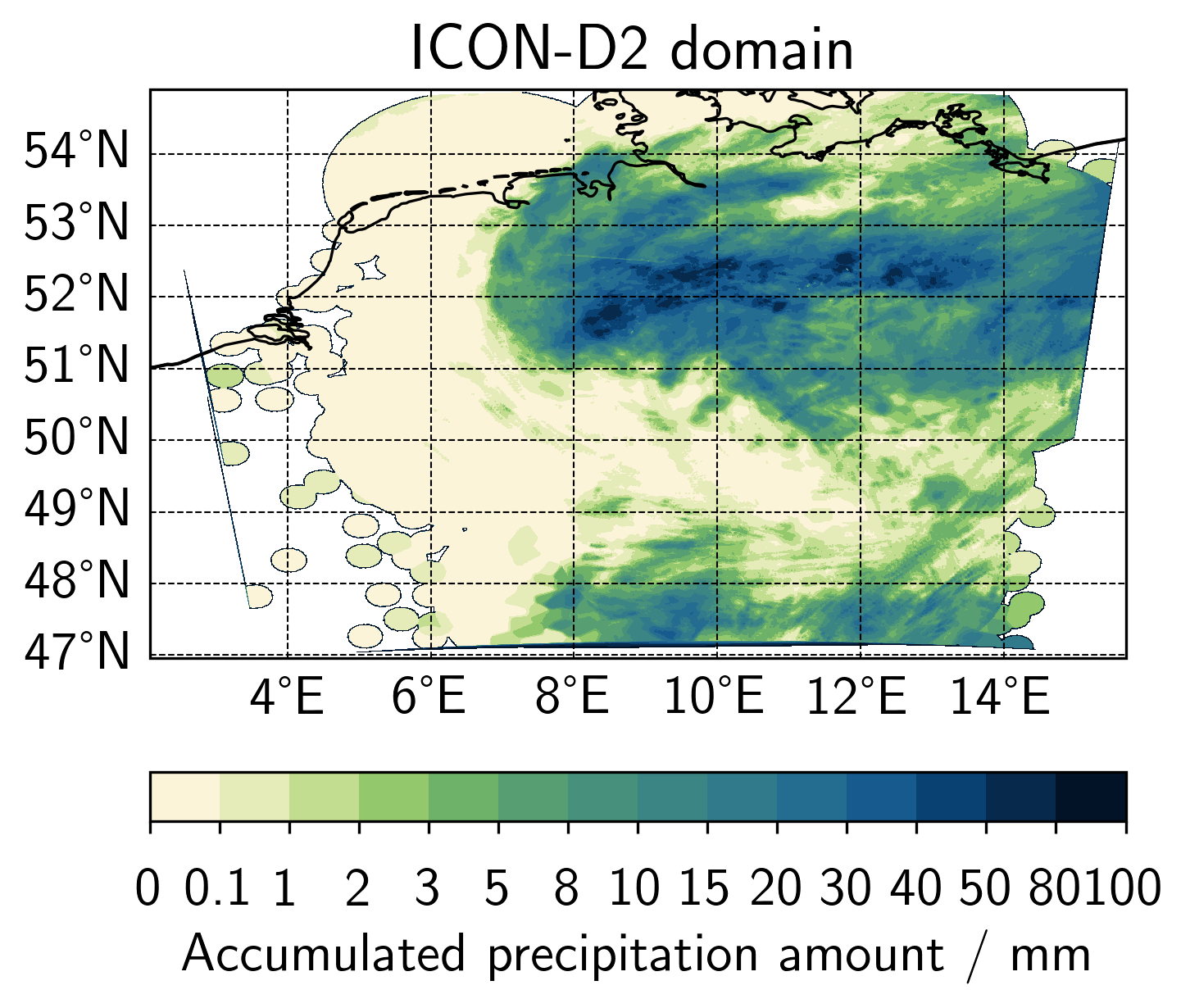}
    \caption{\unit{24\,hour}-accumulated precipitation amount on 23 June 2023 from RADOLAN \citep{bartelsRadolan2004} on the ICON-D2 domain.}
    \label{fig:icond2_radolan}
\end{figure}

From 22 to 23 June 2023, the low-pressure system ``Lambert'' passed eastward from France across north-western Germany, accompanied by heavy rainfall and, locally, hail, thunderstorms and severe winds. On 23 June 2023, heavy precipitation occurred in central and eastern Germany. In Fig~\ref{fig:icond2_radolan}, we show the observed 24-hour accumulated precipitation from the precipitation analysis algorithm Radar Online Adjustment (RADOLAN) \citep{bartelsRadolan2004}.

The ICON model simulations run for 24 hours from 23 June 2023, 00:00 UTC to 24 June 2023, 00:00 UTC. For further details on the case we refer to \citet[][in german]{dwdsturmtieflambert}.

\subsubsection*{Korean Peninsula: ICE-POP campaign 2017/2018}
The ICE-POP campaign took place from November 2017 to April 2018 over the north-eastern region of the Korean Peninsula, coinciding with the PyeongChang 2018 Olympic and Paralympic Winter Games \citep{gehringMicrophysicsDynamicsSnowfall2020,gehringRadarGroundlevelMeasurements2021,jeoungMicrophysicalPropertiesThree2020,limEvaluationSimulatedWinter2020,kimImpactWindPattern2021}. The main scientific objective was to study precipitation processes over complex terrain, particularly heavy snowfall, with intensive microphysical observations \citep{kimImpactWindPattern2021}.

\begin{figure}[t]
    \centering
    \includegraphics[width=8.cm]{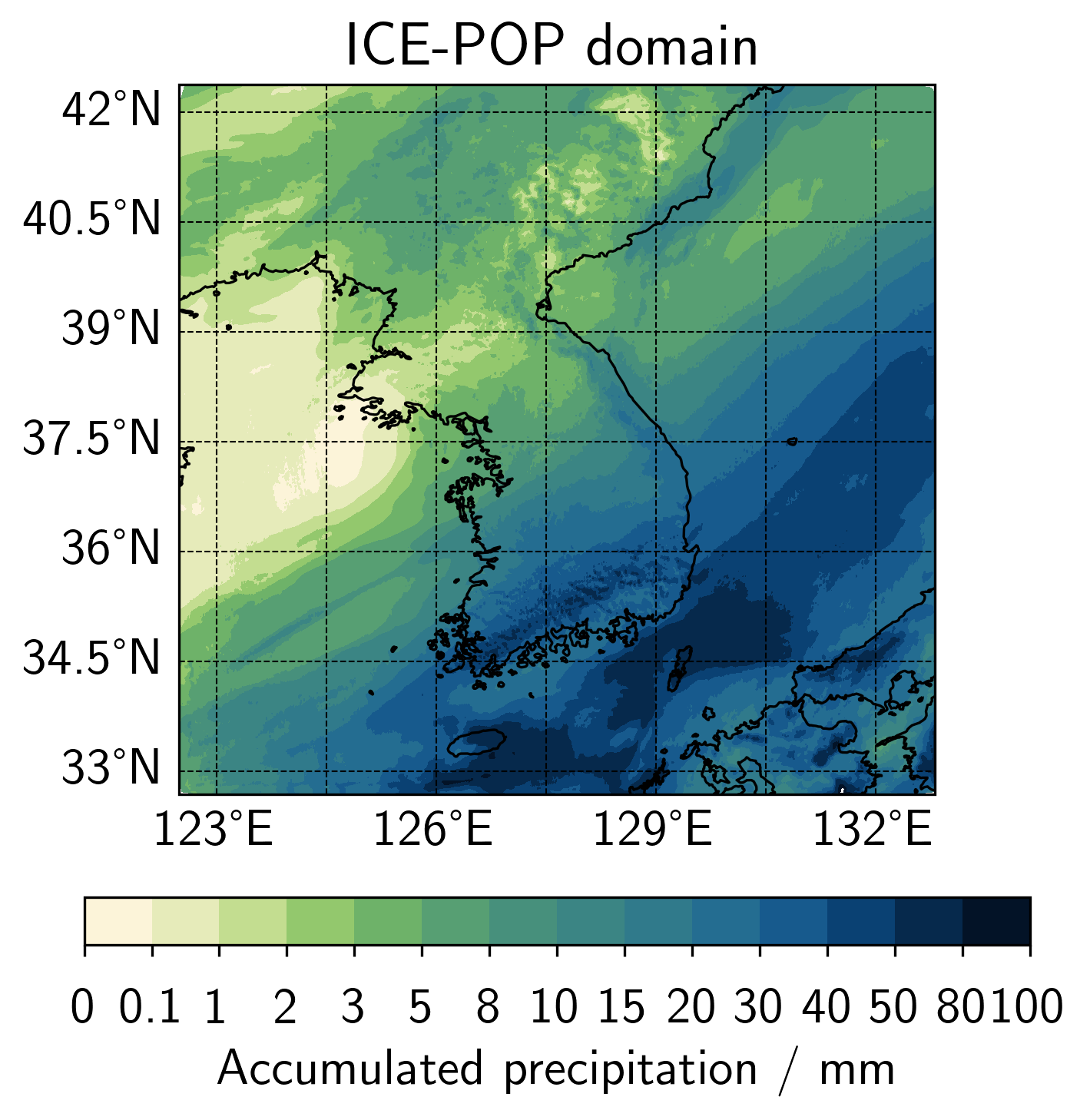}
    \caption{Simulation domain and \unit{24\,hour}-accumulated precipitation amount on 7 March 2018 over the Korean Peninsula for the ICE-POP campaign.}
    \label{fig:icepop_domain}
\end{figure}

Here, we focus on a heavy snowfall event that occurred on 7 and 8 March 2018 \citep[Table~1, Case~6]{koSimulatedMicrophysicalProperties2022}. The event is associated with the passage of a low-pressure system and classified as a ``warm-low'' type, for further details see \citet{kimImpactWindPattern2021}. The simulation is run for 24 hours from 7 March 2018, 00:00 UTC to 8 March 2018, 00:00 UTC. The model domain extends from $122.45^\circ\,$E to $132.81^\circ\,$E and from $32.69^\circ\,$N to $42.37^\circ\,$N and is shown in Fig.~\ref{fig:icepop_domain}. Further details on the simulation case and the domain can be found in \citet{parkIntroducingGraupelDensityPrediction2024, parkAssesingRelativeImportance2026, grzegorczykImplementationPredictedRimeMass2025}. We run the simulations with the modeling setup described in Sect.~\ref{sec:model_simulations} with adjusted initial and boundary conditions.

\subsubsection*{Arctic: MOSAiC campaign 2019/2020}
\begin{figure}[t]
    \centering
    \includegraphics[width=8.3cm]{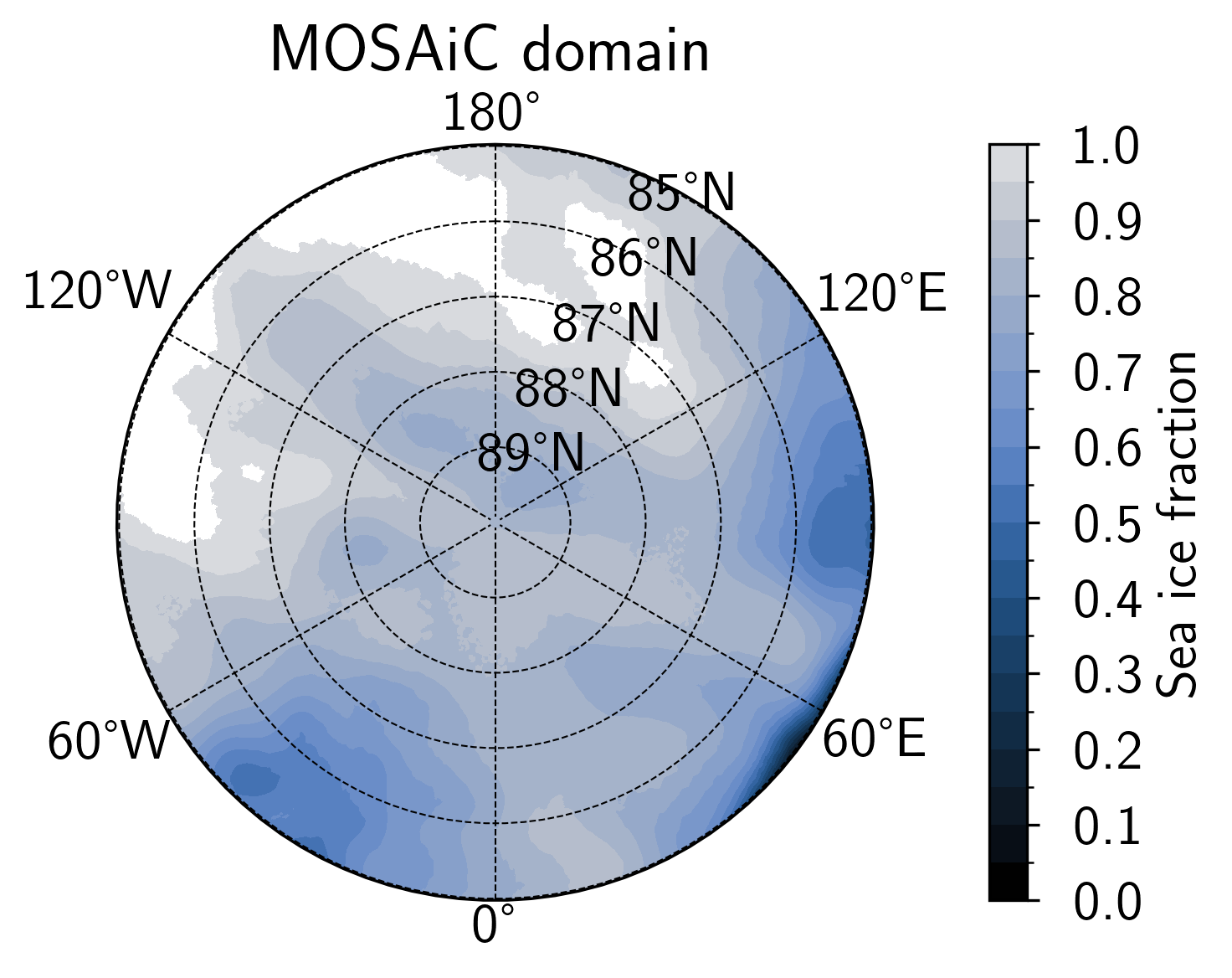}
    \caption{Simulation domain and sea ice fraction on 3 September 2020 over the Arctic for the MOSAiC campaign.}
    \label{fig:mosaic_domain}
\end{figure}

The MOSAiC campaign was a year-long expedition in the central Arctic from September 2019 to October 2020 \citep{shupeOverviewMosaicExpedition2022}. During the campaign, the research icebreaker RV \textit{Polarstern} drifted with the ice across the Arctic Ocean \citep{knustPolarResearchSupply2017}. One of the main focus areas of the campaign was to study Arctic mixed-phase clouds \citep{shupeOverviewMosaicExpedition2022}. We run simulations for 12 hours from 3 September 2020, 04:00 UTC to 3 September 2020, 16:00 UTC. A more detailed description of this case and the model setup can be found in \citet{wallentinSensitivitiesSimulatedMixedphase2025}. Important differences to the model setup used for the simulations with the ICON-D2 and ICE-POP domain are that the fast-physics time step is set to $t^\text{fast} = 12$\unit{\,s}, convection is modeled explicitly, and the grid spacing is set to $\overline{\Delta x} \approx 1.6$\,\unit{km}. We use a circular domain with a radius of $5^\circ\,$ around the North Pole as shown in Fig.~\ref{fig:mosaic_domain}. Using a different ICON model setup allows us to not only study the transferability of the ML models to the Arctic region, but also the sensitivity to the namelist settings for the ICON model run. 

\section{Methods}\label{sec:methods}
In the following, we introduce the framework developed to predict the microphysical process rates (Sect. \ref{sec:twostep}) and outline the hyperparameter tuning procedure (Sect. \ref{sec:ml_algorithms}). In Sect.~\ref{sec:uncertainty_quantification}, we describe how CQR is used to supplement the ML predictions with prediction intervals.

\subsection{Two-step classification-regression approach} \label{sec:twostep}
Following the work of \citet{gettelmanMachineLearningWarmRainProcess2021}, we implement a two-step classification-regression approach to predict the values of the process rates, as depicted in Fig.~\ref{fig:data_generation_flowchart}. We train separate models for each process rate\footnote{That is, we do not construct one single model that predicts the values of all process rates for a given set of input variables simultaneously. Note that this would only be feasible without the two-step approach and without the threshold-based pre-processing (Eq.~\eqref{eq:threshold_qx}). Therefore, it is not explored further in this work.}. This allows us to select a specific set of input variables for each process rate and to tune the models individually. Furthermore, it also allows us to use the best performing ML algorithm for each process rate.

In the first step, a classification model is trained to identify whether the value of the process rate is above or below a pre-defined threshold for each grid cell. This threshold is defined as
\begin{equation}\label{eq:threshold_classification}
\begin{split}
    &\left|\partial_t q_k^X\right| = 10^{-12} \, \unit{kg\,kg^{-1}\,t^{-1}} \, , \\ &\left|\partial_t n_k^X\right| = 1 \, \unit{kg^{-1}\,t^{-1}} \, ,
\end{split}
\end{equation}
corresponding to the constraints in Eq.~\eqref{eq:threshold_regression}. Most of the process rates are strictly non-negative so that the classification models perform binary classification. For process rates which can take positive and negative values, i.e. raindrop self-collection and the deposition rates, we train the models to perform multi-class classification with three classes. In the first and third class, the values of the process rates are larger than the threshold (Eq.~\eqref{eq:threshold_classification}) with positive and negative sign, respectively, whereas in the second class, the values lie in the range $[-10^{-12}, 10^{-12}]$ or $[-1, 1]$. If the classification model predicts that the condition in Eq.~\eqref{eq:threshold_classification} is not satisfied, the value of the corresponding process rate is set to zero. Otherwise, the second step follows, in which a regression model predicts the value of the process rate. Between the classification and the regression step, we re-scale the features with the minimum and maximum values computed during pre-processing (Sect.~\ref{sec:preprocessing}) to correct for the distribution shift in the input features induced by the threshold.

Note that the cloud droplet self-collection rate takes only negative values. In this case, the regression model is trained to predict the absolute value, and the predicted value is multiplied by $-1$ to recover the correct sign at inference time. For process rates that can take both positive and negative values, two regression models are trained on the subset of the training data containing only positive or negative samples. For models trained on the subset containing only the negative samples, we proceed analogously to the approach for cloud droplet self-collection. Thus, depending on the predicted class, either the regression model trained on the subset containing only positive samples or the model trained on the subset with only negative samples is applied to predict the value of the process rate. At inference time, the subsets are combined to obtain the final prediction.

\subsection{Machine learning algorithms and hyperparameter tuning} \label{sec:ml_algorithms}
For each step, we implement three different ML algorithms, namely a random forest (RF) model, a gradient boosting model (XGB) and a feed-forward neural network (NN). Furthermore, we implement logistic regression and linear regression as a baseline for the classification and regression models, respectively. We use the \texttt{scikit-learn} library \citep{pedregosaScikitlearnMachineLearning2011} for linear and logistic regression and for the RF models, the \texttt{xgboost} library \citep{chenXGBoostScalableTree2016} for the gradient boosting models and the \texttt{pytorch} library \citep{paszkePyTorchImperativeStyle2019} for the NNs. We perform a random search \citep{bergstraRandomSearchHyperparameter2012} to identify suitable hyperparameters for each model. An overview of the hyperparameter space is given in Table~\ref{table:hyperparameter_space}. To reduce computational cost, hyperparameter tuning is performed on a random 50\% subsample of the training data. For training the gradient boosting model, we use early stopping with 30 early stopping rounds. We do the same for the NNs with a patience of 15 epochs. We use the Adam optimizer \citep{kingmaAdamMethodStochastic2017} with a learning rate scheduler with a patience of 5 epochs, a decrease factor of 0.5 and a minimum learning rate of $10^{-6}$. The maximum number of epochs is 150. For the regression models, the loss function is treated as a tunable hyperparameter; for classification, we use categorical cross-entropy.

\subsection{Prediction intervals with conformalized quantile regression}\label{sec:uncertainty_quantification}
Following \citet{simm2026calibratedconformalpredictionintervals}, we employ CQR to supplement the deterministic predictions of the regression models with prediction intervals. For this purpose, we train quantile RF (QRF) models using the \texttt{quantile\_forest} library \citep{johnsonQuantileForest2024,meinshausenQuantileRegressionForests2006}, quantile gradient boosting (QXGB) models and quantile neural networks (QNNs). We do not repeat the hyperparameter tuning but use the same set of hyperparameters as for the deterministic regression models. However, for the QNN training, we use the ReLU activation function, double the patience for early stopping from 15 to 30 and use the learning rate scheduler with a patience of 10 epochs and a decrease factor of 0.75 to stabilize the training.

In standard CP, the calibration set is typically obtained by splitting the training dataset. In our framework, however, CP is applied at the end of the two-step classification-regression framework to obtain prediction intervals for the regression outputs. The quantile regression models are therefore trained on the regression training datasets, subject to the constraint (Eq.~\eqref{eq:threshold_regression}). 

To avoid a distribution shift between a calibration dataset and samples encountered at inference time, we construct the calibration set in a different way. A direct split from the training dataset would yield a calibration set that contains only samples with $\left|\partial_t q_k^X\right| \geq 10^{-12} \, \unit{kg\,kg^{-1}\,t^{-1}}$ and $\left|\partial_t n_k^X\right| \geq 1 \, \unit{kg^{-1}\,t^{-1}}$. However, at inference time this constraint is not strictly enforced due to classification errors, which may lead to a different sample distribution. We therefore generate the calibration dataset by first applying the classification model to ICON model output from which the training datasets are sampled (see Sect.~\ref{sec:sampling}) and then selecting $1.25 \times 10^6$ samples where the classification model predicts $\left|\partial_t q_k^X\right| \geq 10^{-12} \, \unit{kg\,kg^{-1}\,t^{-1}}$ and $\left|\partial_t n_k^X\right| \geq 1 \, \unit{kg^{-1}\,t^{-1}}$, according to the threshold in Eq.~\eqref{eq:threshold_classification}.
We set $\alpha = 0.1$ and compute non-conformity scores $E_i$
\begin{equation}\label{eq:nonconf_score_cqr}
    E_i := \max \{\hat{Q}_{\alpha_\text{lo}} (X_i) - Y_i,\, Y_i - \hat{Q}_{\alpha_\text{hi}} (X_i)\} \, ,
\end{equation}
on the calibration set with $n$ samples $\{(X_i, Y_i)\}_{i=1}^n$, where $\hat{Q}_{\alpha_\text{lo}}$ and $\hat{Q}_{\alpha_\text{hi}}$ denote the predicted lower and upper quantile. 

Furthermore, we round the predicted and calibrated interval bounds to the nearest order of magnitude to obtain intervals $\widetilde{\mathcal{C}}(X) = [L_\text{rounded}, U_\text{rounded}]$ with lower and upper interval bounds $L_\text{rounded}$ and $U_\text{rounded}$ computed as
\begin{equation}\label{eq:interval_bounds_cqr_rounded}
\begin{split} 
    L_\text{rounded} &= \begin{cases}
        10^{\lfloor \log_{10}(\hat{Q}_{\alpha_\text{lo}} (X_{n+1}) - Q_{1-\alpha}) \rfloor}, &\text{if } Q_{1-\alpha} < \hat{Q}_{\alpha_\text{lo}} (X_{n+1}) \, ,\\
        0.0, &\text{if } Q_{1-\alpha} \geq \hat{Q}_{\alpha_\text{lo}} (X_{n+1}) \, , 
    \end{cases} \\ 
    U_\text{rounded} &= 10^{\lceil \log (\hat{Q}_{\alpha_\text{hi}}(X_{n+1}) + Q_{1-\alpha}) \rceil}\, ,
\end{split}
\end{equation}
which results in more conservative intervals with increased interval width. Here, $Q_{1-\alpha}$ is the empirical quantile of the non-conformity scores $E_i$.  While this adaptation does not preserve the theoretical conformal coverage guarantee \citep[Theorem 1]{romanoConformalizedQuantileRegression2019}, it improves interpretability and, for our specific application, empirical coverage. Instead of providing information about the exact range of values that the true value most likely falls in, we now provide information about its order of magnitude with empirically calibrated prediction intervals.

\subsection{Computational setup, runtime, and storage requirements}\label{sec:runtime_memory_consumption}
All computations and simulations are carried out on the ``Hochleistungsrechner Karlsruhe'' (HoreKa) high-performance computing system at Karlsruhe Institute of Technology. Each node has two Intel Xeon Platinum 8368 processors with at least 256 GB of local memory. The GPU nodes feature four NVIDIA A100 GPUs per node with 512 GB of local memory. All nodes are interconnected through InfiniBand 4X HDR. The system uses  an IBM Spectrum Scale (GPFS) parallel file system.

The ICON model simulations are performed on 30 CPU nodes. In Table~\ref{table:runtime_icon_model}, we summarize the run-time of the simulations of 23 June 2023 on the ICON-D2 domain for different output time steps with the process rates included in the model output. The runtimes of the simulations of other dates are comparable.
\begin{table*}[t]
    \centering
    \caption{Total model runtime $t_\text{total}$ (excluding initialization and finalization) and time spent in the microphysics scheme $t_\text{2Mom}$ of ICON model simulations of 23 June 2023 on the ICON-D2 domain for simulations with the process rates included in the model output.}
    \label{table:runtime_icon_model}
    \begin{tabular}{c c c c c c c c}
        \tophline
            \multicolumn{2}{c}{Inst., \unit{10\,min}} & \multicolumn{2}{c}{Acc., \unit{1\,min}} & \multicolumn{2}{c}{Acc., \unit{10\,min}} & \multicolumn{2}{c}{Acc., \unit{30\,min}} \\
            \middlehline
            {$t_\text{total}$} & {$t_\text{2Mom}$} & {$t_\text{total}$} & {$t_\text{2Mom}$} & {$t_\text{total}$} & {$t_\text{2Mom}$} & {$t_\text{total}$} & {$t_\text{2Mom}$} \\
            \middlehline
            {\unit{13\,m\,33\,s}} & {\unit{273.835\,s}} & {01\,h\,33\,m} & {\unit{383.593\,s}} & {\unit{13\,m\,46\,s}} & {\unit{280.410\,s}} & {\unit{13\,m\,26\,s}} & {\unit{265.751\,s}}\\
        \bottomhline
    \end{tabular}
\end{table*}
We present a similar comparison of the size of the model output in Table~\ref{table:storage_icon_model}. The output is stored in netCDF4 files, one file for each output time step. Note that already in this case of a limited-area configuration, the total size of the process rate output is larger than that of the standard output variables. For configurations with a larger domain or for longer simulations, it therefore becomes impractical to include the microphysical process rates in the model output.
\begin{table*}[t]
    \centering
    \caption{Total output size of ICON model simulations of 23 June 2023 on the ICON-D2 domain.}
    \label{table:storage_icon_model}
    \begin{tabular}{l c c c c c c c c}
        \tophline
            {} & {Inst., \unit{10\,min}} & {Acc., \unit{1\,min}} & {Acc., \unit{10\,min}} & {Acc., \unit{30\,min}} \\
            \middlehline
            {Output variables} & {\unit{362\,GB}} & {\unit{3.9\,TB}} & {\unit{362\,GB}} & {\unit{123\,GB}} \\
            {Process rates} & {\unit{551\,GB}} & {\unit{5.4\,TB}} & {\unit{551\,GB}} & {\unit{186\,GB}} \\
        \bottomhline
    \end{tabular}
\end{table*}

We train the NNs on a single GPU node and the RF and gradient boosting models on a single CPU node. The total time needed for training the machine learning models strongly depends on the architecture, the number of input features and the early stopping mechanism and ranges from a few minutes to a few hours for training the RF and gradient boosting models and training the NNs, respectively. The pre-processing and hyperparameter tuning is performed on a single CPU node. 

The recovery of the microphysical process rates consists of a sequence of steps (Fig.~\ref{fig:data_generation_flowchart}) which are performed on a single node. We use the Dask Python library \citep{dask2016} to effectively handle the large amounts of data. First, the ICON model output is read in and pre-processed. In this step, the data is split into two datasets. One dataset contains all the grid cells in which the threshold (Eq.~\eqref{eq:threshold_qx}) is not exceeded and thus only needs to be stored to eventually recover the full dataset. The second dataset is further processed, i.e. the classification and regression steps follow, either on a single CPU or GPU node. Afterwards, the datasets are combined, and the predicted process rates are stored in parquet-files, one file for each process rate and time step. Each file is about $\mathcal{O}(\unit{600\,MB})$ in size. Recovering the values of the autoconversion rate (QR\_AC) accumulated over a 10-minute output time step for 23 June 2023 on the ICON-D2 domain takes about 80 minutes on a single CPU node (152 cores), of which 42 minutes are the data handling and 38 minutes the prediction.

\section{Results}\label{sec:results}
In this section, we evaluate the developed models. For clarity, we present results for 10 representative process rates: autoconversion (QR\_AC), accretion (QR\_ACC), cloud droplet and raindrop self-collection (QNC\_SC, QNR\_SC), evaporation of rain (QR\_EVAP), heterogeneous ice nucleation (QI\_HET), rain freezing to snow, graupel and hail (QR\_RF), melting of frozen hydrometeors to rain (QR\_MELT), total riming (QX\_RIMING) and vapor deposition on ice (QI\_DEP). Additional figures and results are provided in Appendix~\ref{sec:appendix_results}. For the instantaneous rates, results are based on input data with an output time step of 10 minutes; for the accumulated rates, results are based on input data with a one-minute, 10-minute and 30-minute output time step. Recall that all ICON output variables used as input features are instantaneous quantities.

We begin with a direct recalculation of the process rates in Sect.~\ref{sec:recalcultation_baseline}, which serves as a performance baseline for the recovery of the process rates with the available ICON model output. The classification model performance is presented in Sect.~\ref{sec:classification}, followed by the evaluation of the regression models in Sect.~\ref{sec:regression}. Additionally, we study the dependence of the model performance on the output time step of the input data in Sect.~\ref{sec:time_step_comparison}. The final results of the combined classification-regression models are presented in Sect.~\ref{sec:combined_results} and Sect.~\ref{sec:all_process_rates}. We evaluate the prediction intervals obtained with CQR in Sect.~\ref{sec:results_conformal_prediction}. Finally, Sect.~\ref{sec:model_output} and Sect.~\ref{sec:results_case_studies} analyze the model output and discuss the results for the case studies.

\subsection{Recalculation baseline for instantaneous process rates}\label{sec:recalcultation_baseline}
Before presenting the ML model results, we assess whether a direct recalculation of the process rates from ICON model output is a feasible alternative.

To begin with, the direct recalculation is only applicable for the instantaneous process rates, as it is not possible to recalculate the summation over multiple time steps for the accumulated process rates (Eq.~\eqref{eq:process_rates_accumulated}), as information about the intermediate states is not available. Moreover, there is a mismatch between the set of output variables obtained for one ICON model time step and the values of the prognostic variables used for the internal computation of the process rates during time stepping (see Sect.~\ref{sec:model_output}), which introduces errors even for the instantaneous recalculations. Beyond that, the recalculation is impractical for certain process rates such as QX\_RIMING due to multiple contributions from various sub-routines in the ICON model code, or even impossible for process rates such as QI\_HET because of missing input information.

We recalculate the process rates from the equations of the two-moment microphysics scheme \citep{seifertbehengTwomomentCloudMicrophysics2006} as they are implemented in ICON, using the samples in the evaluation test set (see Sect.~\ref{sec:preprocessing}). The results are given in Table~\ref{table:results_recalculation}, where we report the coefficient of determination $R^2$, the mean absolute error (MAE) and the root mean square error (RMSE).
\begin{table*}[t]
    \centering
    \caption{Recalculation results for the instantaneous process rates.}
    \label{table:results_recalculation}
    \begin{tabular}{l c c c}
    \tophline
        Process rate & $R^2$ & {MAE / \unit{kg}\,\unit{kg}$^{-1}\,(20\,\text{s})^{-1}$} &  {RMSE / \unit{kg}\,\unit{kg}$^{-1}\,(20\,\text{s})^{-1}$}\\
        \middlehline
        Autoconversion (QR\_AC) & $-2.74$ & $4.09 \times 10^{-9}$ & $5.52 \times 10^{-7}$ \\
        Accretion (QR\_ACC) & $0.96$ & $1.04 \times 10^{-8}$ & $5.05 \times 10^{-7}$ \\
        Cloud droplet self-collection (QNC\_SC) & $0.99$ & $342.78\,\unit{kg}^{-1}$ & $3.85 \times 10^3\,\unit{kg}^{-1}$\\
        Raindrop self-collection (QNR\_SC) & $0.99$ & $13.11\,\unit{kg}^{-1}$ &  $9.42\,\unit{kg}^{-1}$ \\
        Rain evaporation (QR\_EVAP) & $0.87$ & $2.07\times 10^{-8}$ & $1.91\times 10^{-7}$ \\
        Het. ice nucleation (QI\_HET) & \multicolumn{3}{c}{No recalculation} \\
        Melting to rain (QR\_MELT) & $-0.12$ & $1.99 \times 10^{-7}$ & $3.22 \times 10^{-6}$ \\
        Rain freezing (QR\_RF) & $0.57$ & $8.40 \times 10^{-10}$ & $4.0361\times 10^{-7}$\\
        Deposition on ice (QI\_DEP) & $0.38$ & $1.90\times 10^{-8}$ & $6.55\times 10^{-7}$ \\
        Total riming (QX\_RIMING) & \multicolumn{3}{c}{No recalculation} \\
        \bottomhline
    \end{tabular}
\end{table*}

The recalculation is highly accurate for accretion (QR\_ACC) with $R^2 = 0.96$ and both self-collection rates (QNC\_SC and QNR\_SC) with $R^2 = 0.99$. Rain evaporation (QR\_EVAP) is also recalculated reasonably well with $R^2 = 0.87$. In contrast, autoconversion (QR\_AC; $R^2 = -2.74$) and melting to rain (QR\_MELT; $R^2 = -0.12$) cannot be recalculated accurately. For QR\_AC, this might be a consequence of the highly nonlinear dependence of the autoconversion rate on $q_c$ \citep[Eq. 4]{seifertbehengTwomomentCloudMicrophysics2006}, so that even small mismatches between the output values of the prognostic variables and the values that were actually used during the internal time stepping lead to large errors in the recalculated rate.

For the additional process rates included in Table~\ref{table:results_recalculation}, the recalculation is less accurate, yielding $R^2 = 0.57$ for the rain freezing rate (QR\_RF) and $R^2 = 0.38$ for the ice deposition rate (QI\_DEP). The recalculation of the total riming rate (QX\_RIMING) is omitted as it includes contributions from multiple sub-routines in the microphysics parameterization in ICON. For heterogeneous ice nucleation (QI\_HET), the recalculation is not possible due to missing input information from $n_\text{in, act}$.

To summarize, the results confirm that a recalculation baseline achieves limited accuracy for several process rates. Importantly, the process rates for which the recalculation is most inaccurate or not possible are those where the ML approach has the largest potential to add value, as the ML models can learn the mapping from the saved ICON output variables to the process rates without requiring access to the exact internal variables used in the internal computation of the process rate. In the following, we thus focus on the recoverability of the process rates with ML methods.

\subsection{Classification performance across process rates and output time steps}\label{sec:classification}
We evaluate the performance of the classification models with the $F_1$ score \citep{van1979information} and the \textit{Matthews correlation coefficient} (MCC) \citep{matthewsMCC1975}. Two factors motivate this choice of metrics. First, we need summary metrics that are applicable to both binary and multi-class classification problems: most process rates are strictly non-negative but the deposition and rain self-collection rates can also be negative, making this a three-class classification problem. Second, there is a strong class imbalance for certain process rates, with the above-threshold class being considerably rarer than the below-threshold class. The $F_1$ score is the harmonic mean of precision and recall, ranging from $0$ to $1$, with higher values indicating better classification performance. In the binary case, it quantifies whether the model correctly identifies grid cells where the value of the process rate exceeds the pre-defined threshold. For multi-class classification, we compute the class-wise $F_1$ score with a one-vs-rest approach and then compute the arithmetic mean. That is, we report the macro-averaged $F_1$ score, where each class is treated equally regardless of its frequency.

The MCC mitigates a limitation of the $F_1$ score under strong class imbalance, where the $F_1$ score can be high despite low classification performance on the majority class (true negatives) and thus does not fully reflect model behavior on the negative class. The MCC simultaneously takes true and false positives and negatives into account and is therefore a more informative summary metric. Essentially, the MCC acts as a correlation coefficient between the true and the predicted class labels. It is defined from $-1$ to $1$, where $1$ represents a perfect prediction, $0$ a random prediction, and $-1$ indicates total disagreement.

However, false positives and false negatives do not have the same physical consequence for the process rate recovery. In the false negative case, the process rate is automatically set to zero (see Sect.~\ref{sec:twostep}), thus resulting in unrecoverable loss of process rate information. In the false positive case, the regression step follows for a grid cell where the true value of the process rate is tiny, thereby potentially inducing a spurious non-zero value. However, these false positives may be corrected by the regression step predicting near-zero values.

\begin{figure*}[t]
    \centering
    \includegraphics[width=12cm]{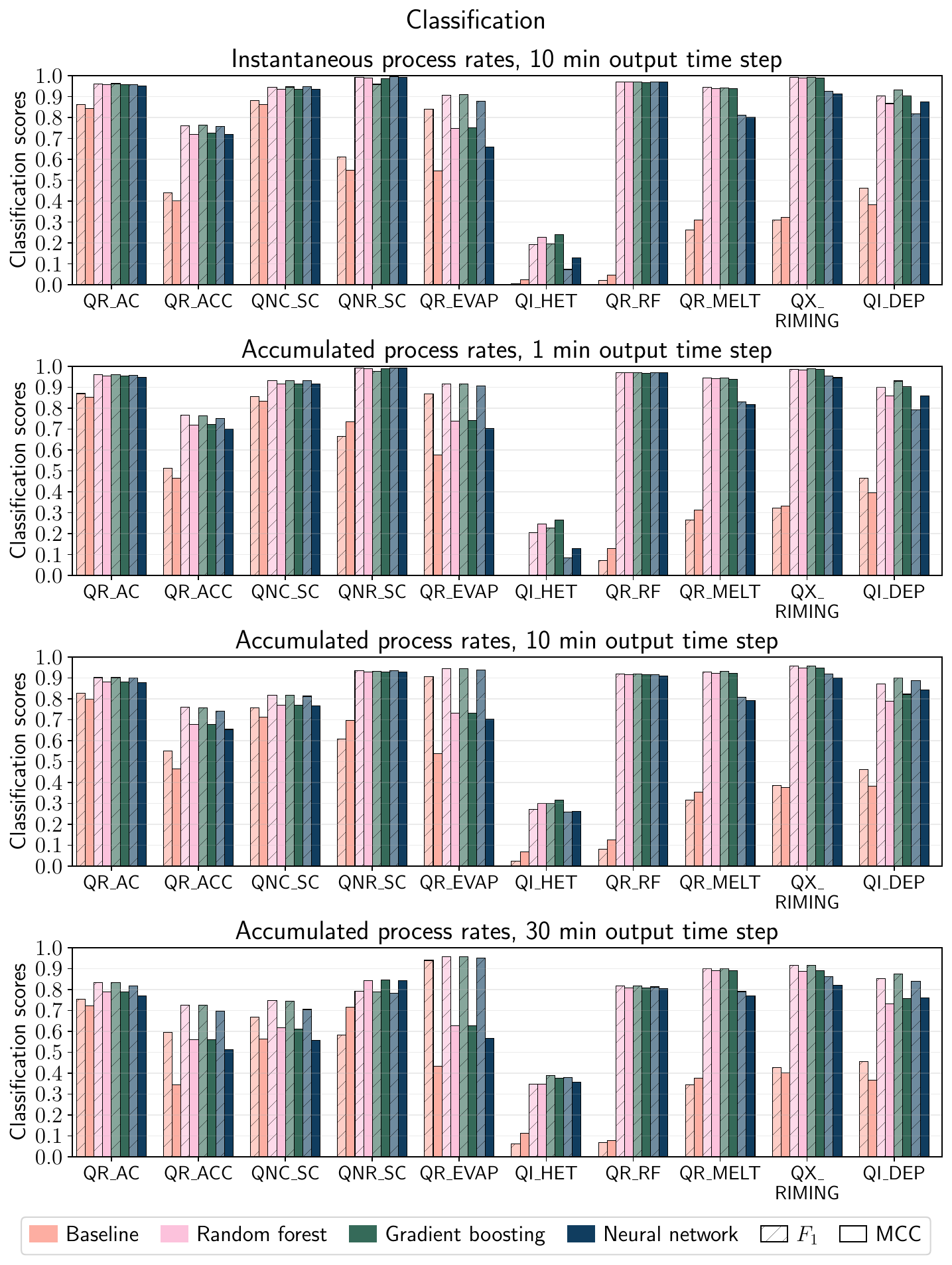}
    \caption{Classification performance (measured by $F_1$ score and MCC) is high for most instantaneous and accumulated process rates but decreases with increasing output time steps; except of heterogeneous ice nucleation (QI\_HET), where low scores slightly increase with increasing output time step. Shown are the $F_1$ score (hatched) and the MCC (solid) for the logistic regression baseline models (orange), the RFs (pink), gradient boosting models (green) and NNs (blue). The results are listed in Table~\ref{table:results_classification}.}
    \label{fig:results_classification_bars}
\end{figure*}

In Fig.~\ref{fig:results_classification_bars}, we present the $F_1$ score and MCC for the logistic regression baseline models, the RFs, gradient boosting models and the NNs, computed on the test dataset for the instantaneous process rates and the process rates accumulated over a one-minute, 10-minute and 30-minute output time step (Table~\ref{table:results_classification}). It is evident that the ML models outperform the logistic regression baseline model for all process rates and time steps. We find no consistent dependence of the predictive performance on the choice of algorithm across process rates and output time steps. For the autoconversion (QR\_AC), rain freezing (QR\_RF), rain melting (QR\_MELT), and total riming (QX\_RIMING) rate as well as both self-collection rates (QNC\_SC, QNR\_SC), all ML approaches consistently yield high $F_1$ and MCC scores. For the accretion (QR\_ACC) and rain evaporation (QR\_EVAP) rate, the scores are slightly lower. Particularly the values of the MCC decrease for longer output time steps while the $F_1$ scores remain high, which indicates that the models perform very well at identifying the grid cells where the value of process rate is above the threshold, but tend to classify too many grid cells as such. Yet, as the MCC still exhibits moderately high values, we conclude that the models yield a balanced prediction overall. 

In contrast, for heterogeneous ice nucleation (QI\_HET) the gradient boosting model yields the highest scores with $F_1 = 0.23$ and $\text{MCC} = 0.27$ for the one-minute output time step. The underlying cause for these comparatively low scores could be the unavailability of the number of activated ice nuclei $n_\text{in,act}$, or the distribution of the input data. Heterogeneous ice nucleation is a rare process and the values of the process rate are very small at most grid cells. This would also explain why the scores increase for longer output time steps, as the accumulated values of the process rates are larger due to the summation over more internal computations. Thus, there are more grid cells where the value of the process rate is non-zero. However, even for the 30-minute output time step, the scores remain low; the highest scores $F_1 = 0.39$ and $\text{MCC} = 0.38$ are achieved with the gradient boosting model.

We summarize the performance of the classification models across different output time steps in Table~\ref{table:mean_results_classification}, where we show the mean $F_1$ score and MCC. Here, the mean is computed over the 10 process rates shown in Fig.~\ref{fig:results_classification_bars}. We conclude that, despite the deficiencies for heterogeneous ice nucleation, the ML models yield satisfactory results overall.

\begin{table*}[t]
    \centering
    \caption{Mean $F_1$ score and MCC of the classification models computed on the test set for the different output time steps. The mean values are computed with the 10 process rates shown in Fig.~\ref{fig:results_classification_bars}.}
    \label{table:mean_results_classification}
    \begin{tabular}{l c c c c c c c c}
    \tophline
        \multirow{2}{*}{Model architecture} & \multicolumn{2}{c}{Inst., \unit{10\,min}} & \multicolumn{2}{c}{Acc., \unit{1\,min}} & \multicolumn{2}{c}{Acc., \unit{10\,min}} & \multicolumn{2}{c}{Acc., \unit{30\,min}} \\
        \cline{2-9} & {$\overline{F_1}$} & {$\overline{\mathrm{MCC}}$} & {$\overline{F_1}$} & {$\overline{\mathrm{MCC}}$} & {$\overline{F_1}$} & {$\overline{\mathrm{MCC}}$} & {$\overline{F_1}$} & {$\overline{\mathrm{MCC}}$} \\
        \middlehline
        Baseline          & $0.469$ & $0.429$ & $0.490$ & $0.463$ & $0.491$ & $0.452$ & $0.490$ & $0.411$ \\
        Random forest     & $0.857$ & $0.834$ & $0.857$ & $0.832$ & $0.831$ & $0.786$ & $0.788$ & $0.710$ \\
        Gradient boosting & $0.857$ & $0.839$ & $0.860$ & $0.838$ & $0.836$ & $0.791$ & $0.794$ & $0.715$ \\
        Neural network    & $0.813$ & $0.794$ & $0.816$ & $0.798$ & $0.811$ & $0.764$ & $0.764$ & $0.676$ \\
        \bottomhline
    \end{tabular}
\end{table*}

\subsection{Regression performance across process rates and output time steps}\label{sec:regression}
The regression models are evaluated with the coefficient of determination $R^2$ and the root mean square error (RMSE). The $R^2$ score ranges from $-\infty$ to $1$, where $1$ indicates a perfect prediction, $0$ a model whose mean squared error equals the variance of the target, and negative values indicate that the model performs worse than predicting the mean. As the regression models are only applied to grid cells for which the classification model predicts an above-threshold value (Eq.~\eqref{eq:threshold_classification}), the training, validation and test datasets for the regression models contain only above-threshold samples.

\begin{figure*}[t]
    \centering
    \includegraphics[width=12cm]{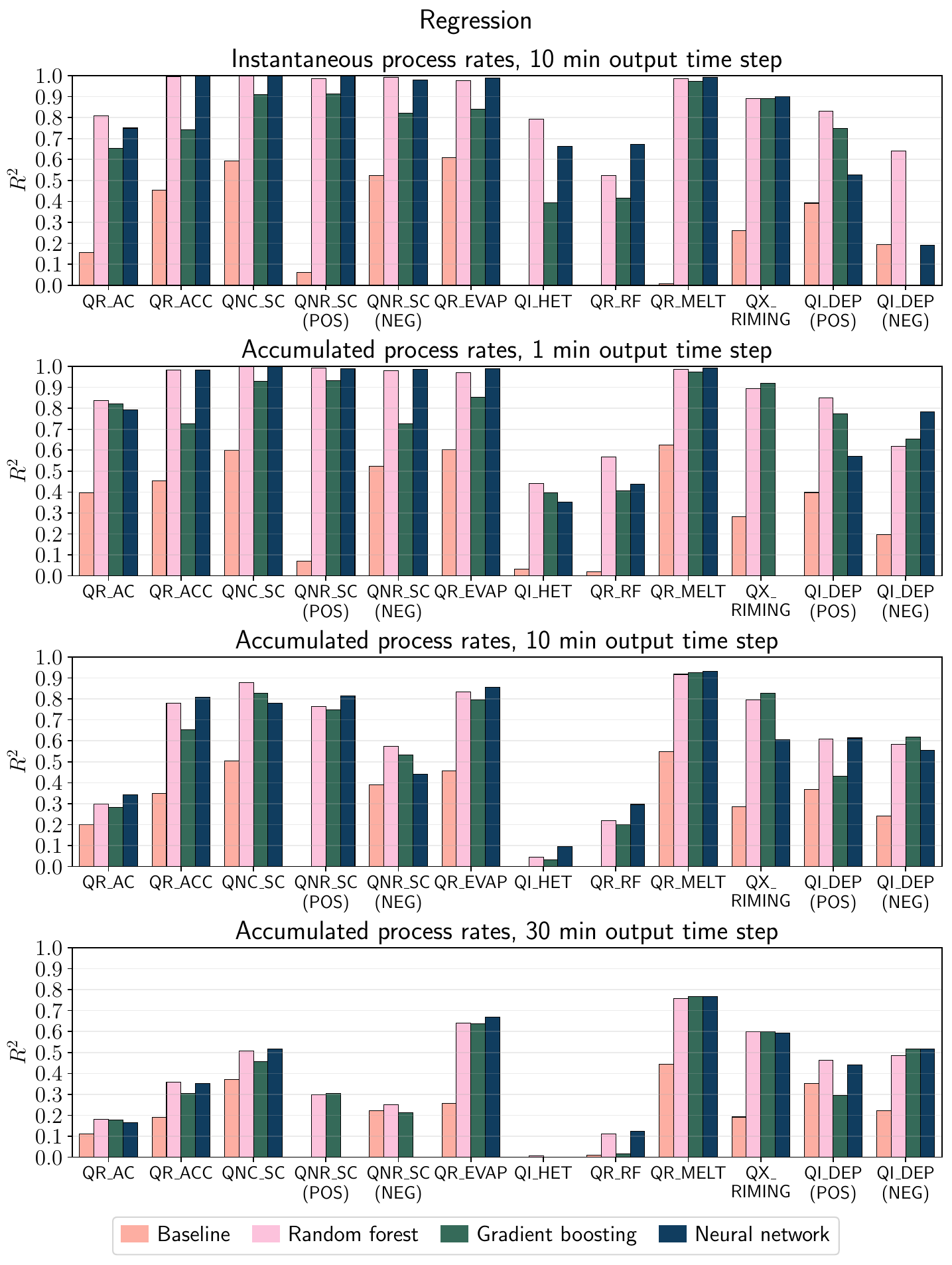}
    \caption{Regression performance (measured by $R^2$ score) is moderate to high in most cases but varies across process rates and model architectures and generally decreases with increasing output time step. Shown is the $R^2$ score for the linear regression baseline models (orange), the RFs (pink), gradient boosting models (green) and NNs (blue). A missing bar indicates $R^2 < 0$. The full results are given in Table~\ref{table:results_regression}.}
    \label{fig:results_regression_bars}
\end{figure*}

The $R^2$ scores and RMSE values for all model architectures are shown in Fig.~\ref{fig:results_regression_bars} and listed in Table~\ref{table:results_regression}. The ML models outperform the linear regression baseline for each process rate and time step. The recovery performance varies considerably across process rates and is linked to the extend to which each process rate is determined by instantaneous state variables in the input. For example, in the microphysics parameterization, the warm-rain process rates autoconversion, accretion and cloud droplet and raindrop self-collection primarily depend on $q_c$, $q_r$, $n_c$ and $n_r$. Consequently, the mapping from input variables to process rates is comparatively well constrained, resulting in high $R^2$ scores. In contrast, the regression performance for the heterogeneous ice nucleation rate (QI\_HET) is very low, consistent with the classification results. A possible explanation for this is that QI\_HET depends on additional quantities that are not included in the input, such as the prognostic variable $n_\text{in, act}$, which represents the number of activated ice nuclei. As a result, the input variables do not uniquely determine the process rate, which makes the prediction task substantially more difficult. Additionally, in the two-moment microphysics scheme, QI\_HET also depends on the vertical velocity $w$, which we do not include in the set of input features. We study the impact of the vertical velocity $w$ and the wind fields $u$ and $v$ as additional input features in App.~\ref{sec:wind_fields}. For comparison, we also conduct this analysis for QI\_HOM and QC\_MELT. Including $u$, $v$ and $w$ yields slightly higher $R^2$ scores for QI\_HET, but does not result in a significant improvement.

In Table~\ref{table:mean_results_regression}, we summarize the results for the regression models and the different output time steps, where we show the mean $R^2$ score, computed with the 10 process rates. The mean $R^2$ scores of the RF are slightly higher than those of the XGB and NN models for all output time steps. Considering Fig.~\ref{fig:results_regression_bars}, we observe that this is not a general trend and that there are sizeable differences both between the performance of the different ML algorithms and the models' ability to predict the individual process rates. It is not evident that one architecture consistently outperforms the others. This motivates the use of the best-performing model architecture on the validation set (see Table~\ref{table:classification_validation_scores} and Table~\ref{table:regression_validation_scores}) for each process rate individually (see Sect.~\ref{sec:twostep}).

\begin{table*}[t]
    \centering
    \caption{Mean $R^2$ score of the regression models computed on the test set for the different output time steps. The mean values are computed the 10 process shown in Fig.~\ref{fig:results_regression_bars}. To avoid distortion towards negative values, if $R^2 < -1$ for a specific process rate, the score is set to $R^2 = -1$ in the computation of the mean. The full results are given in Table~\ref{table:results_regression}.}
    \label{table:mean_results_regression}
    \begin{tabular}{l c c c c}
    \tophline
        \multirow{2}{*}{Model architecture} & {Inst., \unit{10\,min}} & {Acc., \unit{1\,min}} & {Acc., \unit{10\,min}} & {Acc., \unit{30\,min}}\\
        \cline{2-5} & {$\overline{R^2}$} & {$\overline{R^2}$} & {$\overline{R^2}$} & {$\overline{R^2}$}\\
        \middlehline
        Baseline          & $0.270$ & $0.350$ & $0.279$ & $0.196$ \\
        Random forest     & $0.868$ & $0.843$ & $0.608$ & $0.388$ \\
        Gradient boosting & $0.615$ & $0.759$ & $0.573$ & $0.285$ \\
        Neural network    & $0.805$ & $0.656$ & $0.595$ & $0.177$ \\
    \bottomhline
    \end{tabular}
\end{table*}

\subsection{Sensitivity of model performance to output time step}\label{sec:time_step_comparison}
\begin{figure*}[t]
    \centering
    \includegraphics[width=12cm]{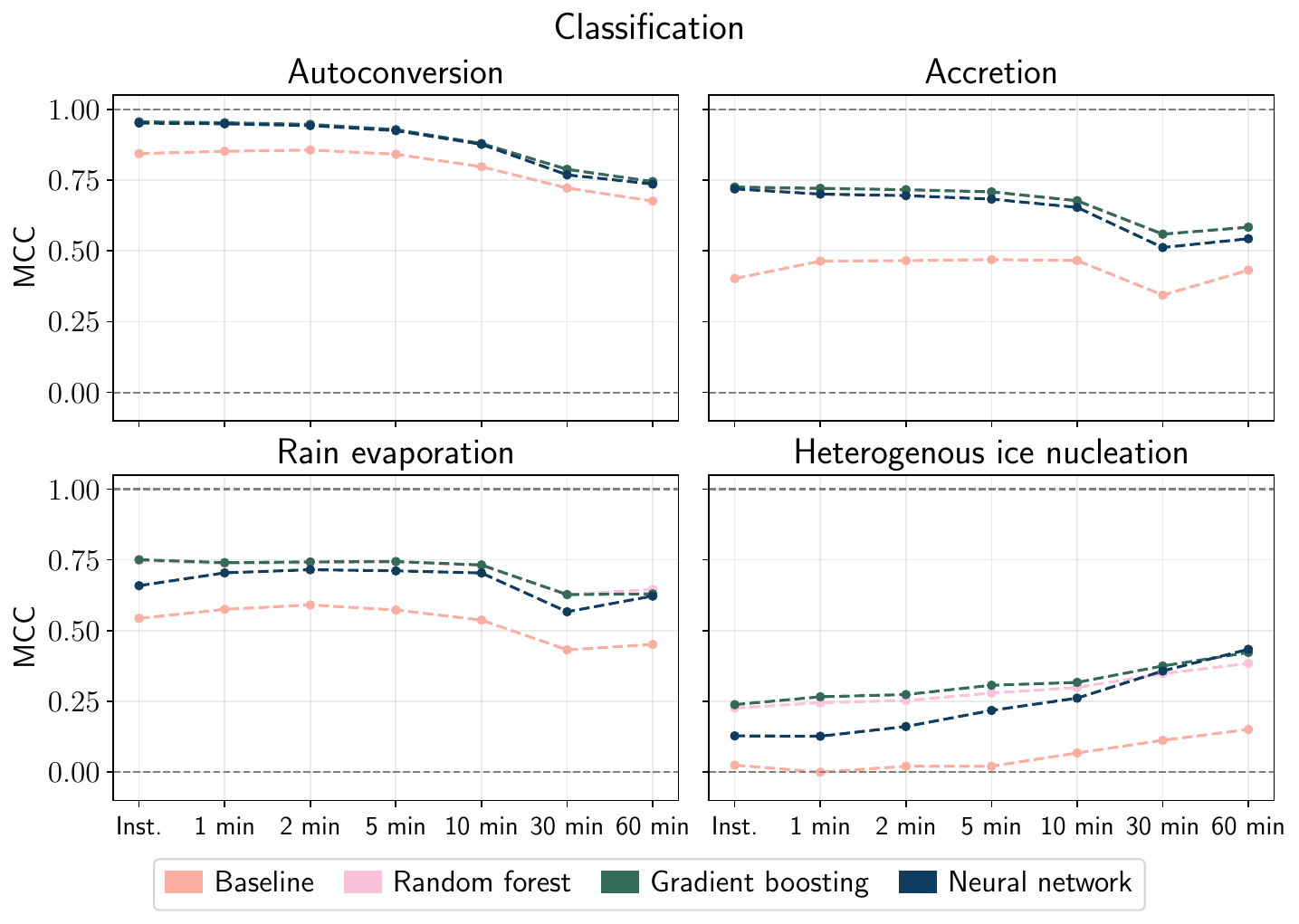}
    \caption{
    The MCC of the classification models remains stable up to an output time step of 10 minutes for autoconversion, accretion, and rain evaporation, but decreases at longer accumulation intervals. Heterogeneous ice nucleation is the main exception, with MCC increasing toward longer output time steps. Shown are MCC values for the logistic regression baseline models (orange), RFs (pink), gradient boosting model (green), and NNs (blue). Note that the x-axis is discrete.}
    \label{fig:time_steps_classification}
\end{figure*}

\begin{figure*}[t]
    \centering
    \includegraphics[width=12cm]{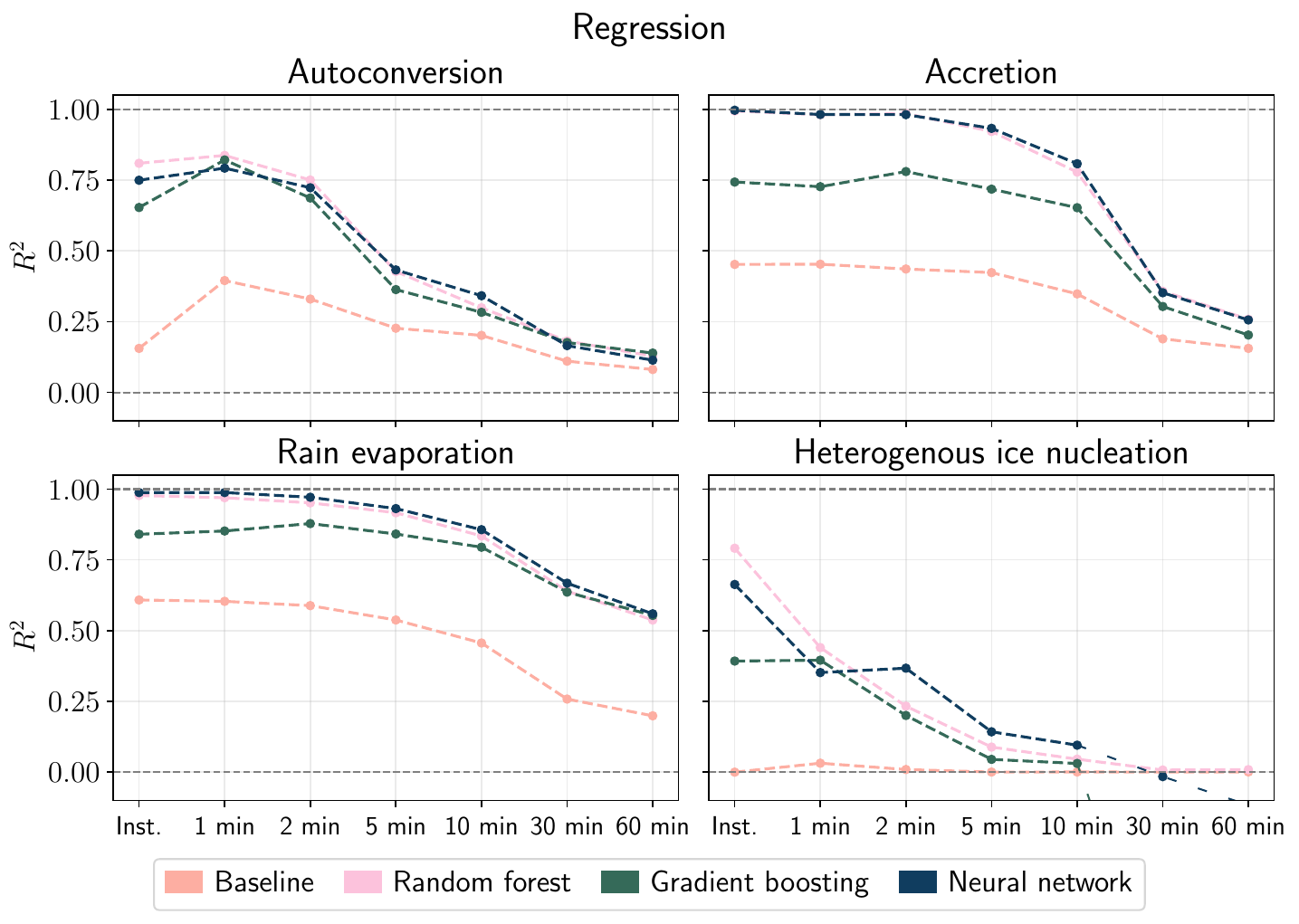}
    \caption{The $R^2$ of the regression models remains high up to an output time step of 2 minutes for autoconversion and 10 minutes for accretion and rain evaporation, but decreases at longer accumulation intervals. Heterogeneous ice nucleation follows the same trend with generally lower scores, except for the instantaneous case. Shown are $R^2$ scores for the linear regression baseline models (orange), RFs (pink), gradient boosting models (green), and NNs (blue). Note that the x-axis is discrete.}
    \label{fig:time_steps_regression}
\end{figure*}

To more carefully analyze the dependence of model performance to the output time step, we repeat the training, including the hyperparameter tuning,  using data with process rates accumulated over two, five and 60 minutes for eight process rates. In Fig.~\ref{fig:time_steps_classification} and Fig.~\ref{fig:time_steps_classification_app}, we show the classification model results and in Fig.~\ref{fig:time_steps_regression} and Fig.~\ref{fig:time_steps_regression_app} the regression model results in terms of the MCC and $R^2$ score, respectively. For comparison, we also include the results of the instantaneous process rates. 

Before discussing the results, note that the comparison of the different output time steps is slightly skewed by the thresholding. The threshold (Eq.~\eqref{eq:threshold_classification}) is defined uniformly regardless of output time step. Longer output time steps lead to slightly higher process rate values on average, as the rates are summed over more internal fast-physics computations, while the distribution of the input features does not change. But, this also leads to a higher number of samples with non-zero but tiny values below and above the threshold, which counteracts this effect to a certain extent. This could lead to a slight improvement in classification performance for longer output time steps. To analyze this effect on the performance of the regression models for the 30-minute output time step, we trained models on input data with an adjusted threshold to predict the three exemplary process rates QR\_AC, QR\_ACC and QR\_EVAP (Fig.~\ref{fig:compare_adjusted_threshold}), finding that the effect is negligible.

The MCC of the classification models is relatively stable across output time steps up to 10 minutes, but we observe a slight decrease in performance from the 10-minute to the 30-minute output time step. For QI\_HET, the MCC increases with longer output time steps, for the reasons discussed in Sect.~\ref{sec:classification}.

The regression performance generally decreases for longer output time steps, which is expected. A longer output time step corresponds to a greater number of internal fast-physics time steps over which the process rates are accumulated. Yet, the ML model input features are only available at the current and previous output time step and contain no information about the intermediate states, rendering the regression task increasingly difficult. Furthermore, we observe that the drop in performance at the 30-minute time step is not equally pronounced for all process rates. For instance, comparing accretion (QR\_ACC) and rain evaporation (QR\_EVAP) between the 10-minute and 30-minute output time step, the $R^2$ of the NN decreases by $0.46$ for QR\_ACC but by only $0.19$ for QR\_EVAP.

For some process rates, e.g. autoconversion, performance improves slightly from the instantaneous case to the accumulated case with the one-minute output time step. The potential reasons for this are twofold: first, it could be a result of the thresholding, as already discussed above. Second, the set of input features for the accumulated rates also includes the values of the mass mixing ratios and number concentrations at the previous output time step (Eq.~\eqref{eq:input_parameters_accumulated}), potentially providing useful additional temporal information that is not available for the instantaneous case (Eq.~\eqref{eq:input_parameters}).

\subsection{Combined classification-regression performance}\label{sec:combined_results}
We evaluate the performance of the combined classification-regression models with the $R^2$ score on a second test set (see Sect.~\ref{sec:preprocessing}). As the true target values span multiple orders of magnitude, the linear $R^2$ score is dominated by large values and can become very negative if the model does not accurately predict the highest values. To understand whether this is a consequence of very few large errors or reflects the overall model performance, we additionally report the $R^2_\mathrm{log}$ score, computed on logarithmically transformed true and predicted values. Since larger process rate values are typically of greater importance, we use the linear $R^2$ score as the main evaluation metric.

\begin{figure*}[t]
    \centering
    \includegraphics[width=12cm]{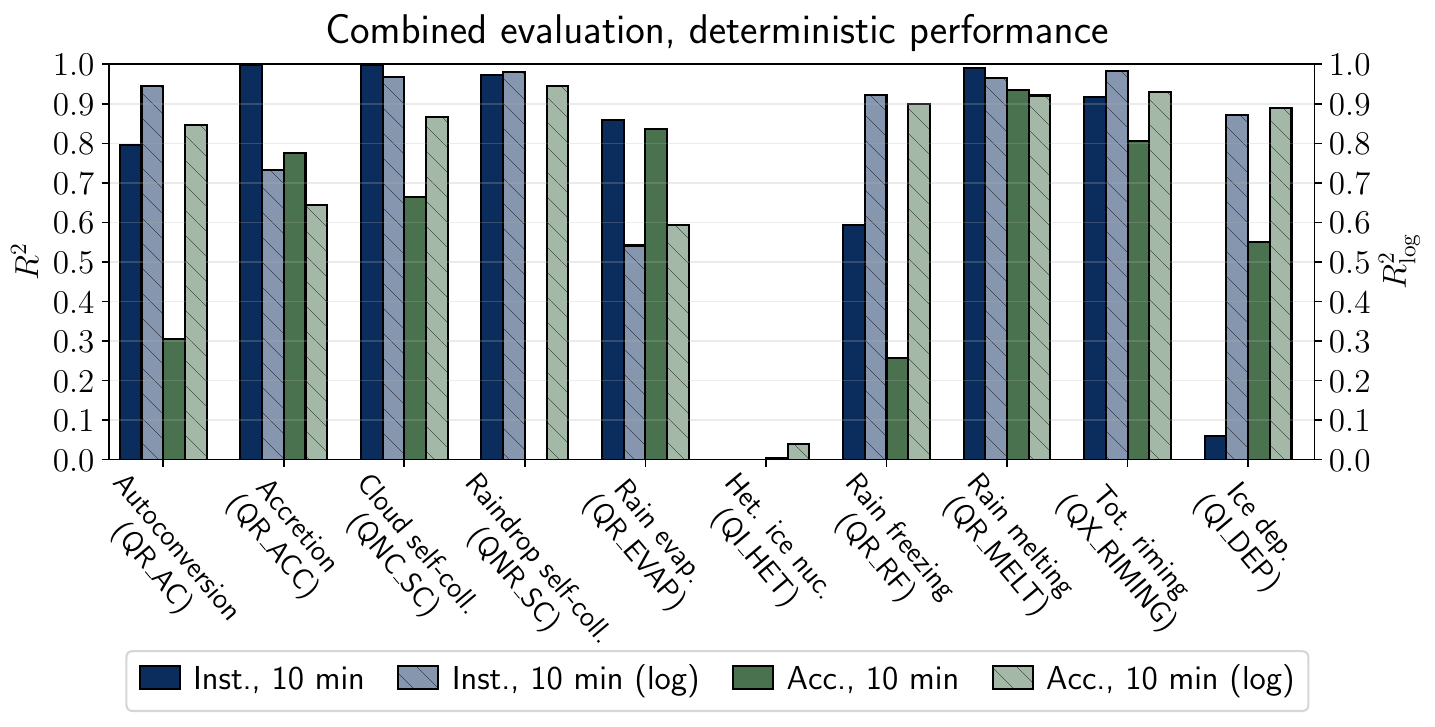}
    \caption{The combined classification-regression models show moderate to high predictive performance, which varies across process rates and is slightly lower for the accumulated process rates. Heterogeneous ice nucleation is the clear exception, with $R^2$ values close to zero for both instantaneous and the accumulated cases, indicating no predictive performance. Shown is the $R^2$ score (solid bars) and $R^2_\mathrm{log}$ score (hatched bars) for the 10-minute output time step for instantaneous process rates (dark blue colors) and accumulated process rates (dark green colors). A missing bar indicates $R^2 < 0$. The results are listed in Table~\ref{table:results_combined_evaluation}.
    }
    \label{fig:results_combined_evaluation_r2}
\end{figure*}

For the prediction of each process rate, we apply the classification and regression model that performed best on the validation set for the instantaneous process rates, as listed in Table~\ref{table:classification_validation_scores}, Table~\ref{table:regression_validation_scores} and Table~\ref{table:features_models}. For simplicity, we use one common architecture for all time steps per process rate. The results are visualized in Fig.~\ref{fig:results_combined_evaluation_r2} and listed in Table~\ref{table:results_combined_evaluation}.
The models for the instantaneous process rates (10-minute output time step) achieve a mean $\overline{R^2} \approx 0.72$ and $\overline{R^2_\mathrm{log}} \approx 0.79$ across all 10 process rates (Table~\ref{table:results_combined_evaluation}). The models for the accumulated process rates (10-minute output time step) yield $\overline{R^2} \approx 0.41$ and $\overline{R^2_\mathrm{log}} \approx 0.76$. For heterogeneous ice nucleation (QI\_HET), the model completely fails, which is consistent with our previous results. Excluding QI\_HET, the mean scores increase to $\overline{R^2}_\mathrm{inst.,\,10\, min} \approx 0.80$, $\overline{R^2_\mathrm{log}}_\mathrm{inst.,\,10\, min} \approx 0.88$, $\overline{R^2}_\mathrm{acc.,\,10\, min} \approx 0.46$ and $\overline{R^2_\mathrm{log}}_\mathrm{acc.,\,10\, min} \approx 0.84$. 

The mean $\overline{R^2}$ for the different output time steps (see Table~\ref{table:results_combined_evaluation}) also suggests that the final model is best suited to predict the process rates accumulated over shorter time steps, $t^\text{out} < \mathrm{10\,min}$, or to predict instantaneous process rates, where the predictive performance does not depend on the output time step. While the $\overline{R^2_\mathrm{log}}$ score is relatively consistent across output time steps, the $\overline{R^2}$ score decreases considerably, indicating that longer output time steps primarily impair the recovery of very high values.

The combined classification-regression models outperform the recalculation baseline (Sect.~\ref{sec:recalcultation_baseline}) for QR\_AC, QR\_MELT and QR\_RF. For the accretion (QR\_ACC) and evaporation (QR\_EVAP) and both self-collection rates (QNC\_SC, QNR\_SC), the results are comparable.

\begin{figure*}[t]
    \centering
    \includegraphics[width=12cm]{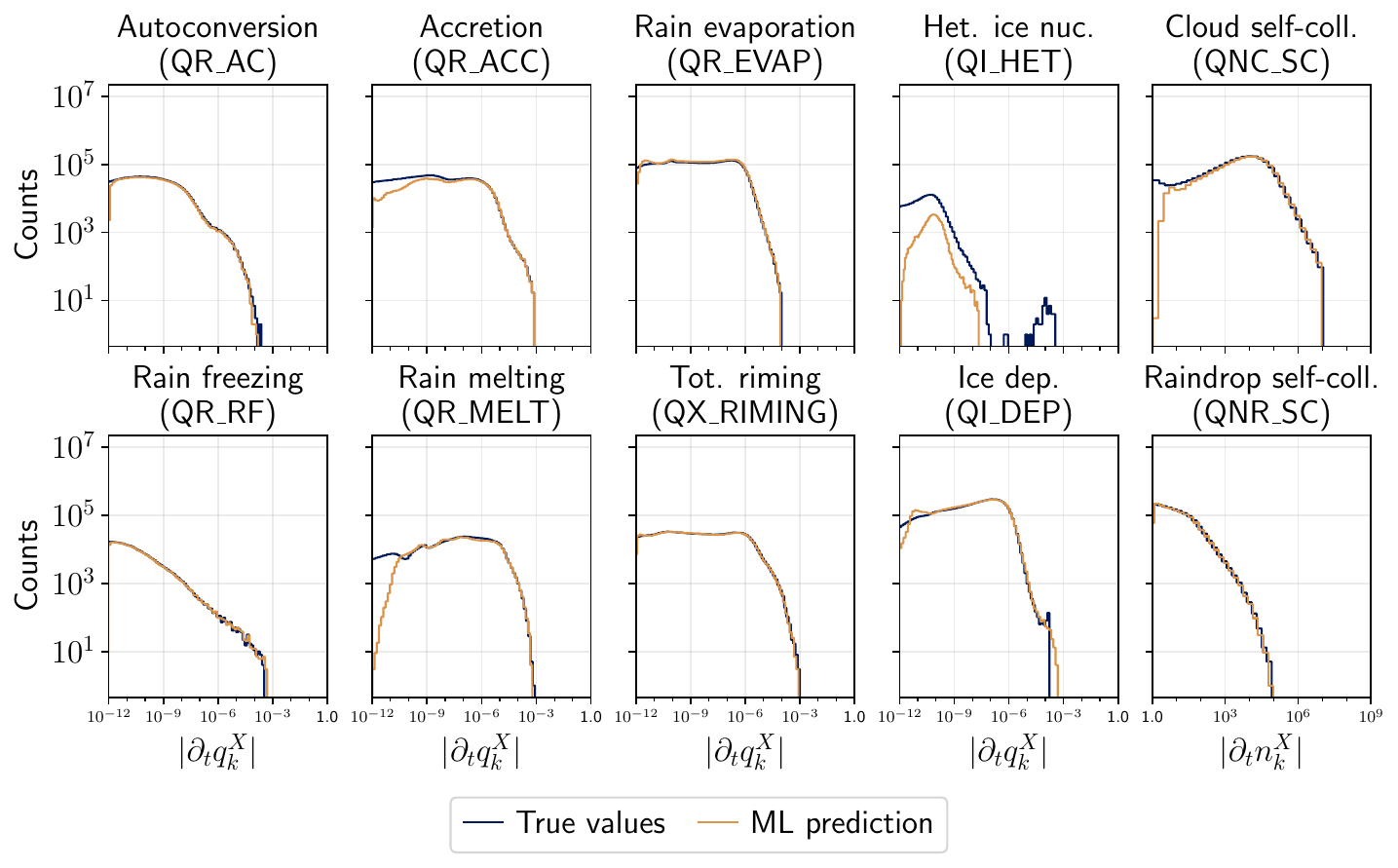}
    \caption{The distributions of the true (dark blue) and predicted (orange) values of the instantaneous microphysical process rates with a 10-minute output time step show general agreement, but the frequency of very small and very large values is slightly underestimated. The predicted distribution of heterogeneous ice nucleation misses the second peak at high true values. Shown are distributions of the true and predicted values in the test set. Absolute values are shown for better visualization.}
    \label{fig:distributions_combined_evaluation_inst_10min}
\end{figure*}

\begin{figure*}[t]
    \centering
    \includegraphics[width=12cm]{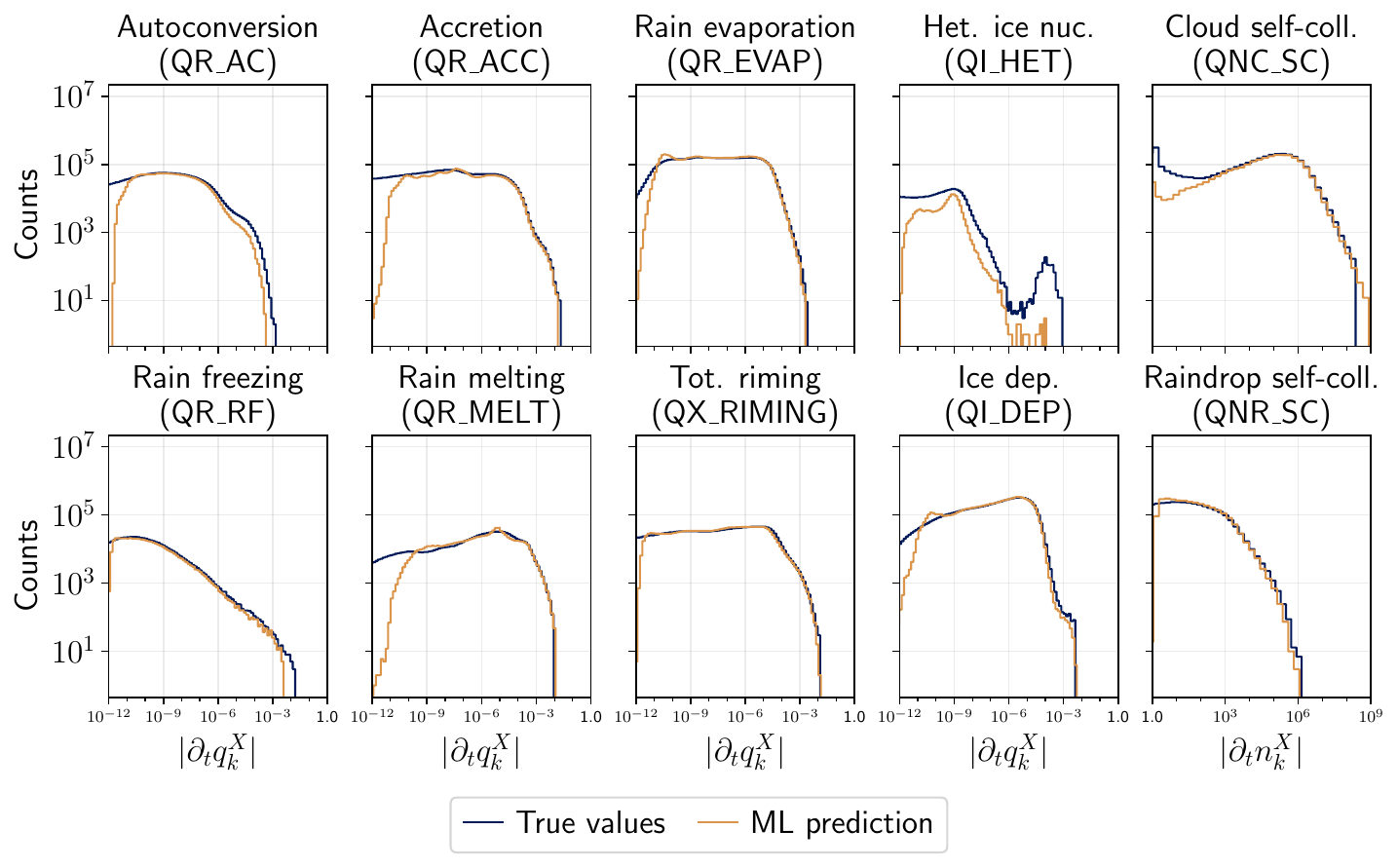}
    \caption{The distributions of the true (dark blue) and predicted (orange) values of the microphysical process rates accumulated over a 10-minute output time step show general agreement, but the frequency of very small and very large values is underestimated. The predicted distribution of heterogeneous ice nucleation severely underestimates the second peak at high true values. Shown are distributions of the true and predicted values in the test set. Absolute values are shown for better visualization.}
    \label{fig:distributions_combined_evaluation_acc_10min}
\end{figure*}

The distributions of true and predicted values for the test set are shown in Fig.~\ref{fig:distributions_combined_evaluation_inst_10min} and Fig.~\ref{fig:distributions_combined_evaluation_acc_10min}, revealing that the predicted distributions largely agree with the ICON model output (the true values). It is evident that the predicted frequency of very small values is too low for several process rates, which might be a result of the threshold (Eq.~\eqref{eq:threshold_classification}) and the classification step. Furthermore, the model for QI\_HET under-predicts the values substantially and misses the second peak. Considering the process rates accumulated over longer output time steps (App.~\ref{sec:output_distributions}) we find that the distribution is slightly narrowed and that both very small and very large values are under-predicted, which is in accordance with the lower $\overline{R^2}$ score reported for longer output time steps.


\subsection{Combined classification-regression performance for the remaining process rates}\label{sec:all_process_rates}
\begin{figure*}[t]
    \centering
    \includegraphics[width=12cm]{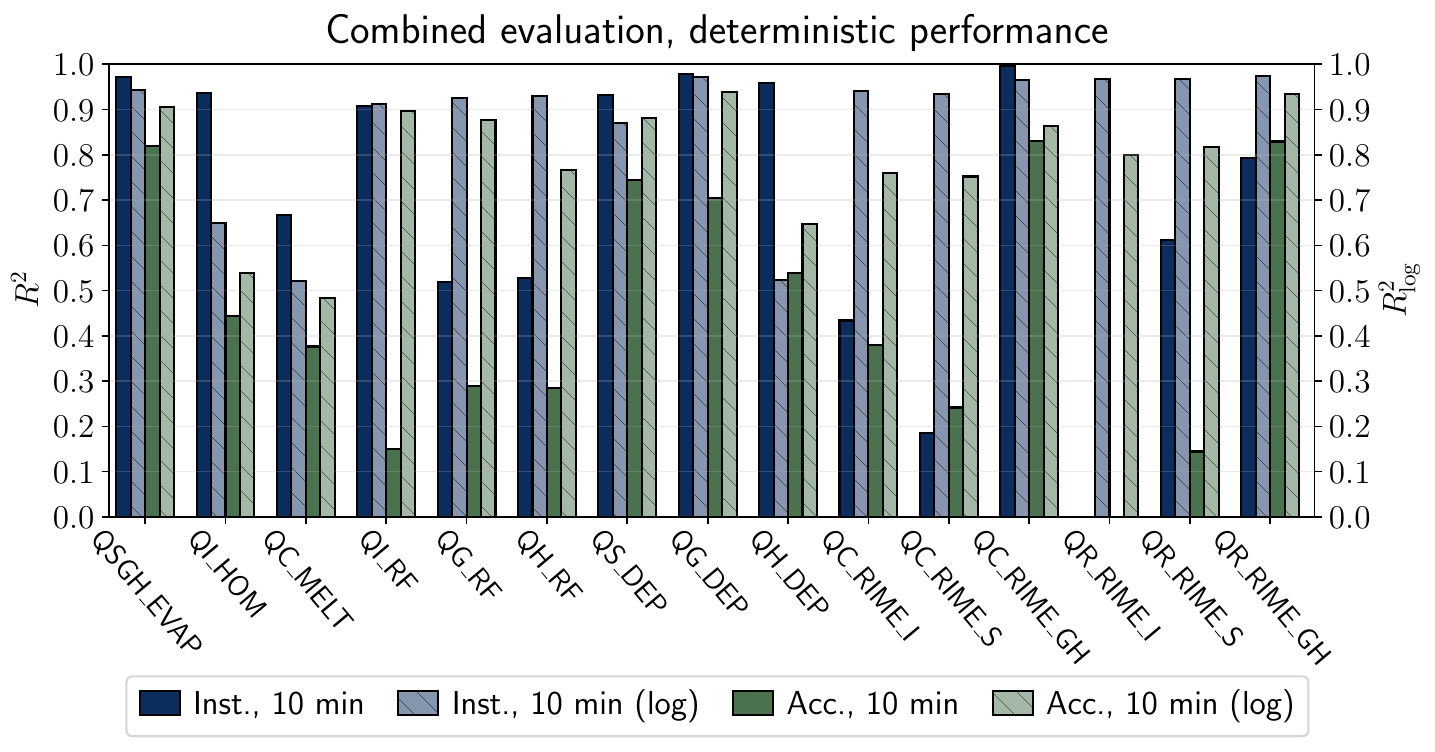}
    \caption{The combined classification-regression models show moderate to high predictive performance (measured by $R^2$ score and $R^2_\mathrm{log}$), which varies across process rates and is slightly lower for the accumulated process rates. Shown is the $R^2$ score (solid bars) and $R^2_\mathrm{log}$ score (hatched bars) for the 10-minute output time step for instantaneous process rates (dark blue colors) and accumulated process rates (dark green colors). A missing bar indicates $R^2 < 0$. The results are listed in Table~\ref{table:results_combined_evaluation_remaining_appendix}.}
    \label{fig:results_additional_process_rates_combined_evaluation}
\end{figure*}

In Fig.~\ref{fig:results_additional_process_rates_combined_evaluation}, we present the results of the combined classification-regression model for the remaining 15 process rates, evaluated for the instantaneous and accumulated process rates and the 10-minute output time step.
The results confirm our previous findings (see Sect.~\ref{sec:combined_results}). Averaged over all 25 process rates, the mean scores are $\overline{R^2}_\text{inst.,\,10\,min} \approx 0.66$ and $\overline{R^2}_\text{acc.,\,10\,min} \approx 0.40$. Without QI\_HET, the mean scores increase to $\overline{R^2}_\mathrm{inst.,\,10\, min} \approx 0.69$ and $\overline{R^2}_\mathrm{acc.,\,10\, min} \approx 0.41$. The results are slightly lower than those computed with the 10 rates in Sect~\ref{sec:combined_results}, reflecting the inclusion of multiple ice process rates. Again, we observe a decrease in performance for the 10-minute output time step. Nonetheless, the models show robust performance overall. The full results are listed in Table~\ref{table:results_combined_evaluation_remaining_appendix}. 

\subsection{Prediction intervals with conformalized quantile regression}\label{sec:results_conformal_prediction}
We evaluate the quality of the prediction intervals with the prediction interval coverage probability (PICP), hereinafter also referred to as coverage, defined as the proportion of true values that fall within the prediction intervals. The PICP is computed as
\begin{equation}
    \text{PICP} = \frac{1}{n} \sum_{i=1}^n \mathbb{1} \left\{Y_i \in \mathcal{C}(X_{i})\right\} \, ,
    \label{eq:picp}
\end{equation}
where $n$ is the size of the test set, $Y_i$ the true value, and $\mathcal{C}(X_{i})$ the prediction interval. With $\alpha = 0.1$, we aim for a nominal coverage level of $\text{PICP} \gtrsim (1 - \alpha)\cdot 100 \% = 90\%$. 

\begin{figure*}[t]
    \centering
    \includegraphics[width=12cm]{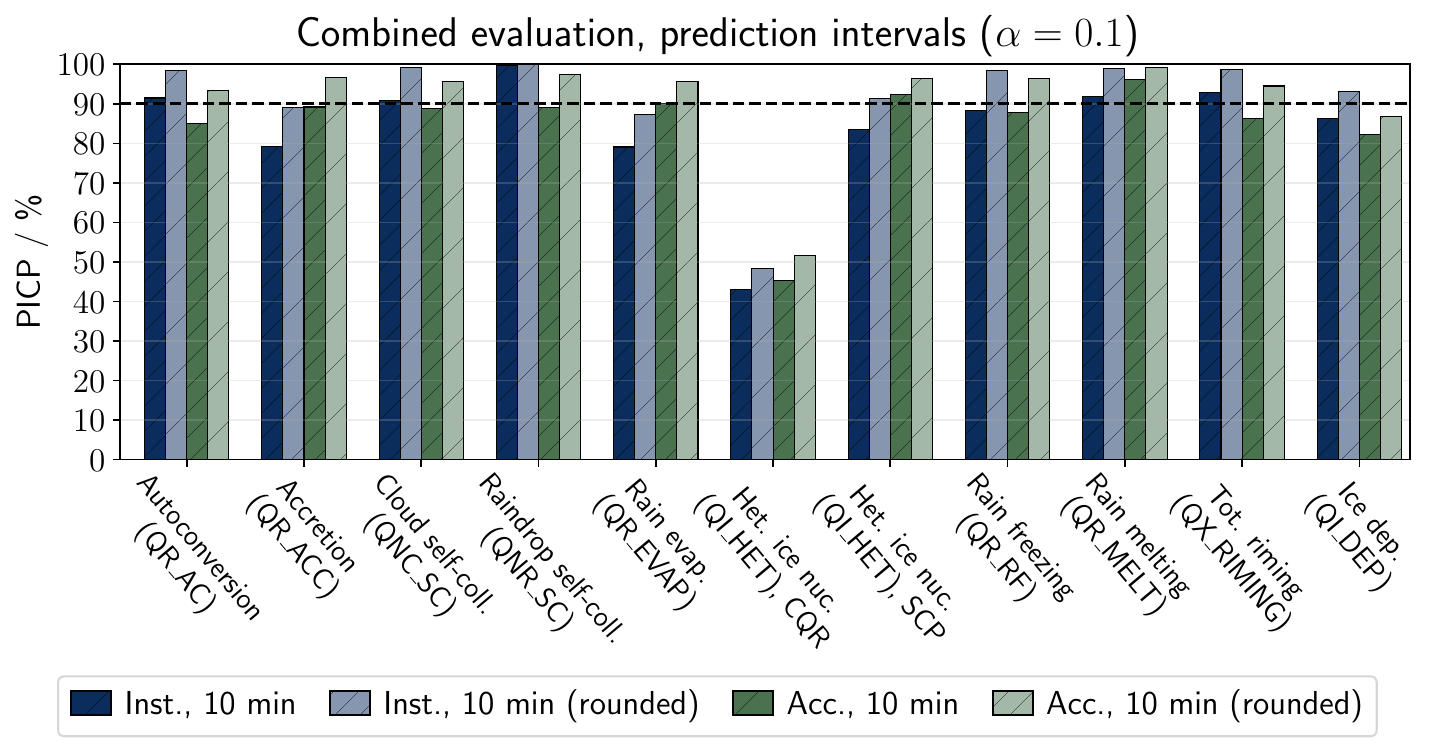}
    \caption{The prediction intervals for the combined classification-regression models achieve high empirical coverage with PICP values close to the nominal $90\%$ level. The rounding operation (Eq.~\eqref{eq:interval_bounds_cqr_rounded}) increases the empirical coverage to above $90\%$ for most process rates. This is except for heterogeneous ice nucleation (QI\_HET) for which $\text{PICP} \approx 45\%$ across different output time steps. In this case, split conformal prediction (SCP) yields more reliable prediction intervals. The corresponding values are listed in Table~\ref{table:results_combined_evaluation_cqr}.
    }
    \label{fig:results_combined_evaluation_cqr}
\end{figure*}

The PICP results for the quantile NN-based conformal prediction intervals are shown in Fig.~\ref{fig:results_combined_evaluation_cqr} and listed in Table~\ref{table:results_combined_evaluation_cqr}, both for the standard conformal prediction intervals and for the rounded prediction intervals ($\text{PICP}_\text{rounded}$), in which the interval bounds are rounded to the nearest order of magnitude (see Sect.~\ref{sec:uncertainty_quantification}). The sharpness of the prediction intervals is evaluated with the normalized mean prediction interval width (NMPIW),
\begin{equation}\label{eq:NMPIW}
    \mathrm{NMPIW} = \frac{1}{n} \sum_{i=1}^n \left(\frac{U(X_i) - L(X_i)}{\max(Y) - \min(Y)}\right) \, ,
\end{equation}
where $U$ and $L$ denote the calibrated upper and lower prediction interval bounds. The results are listed in Table~\ref{table:results_combined_evaluation_cqr_nmpiw}.

It is apparent that the PICP is close to $90\%$ for most process rates. Before discussing the results in more detail, it is useful to consider the possible reasons for an empirical coverage probability below the nominal level of $90\%$. 
First, the coverage guarantee of conformal prediction is marginal, i.e. averaged over all training, calibration and test samples \citep{romanoConformalizedQuantileRegression2019}. Consequently, the empirical coverage can fall below the nominal level for subsets of the data, e.g. certain regimes. Furthermore, for a calibration set of size $n_\mathrm{cal}$, the PICP fluctuates around the nominal coverage level with deviations scaling as $\mathcal{O}({1}/{\sqrt{n_\mathrm{cal}}})$ \citep{angelopoulosTheoreticalFoundationsConformal2025}. Second, conformal prediction relies on the assumption that the training, calibration and test samples are exchangeable, i.e. drawn from the same joint probability distribution. This assumption is slightly weaker than the independent and identically distributed assumption commonly used in machine learning \citep{gopakumarUncertaintyQuantificationSurrogate2025, shafer2007tutorialconformalprediction}. Deviations of the PICP from the nominal $90\%$ level for some process rates therefore may indicate slight violations of exchangeability, potentially resulting from spatio-temporal structure in the data\footnote{See \citet[Sect. A.3]{simm2026calibratedconformalpredictionintervals} for a discussion on exchangeability for microphysical process rate recovery.}.

\begin{figure*}[t]
    \centering
    \includegraphics[width=12cm]{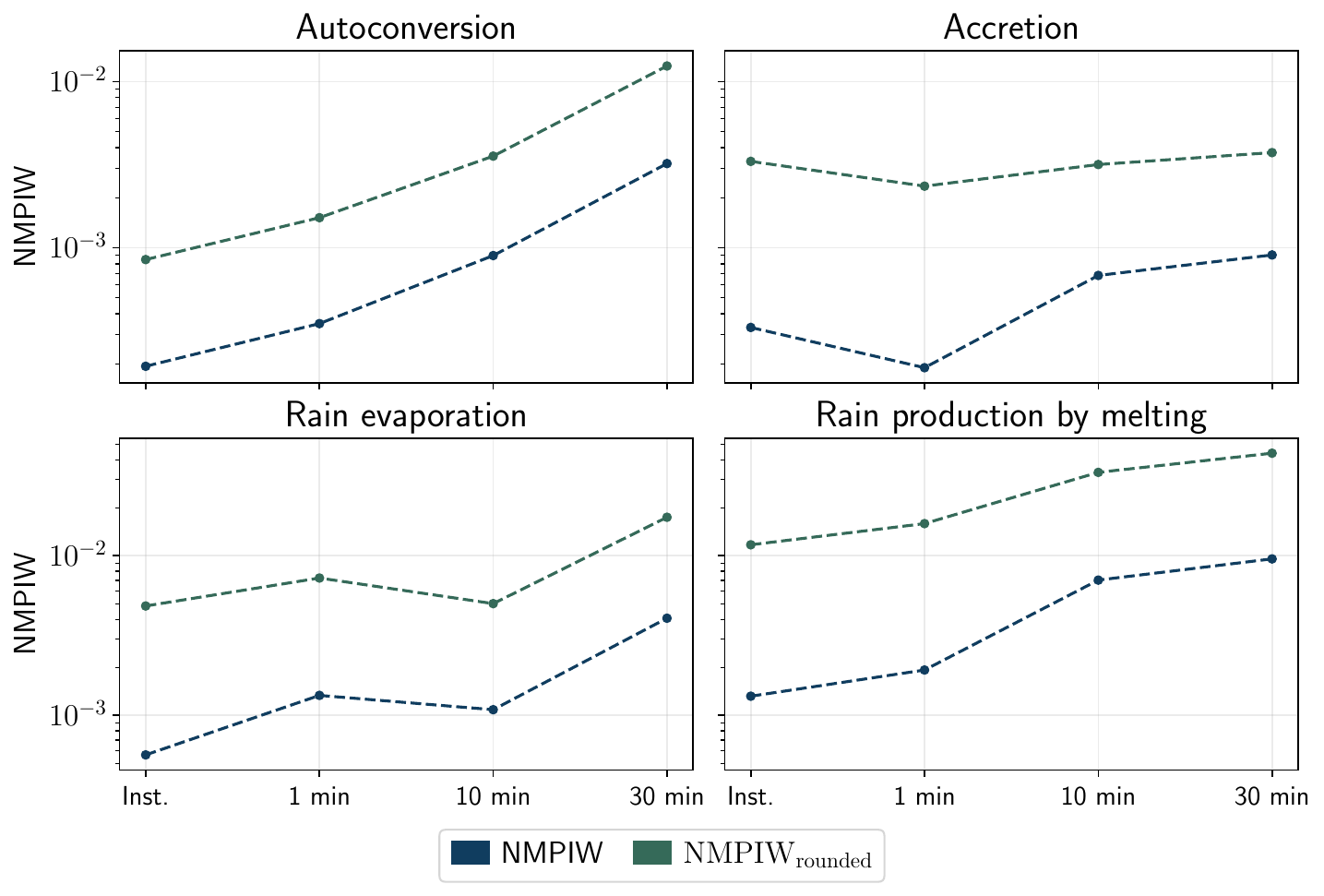}
    \caption{The normalized mean prediction interval width (NMPIW) generally increases for longer output time steps, reflecting increasing predictive uncertainty. The improved empirical coverage (Fig.~\ref{fig:results_combined_evaluation_cqr}) due to the rounding operation ($\mathrm{NMPIW}_\mathrm{rounded}$) is accompanied by wider prediction intervals. The corresponding values are listed in Table~\ref{table:results_combined_evaluation_cqr_nmpiw}.
    }
    \label{fig:results_nmpiw}
\end{figure*}

Computed across all 10 process rates and quantile regression models (using SCP for QI\_HET, see below), the mean PICP values are $88.09\%$ and $86.95\%$ for the instantaneous and accumulated process rates and a 10-minute output interval, respectively. After the rounding operation, the values increase to $95.31\%$ and $94.19\%$. Overall, the uncertainty estimates are reasonably well calibrated and close to the nominal level of 90\%. While the calibrated prediction intervals occasionally yield PICP values slightly below $90\%$, the rounding operation improves the $\mathrm{PICP}_\mathrm{rounded}$ to exceed 90\% for most process rates. This systematic improvement in empirical coverage is accompanied by wider intervals, as reflected by the larger NMPIW values for the rounded intervals in Fig.~\ref{fig:results_nmpiw} and Table~\ref{table:results_combined_evaluation_cqr_nmpiw}.

An exception are the results for heterogeneous ice nucleation (QI\_HET), for which the CQR-calibrated intervals yield very low coverage, PICP $\approx 43\% - 48\%$, across different output time steps. This likely indicates that the quantile regression model does not accurately predict the upper and lower quantiles, limiting the effectiveness of the conformal calibration step. To address this issue, we instead apply split conformal prediction (SCP), which calibrates prediction intervals for deterministic point predictions using the absolute error residual $R(x,y) = |y-\hat{\mu}(x)|$ as the non-conformity score for calibration samples $\{(X_i, Y_i)\}_{i=1}^n$ and deterministic predictions $\hat{\mu}$. Further details are provided in \citet{simm2026calibratedconformalpredictionintervals}. For QI\_HET, SCP yields substantially improved PICP values across all output time steps (Table~\ref{table:results_combined_evaluation_cqr}) with $\mathrm{PICP}_\mathrm{inst., 10 min} = 83.43\%$ and $\mathrm{PICP}_\mathrm{acc., 10 min} = 92.26\%$. Importantly, the SCP calibrated intervals for QI\_HET are also considerably sharper (Table~\ref{table:results_combined_evaluation_cqr_nmpiw}).  These results indicate that SCP provides substantially more reliable and sharper prediction intervals for QI\_HET, and we therefore adopt SCP for QI\_HET in the following. 

Comparing the results for the different output time steps, the prediction intervals generally adapt to the decrease in deterministic model performance (see Sect.~\ref{sec:time_step_comparison}) with wider intervals (Table~\ref{table:results_combined_evaluation_cqr_nmpiw}). Despite this, the PICP remains approximately valid across output time steps. For the 30-minute output time step, deviations from $90\%$ are larger for most process rates. However, except for QR\_ACC and QI\_DEP, where the $\mathrm{PICP}_\mathrm{rounded}$ drops to $73.63\%$ and $84.20\%$, respectively, the rounding operation restores $\mathrm{PICP}_\mathrm{rounded}$ close to or above $90\%$.

\subsection{Spatiotemporal structure of the recovered process rates for the 23 June 2023 ICON-D2 case}\label{sec:model_output}
\begin{figure*}[t]
    \includegraphics[width=12cm]{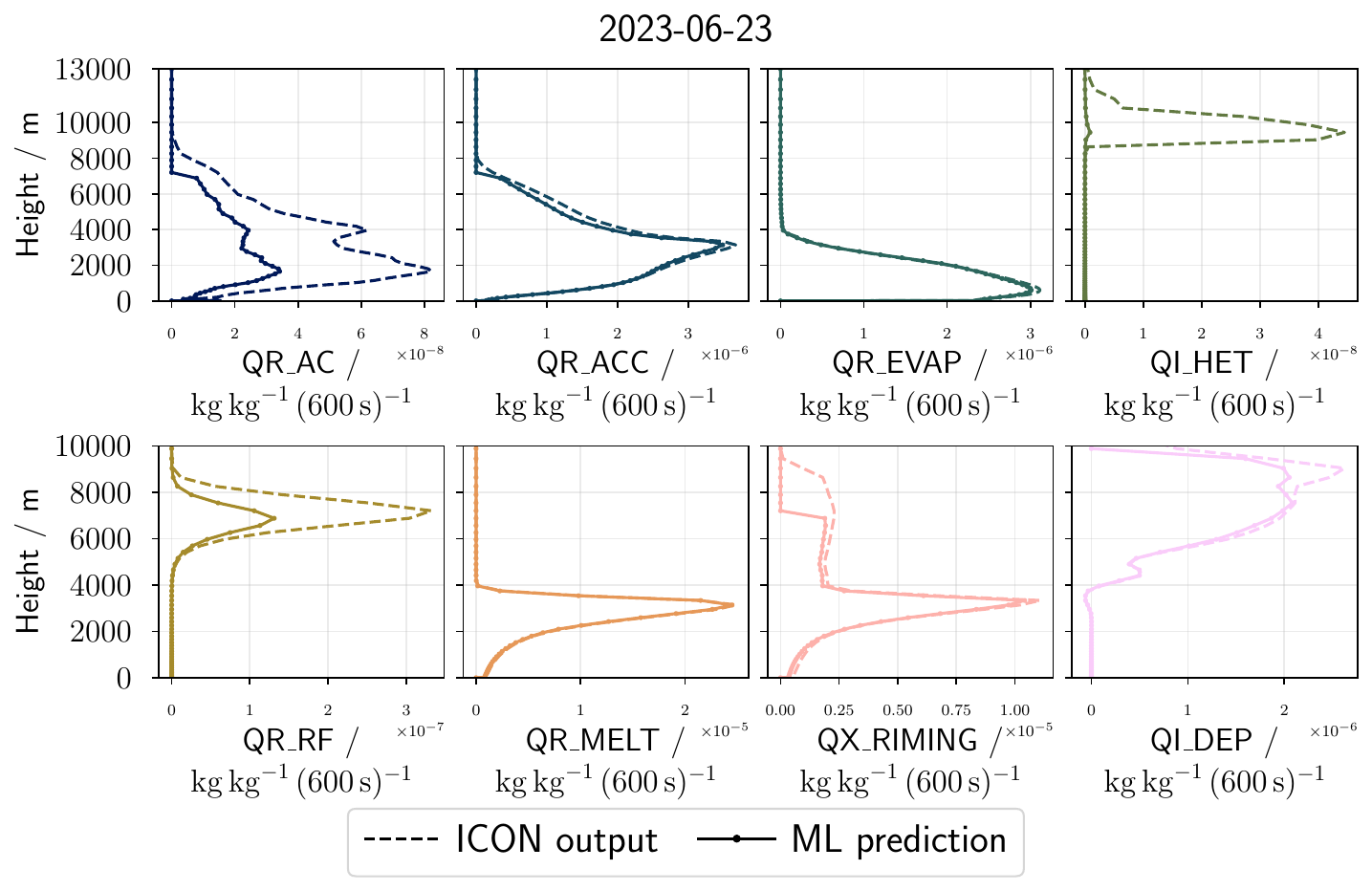}
    \caption{The vertical profiles of the spatio-temporal mean microphysical process rates show overall agreement between the ML predictions and the ICON model output, except for heterogeneous ice nucleation (QI\_HET). For autoconversion (QR\_AC) and rain freezing (QR\_RF), the predicted mean values are substantially lower than the ICON output across height levels. Shown are vertical profiles of eight process rates accumulated over a 10-minute output time step for 23~June~2023, averaged over 24 hours and all grid points at each level. The solid-dotted line depicts the ML prediction and the dashed line the ICON model output.}
    \label{fig:vertical_profiles_20230623}
\end{figure*}

Let us now study the model output in more detail. To this end, we will first consider the prediction of the process rates for 23~June~2023 on the ICON-D2 domain, a date which was also included in the test set (see Table~\ref{table:full_training_data_overview}). Figure~\ref{fig:vertical_profiles_20230623} shows the vertical profiles of spatio-temporal averages of eight microphysical process rates, accumulated over a 10-minute output time step, computed over 24 hours and all grid points at each level. We observe that the profiles computed from the ICON model output and the ML predictions largely agree in most cases, confirming that the models successfully recover the mean vertical structure of the process rates. However, for autoconversion (QR\_AC) and rain freezing (QR\_RF), the model under-predicts the values, consistent with the slight under-prediction of high values in the predicted distribution (Fig.~\ref{fig:distributions_combined_evaluation_acc_10min}). This highlights that the domain mean of a process rate is primarily controlled by the tail of its distribution, where even a moderate under-prediction of very high values leads to a disagreement between the predicted vs. the true spatially averaged vertical profile. For QI\_HET, the model does not capture the peak at $\approx$ \unit{9\,km}, also consistent with Fig.~\ref{fig:distributions_combined_evaluation_acc_10min}. 

\begin{figure*}[t]
    \centering
    \includegraphics[width=12cm]{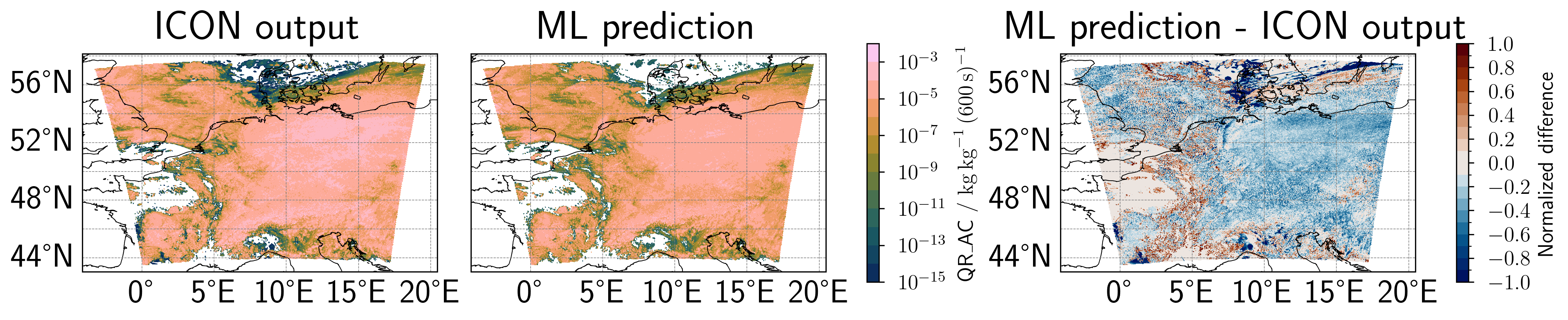}\\
    \includegraphics[width=12cm]{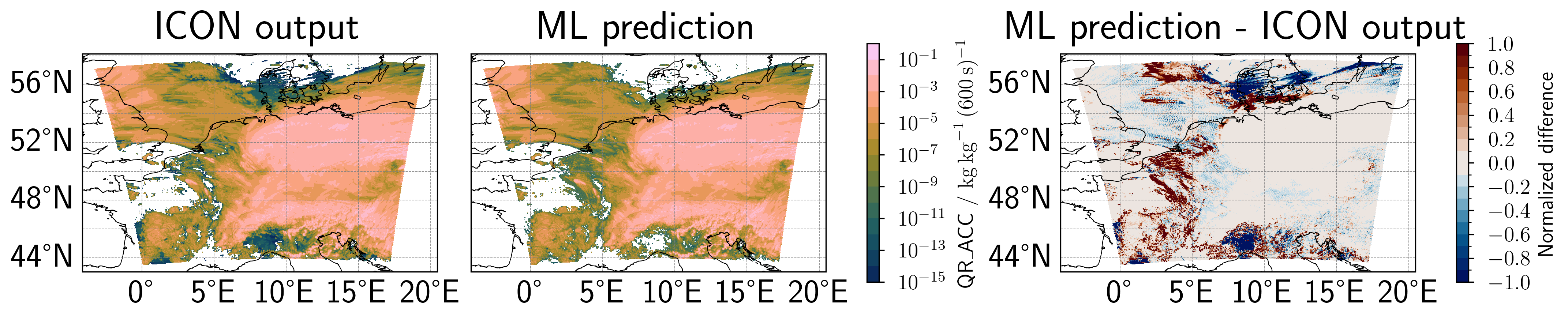}
    \caption{The spatial distributions of autoconversion (QR\_AC) and accretion (QR\_ACC), accumulated over a 10-minute output time step for 23 June 2023 with the ICON-D2 domain, show similar patterns between the ICON model output (left) and the ML prediction (center) with small to moderate regional differences, normalized to the respective range of the process rate (right). The spatial distributions are averaged over all height levels and the \unit{24\,hour}-simulation interval. Additional figures are provided in App.~\ref{sec:spatial_distribution}.}
    \label{fig:2d_plots_20230623}
\end{figure*}

The spatial distribution of the average predicted and simulated microphysical process rates for autoconversion and accretion accumulated over a 10-minute output time step is shown in Fig.~\ref{fig:2d_plots_20230623} (Fig.~\ref{fig:2d_plots_20230623_additional_rates} for additional process rates), revealing a good agreement between the ICON model output and the ML prediction and supporting previous results in Fig.~\ref{fig:results_combined_evaluation_r2}. It is apparent that the ML models miss the grid cells where the value of the process rate is very low, which could be a consequence of the thresholding and the two-step approach. The differences between the ICON model output and the ML prediction are larger for QI\_HET and QI\_DEP. For QI\_HET, the model severely under-predicts the values of the process rates, whereas the model both over- and under-predicts QI\_DEP. Yet, the spatial pattern of the prediction of QI\_DEP appears to be broadly similar to the ICON model output.

\subsection{Spatial transferability to other domains and ICON configurations}\label{sec:results_case_studies}
\begin{figure*}[t]
    \includegraphics[width=12cm]{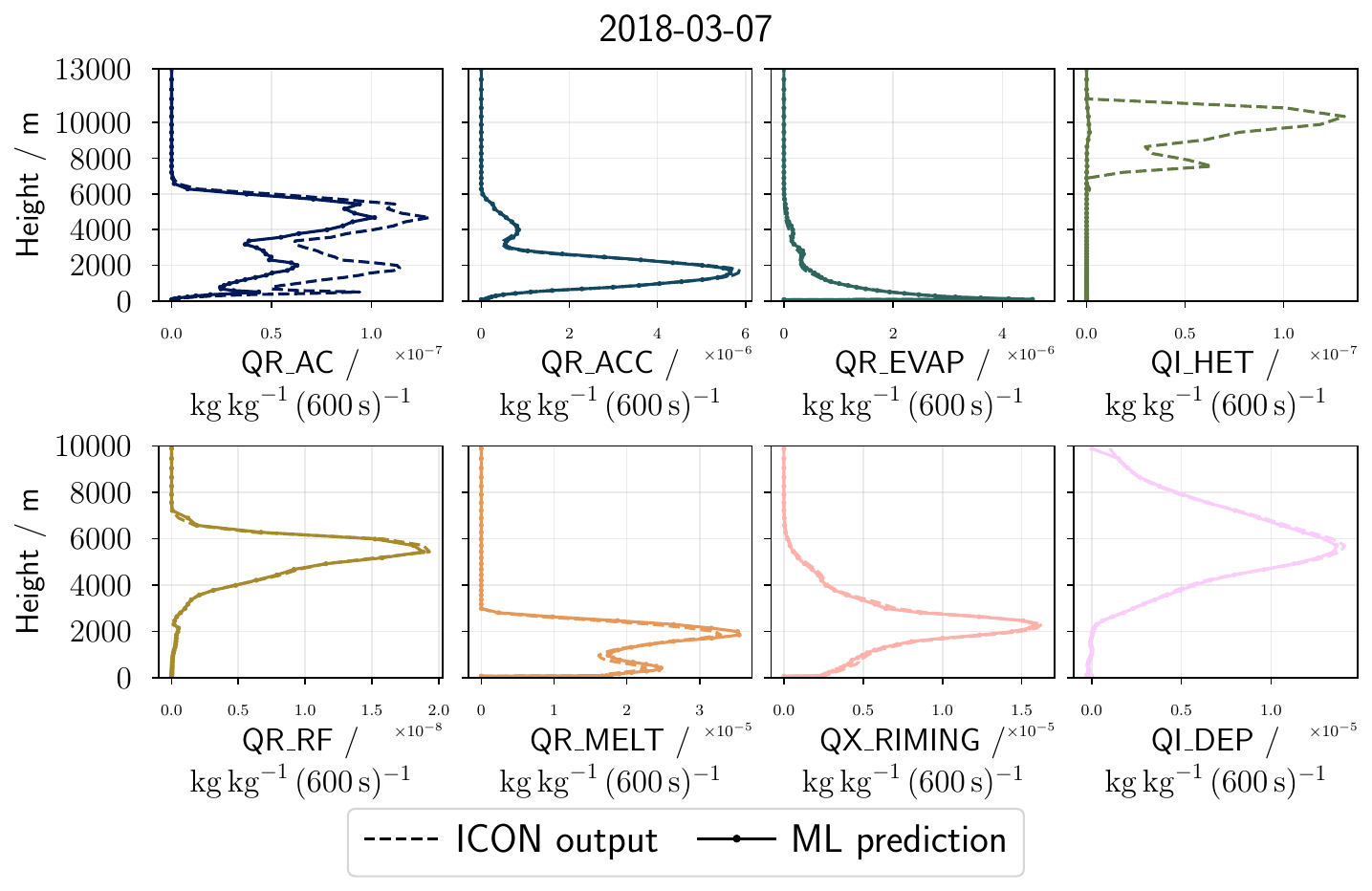}
    \caption{The vertical profiles of the spatio-temporal mean microphysical process rates show overall agreement between the ML predictions and the ICON model output, except for heterogeneous ice nucleation (QI\_HET). For autoconversion (QR\_AC) the predicted mean values are substantially lower than the ICON output across height levels. Shown are vertical profiles of eight process rates accumulated over a 10-minute output time step for 7 March 2018 with the ICE-POP domain (Fig.~\ref{fig:icepop_domain}), averaged over 24 hours and all grid points at each level. The solid-dotted line depicts the ML prediction and the dashed line the ICON model output.}
    \label{fig:vertical_profiles_icepop}
\end{figure*}

\begin{figure*}[t]
    \centering
    \includegraphics[width=12cm]{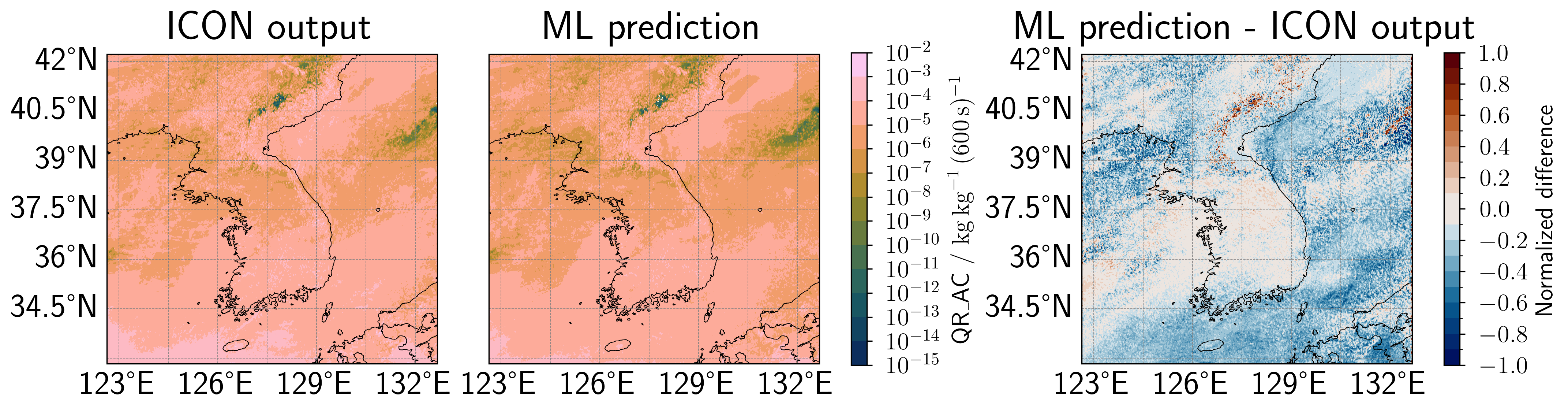}\\
    \includegraphics[width=12cm]{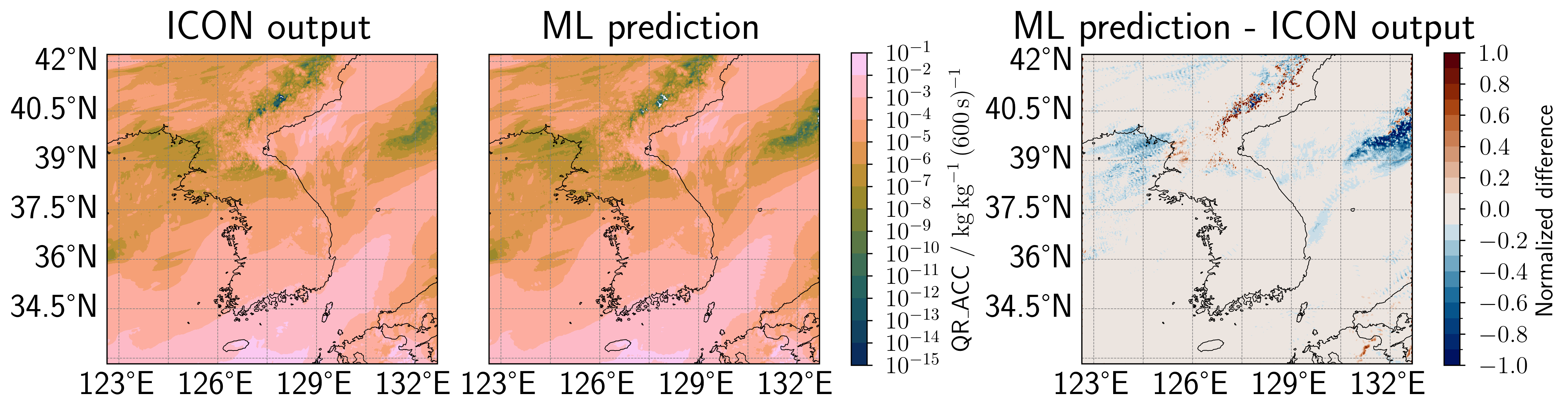}
    \caption{The spatial distributions of autoconversion (QR\_AC) and accretion (QR\_ACC), accumulated over a 10-minute output time step for 7 March 2018 with the ICE-POP domain (Fig.~\ref{fig:icepop_domain}), show similar patterns between the ICON model output (left) and the ML prediction (center) with small regional differences, normalized to the respective range of the process rate (right). The spatial distributions are averaged over all height levels and the \unit{24\,hour}-simulation interval. Additional figures are provided in App.~\ref{sec:spatial_distribution}.}
    \label{fig:2d_plots_icepop}
\end{figure*}

We repeat the previous analysis for the output from simulations of two case studies (see Sect.~\ref{sec:transferability_other_regions}) on different domains. In Fig.~\ref{fig:vertical_profiles_icepop}, we show the vertical profiles of spatio-temporal averages for the microphysical process rates for 7 March 2018 on the ICE-POP domain (Fig.~\ref{fig:icepop_domain}). Additionally, Fig.~\ref{fig:2d_plots_icepop} and Fig.~\ref{fig:2d_plots_icepop_additional_rates} show the corresponding spatial distributions. We find a good agreement between the ICON model output and the ML prediction, except for QI\_HET. For QR\_AC, the ML model slightly under-predicts. Interestingly, compared with the results for 23 June 2023, the agreement between model output and ML prediction appears to be even better, despite the ICE-POP model domain being outside the training domain. This indicates that the ML models generalize well across domains.

\begin{figure*}[t]
    \centering
    \includegraphics[width=12cm]{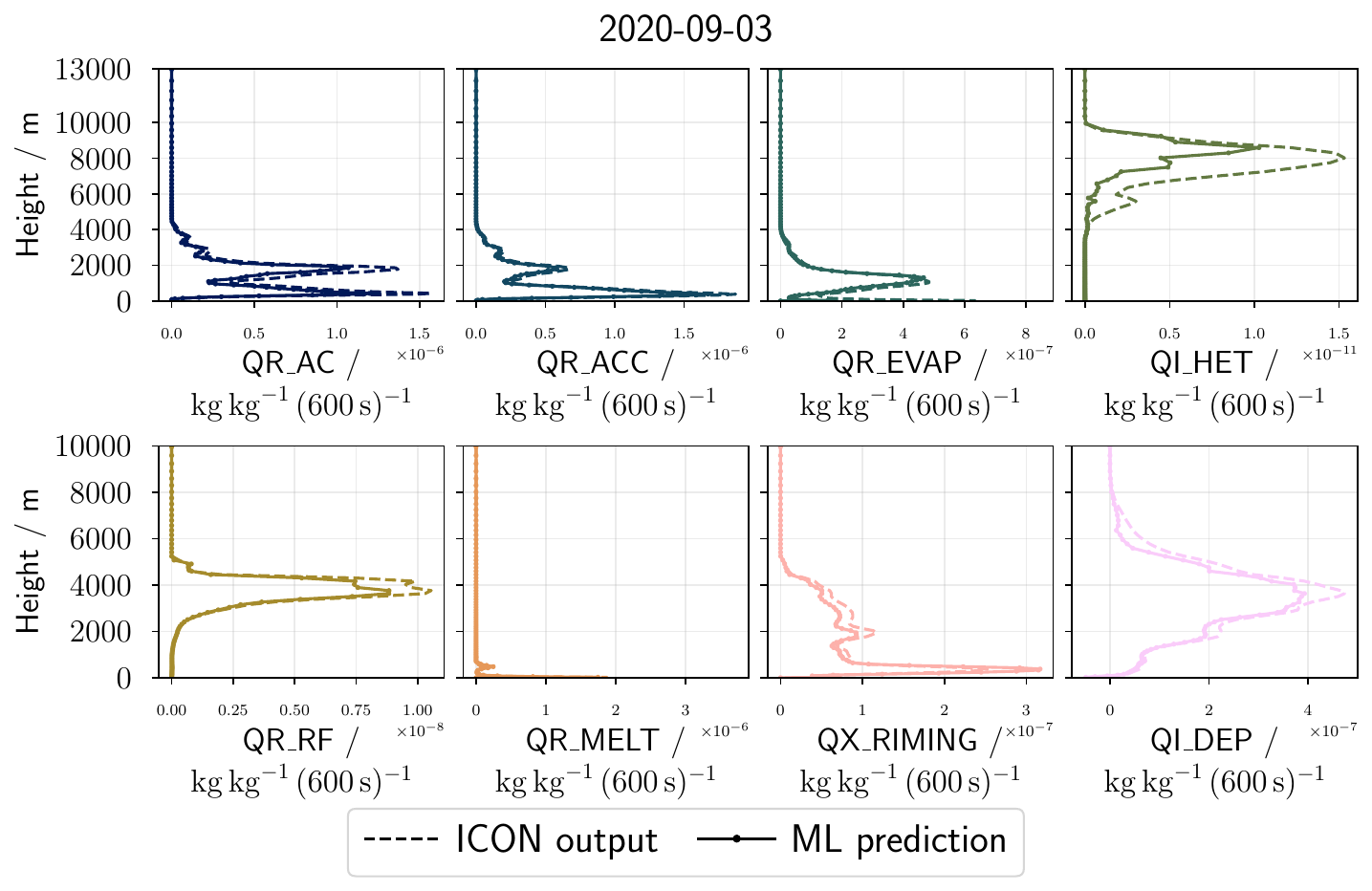}
    \caption{The vertical profiles of the spatio-temporal mean microphysical process rates show overall agreement between the ML predictions and the ICON model output. For heterogeneous ice nucleation (QI\_HET) the predicted mean values are substantially lower than the ICON output across height levels. Shown are vertical profiles of eight process rates accumulated over a 10-minute output time step for 3 September 2020 with the MOSAiC domain (Fig.~\ref{fig:mosaic_domain}), averaged over 12 hours and all grid points at each level. The solid-dotted line depicts the ML prediction and the dashed line the ICON model output.}
    \label{fig:vertical_profiles_mosaic}
\end{figure*}

\begin{figure*}[t]
    \centering
    \includegraphics[width=12cm]{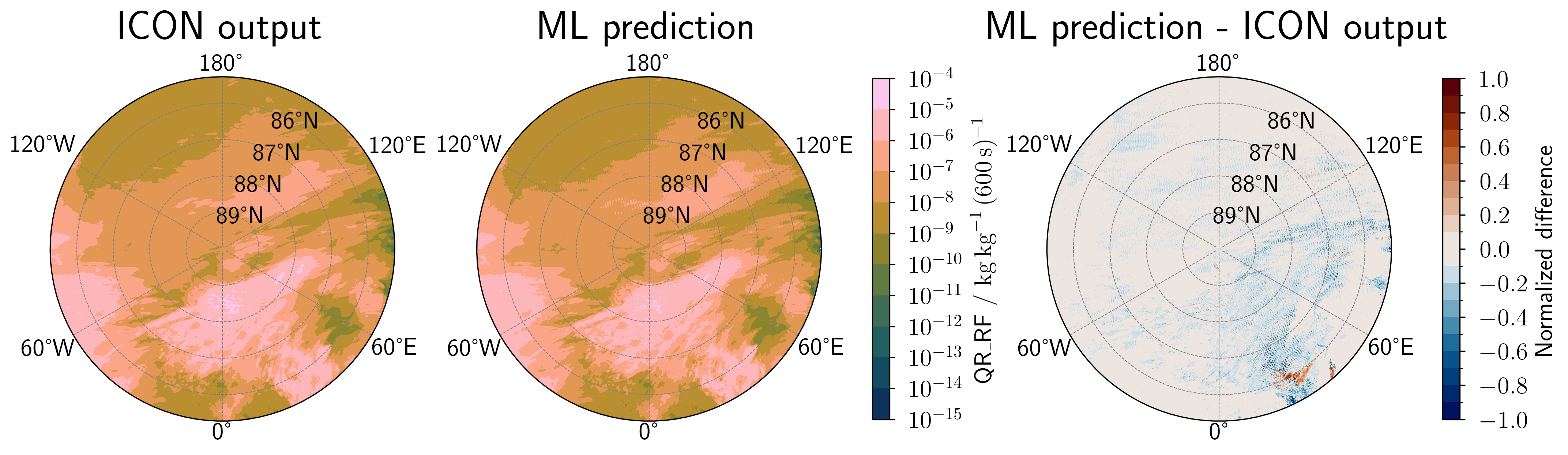}\\
    \includegraphics[width=12cm]{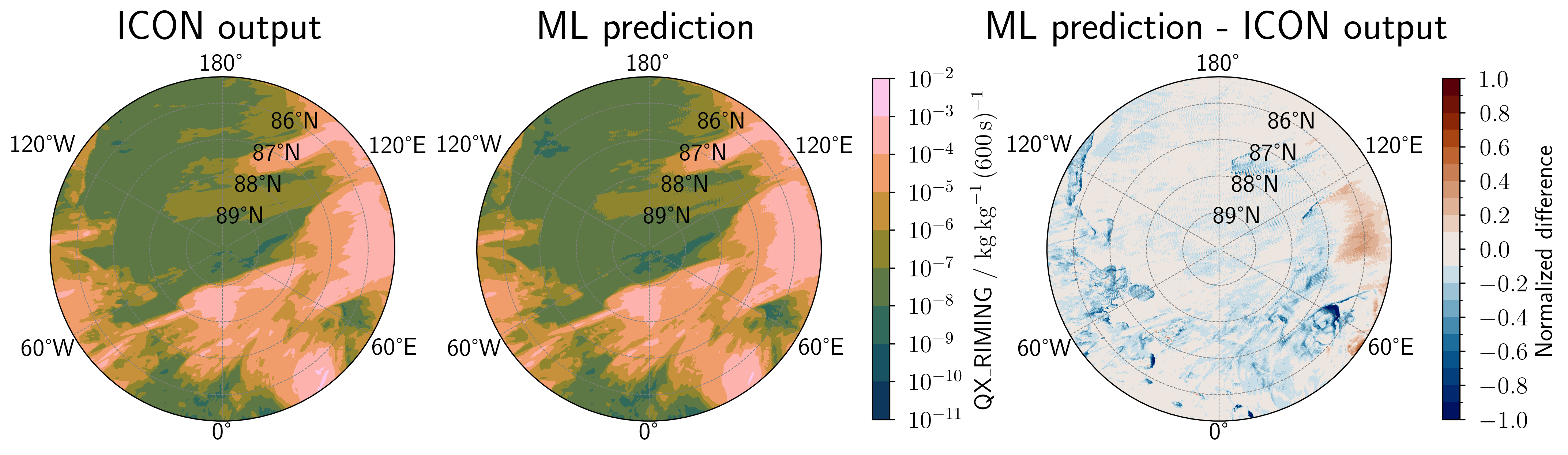}
    \caption{The spatial distributions of rain freezing (QR\_RF) and the total riming rate (QX\_RIMING), accumulated over a 10-minute output time step for 3 September 2020 with the MOSAiC domain (Fig.~\ref{fig:mosaic_domain}), show similar patterns between the ICON model output (left) and the ML prediction (center) with small regional differences, normalized to the respective range of the process rate (right). The spatial distributions are averaged over all height levels and the \unit{12\,hour}-simulation interval. Additional figures are provided in App.~\ref{sec:spatial_distribution}.}
    \label{fig:2d_plots_mosaic}
\end{figure*}

This is also the case for the ML predictions for 3 September 2020 with the MOSAiC domain (Fig.~\ref{fig:mosaic_domain}). In Fig.~\ref{fig:vertical_profiles_mosaic} and in Fig.~\ref{fig:2d_plots_mosaic} and Fig.~\ref{fig:2d_plots_mosaic_additional_rates}, we show the vertical profiles and the spatial distribution, respectively, of averages of the microphysical process rates. Again, there is a strong agreement between ICON model output and ML prediction. In contrast to the previous two cases, the ML model prediction for QI\_HET captures the main features of the vertical profile, yet the model under-predicts the true values. Importantly, this case was simulated with different ICON namelist settings than the training data (i.e. a smaller fast-physics time step and grid spacing and explicit convection) and the ICE-POP simulations. The fact that the models generalize well despite these differences demonstrates that PRecover is robust to moderate changes in the ICON model setup and is not overfit to the specific configuration used during training.

\conclusions[Conclusions] \label{sec:conclusions}
We have introduced PRecover, a data-driven approach for the recovery of microphysical process rates from output of the ICON model. This approach is particularly useful when such rates are not archived because of storage constraints, e.g., in global storm-resolving simulations. It may also prove useful for observational applications, where the coarse atmospheric state is easier to measure than the associated microphysical process rates, which are often difficult, if not impossible, to observe directly.

In this study, we have focused on microphysical process rates related to the \citet{seifertbehengTwomomentCloudMicrophysics2006} two-moment microphysics parameterization implemented in ICON. Our approach combines ML-based classification and regression models in a sequential two-step framework to emulate the computation of both instantaneous and accumulated warm-rain and ice microphysical process rates. For this purpose, we trained random forests (RFs), gradient boosting models and feed-forward neural networks (NNs) on data from high-resolution simulations in a limited-area configuration on the ICON-D2 domain, and evaluated the framework across output time steps ranging from one minute to 60 minutes and different model domains and configurations. To supplement the deterministic predictions with uncertainty estimates, we introduced a modified version of conformalized quantile regression (CQR) that yields prediction intervals rounded to the nearest order of magnitude, thereby providing information about the order of magnitude that the true value most likely falls in. We applied split conformal prediction (SCP) as an alternative where CQR proves unreliable.

We first assessed whether direct recalculation of the process rates from ICON model output is a feasible alternative to ML-based recovery. While only applicable to instantaneous process rates, the recalculation is successful for accretion (QR\_ACC) and both self-collection rates (QNC\_SC and QNR\_SC) but fails for other rates, such as autoconversion (QR\_AC) and melting to rain (QR\_MELT). The rain evaporation rate (QR\_EVAP) is recalculated reasonably well, while the recalculation of the rain freezing (QR\_RF) and vapor deposition on ice (QI\_DEP) rates are less accurate. Other process rates, such as the heterogeneous ice nucleation (QI\_HET) or the total riming (QX\_RIMING) rate cannot be recalculated at all, motivating the ML-based approach. 

The classification models achieve mean $\overline{F_1} \approx 0.85$ and $\overline{\text{MCC}} \approx 0.84$ for the instantaneous process rates, and $\overline{F_1} \approx 0.82$ and $\overline{\text{MCC}} \approx 0.80$ for process rates accumulated over a 10-minute output time step, demonstrating that the models reliably identify grid cells where process rate values exceed the pre-defined threshold. Notably, no ML architecture consistently outperforms the others. The regression models achieve $\overline{R^2} \approx 0.77$ for the instantaneous rates and $\overline{R^2} \approx 0.53$ for the accumulated process rates, capturing most of the variance for the instantaneous rates and substantial fraction of the variance for the accumulated rates.

The final classification-regression models achieve mean $\overline{R^2}_{\mathrm{inst.},\,\unit{10\,min}} \approx 0.66$ and $\overline{R^2}_{\mathrm{acc.},\,\unit{10\,min}} \approx 0.40$, averaged over all 25 process rates, demonstrating successful recovery of most process rates. Excluding QI\_HET, the mean scores increase to $\overline{R^2}_\mathrm{inst.,\,10\, min} \approx 0.69$ and $\overline{R^2}_\mathrm{acc.,\,10\, min} \approx 0.41$. Importantly, the ML-models clearly outperform the recalculation baseline for QR\_AC, QR\_RF and QI\_DEP. For QR\_ACC, QR\_EVAP and both self-collection rates (QNC\_SC, QNR\_SC), the ML results are comparable to the recalculation baseline. The $R^2_\mathrm{log}$ scores are notably more consistent across output time steps than the linear $R^2$ scores, which indicates that accumulation over longer output intervals primarily leads to a lower recover-ability of very high process rate values.

The analysis of the sensitivity of model performance to the output time step shows that the approach is most effective for instantaneous process rates and for rates accumulated over output time steps up to 10 minutes. Performance degrades for longer accumulation intervals because intermediate states are not available, which increases the difficulty of the prediction task. Improving the performance of the classification-regression models in a way that allows to recover process rates that are accumulated over longer output time steps could be achieved by compiling datasets from simulations with multiple output time steps, or incorporating additional output variables in the set of input features that allow the models to extract information about the intermediate states. Yet, we emphasize that this limitation is specific to the accumulated process rates, whereby the recovery of instantaneous process rates is independent of the output time step. 

Heterogeneous ice nucleation (QI\_HET) stands out as the process rate that cannot be reliably recovered, as reflected in consistently low MCC, $F_1$ and $R^2$ scores. The classification models achieve $F_1 = 0.07 - 0.39$ and $\mathrm{MCC} = 0.13 - 0.38$ depending on the output time step and model architecture, and the regression models yield $R^2$ scores close to zero. Furthermore, the analysis of the dependence of model performance on the output time step shows that the scores slightly increase with longer output time steps. The low performance can be explained by a combination of factors, most importantly the unavailability of the number of activated ice nuclei $n_\mathrm{in, act}$ as an input feature, as well as properties of the training, validation and test datasets. QI\_HET is characterized by very low values, which suggest that this regime is not well represented in the datasets. Including wind components $u$, $v$ and $w$ as additional input features does not significantly improve performance.

More generally, the recovery of some ice process rates remains challenging, in particular certain riming, melting and freezing rates, i.e. QC\_MELT, QG\_RF and QR\_RIME\_S. This could point towards missing input information and limitations of the datasets used for training the models. As we focus on multiple process rates simultaneously, all process rates are learned within a common framework and training setup, thus, process rate-specific limitations, for example in the training data, are not explicitly accounted for. While we apply thresholds for each individual process rate during pre-processing, future work targeting ice process rates such as QI\_HET specifically could benefit from simulations designed to better sample the respective regime. Furthermore, it would be instructive to approach the selection of input features in a more systematic way, e.g. by sequential feature selection, which was not conducted in this work due to limited computational resources.

CQR provides well-calibrated prediction intervals for most process rates with mean PICP values of $88.09\%$ and $86.95\%$ for the instantaneous and accumulated process rates, respectively (10-minute output time step), close to the nominal $90\%$ level. Rounding the prediction intervals to the nearest order of magnitude consistently improves the empirical coverage to above $90\%$, which comes at the cost of wider intervals. However, we argue that this trade-off is acceptable as this adjustment yields more reliable intervals, which provide information on the order of magnitude of the true process rate. In cases where the PICP values are slightly lower than $90\%$, this can be attributed to finite-sample variability and mild violations of the exchangeability assumption fundamental to conformal prediction, due to spatio-temporal structure of the data. The width of the prediction intervals increases with longer output time steps, reflecting the increasing uncertainty of the deterministic prediction at larger accumulation intervals. However, CQR fails for QI\_HET, yielding $\mathrm{PICP} \approx 43\% - 48\%$. It seems possible that this is due to the underlying quantile NN, therefore, we instead apply SCP to obtain prediction intervals from deterministic point predictions, which proves to be substantially more reliable. With SCP, we obtain prediction intervals that are considerably sharper, with PICP values close to or above the nominal $90\%$ level. 

The models trained on the ICON-D2 domain generalize well to domains unseen during training, as demonstrated by two case studies based on atmospheric measurement campaigns: a wintertime orographic precipitation case over Korea (ICE-POP domain, 7 March 2018) and an Arctic case (MOSAiC domain, 3 September 2020) run with different ICON namelist settings. In both cases, we observe a strong agreement between the ML predictions and the ICON model output for most process rates. The generalization capabilities are particularly important for the intended application of PRecover to global storm-resolving simulations, which employ different model configurations.

We do want to address a structural limitation of the presented approach. By training directly on output of simulations with a numerical weather prediction model, an implicit limitation to our approach lies in the fact that the models can only be as accurate in describing the underlying physics as the microphysics parameterization that was used in the simulations that provide the training data. As a bulk moment parameterization, the two-moment microphysics scheme involves various known simplifications, such as an artificial separation of the droplet size spectrum and an overly simplistic representation of cloud microphysics in general \citep{beuclerMachineLearningClouds2023}. We note, however, that this limitation applies to the specific models trained in this work, not to the framework of PRecover itself. In principle, PRecover could be re-trained with output from different microphysics schemes that are more physically accurate, in order to recover their associated process rates. Alternative parameterization schemes, such as bin microphysics and Lagrangian particle-based schemes would provide more physically consistent training data. While they are not yet computationally efficient enough to be used in global high-resolution simulations \citep{morrisonConfrontingChallengeModeling2020}, the recent success of an ML-based emulator trained on Lagrangian superdroplet simulations \citep{sharmaSuperdropnetStableAccurate2025} demonstrates the potential of these schemes as training data sources for ML models.

Regarding computational performance, data handling constitutes the primary bottleneck. While the ML inference itself is rather fast, the data handling is time-consuming, up to the point where the direct ICON model simulation could even be faster for short simulations over small modeling domains. The total runtime for the prediction of one process rate for a full date strongly depends on the model output time step, the number of vertical levels and the horizontal and temporal extent of the simulation and domain. The recovery of the autoconversion rate (QR\_AC) for 23 June 2023 for 10-minute output intervals on the ICON-D2 domain on a single CPU node (152 cores) takes approximately $80$ minutes. We notice that the most resource-intensive and time-consuming part of the recovery pipeline is the processing of the ICON simulation output. We believe that improvements in runtime are attainable through technical optimization of the data processing steps, but this is beyond the scope of this work.

To conclude, our approach helps obtain information about microphysical process rates in a more resource-efficient and flexible way, opening the possibility to make use of extensive high-resolution simulations for studying cloud microphysics. While we have focused on the ICON model with the \citet{seifertbehengTwomomentCloudMicrophysics2006} two-moment microphysics scheme, the general PRecover framework is transferable to other numerical models and microphysics parameterizations. It can be used for the recovery of quantities from the output of numerical models that have not been archived due to storage constraints, enabling in-depth studies of small-scale processes based on existing model output even when a different scheme was employed, and potentially extending to directly using atmospheric state observations.

\codedataavailability{The exact version of the model used to produce the results used in this paper is archived on Zenodo under DOI \url{https://doi.org/10.5281/zenodo.19114989}, as are input data and scripts to run the model and produce the plots for all the simulations presented in this paper. The code is also publicly available in the GitHub repository \url{https://github.com/miriamsimm/PRecover}. The source code of the ICON model is available under the BSD-3-Clause open source license at \url{https://www.icon-model.org}.}

\authorcontribution{Conceptualization: MS, TB, CH, Data curation: MS, Formal analysis: MS, Investigation: MS, Methodology: MS, TB, CH, Visualization: MS, Writing - original draft: MS, Writing - review \& editing: MS, TB, CH. All authors approved the final submitted draft.}
\competinginterests{The authors declare that they have no conflict of interest.}
\begin{acknowledgements}
The authors gratefully acknowledge the work of J.~Hesemann (Karlsruhe Institute of Technology), which has motivated the presented research. MS thanks C.~Barthlott (Karlsruhe Institute of Technology) for support with the ICON model simulations and the process rate output. MS thanks G.~Wallentin (Karlsruhe Institute of Technology) for helpful comments and support with the ICON model simulations on the Arctic domain. MS thanks P.~Stier (University of Oxford) for helpful discussions. The GitHub Copilot extension for Visual Studio Code was used to support the generation of the model code with code completions, which were carefully reviewed by the authors. Figure 1 was created with the \LaTeX package Ti\textit{k}Z \citep{tikzTantau2013}. All figures in this manuscript use the Scientific color maps 7.0 \citep{crameriMisuseColourScience2020}.

This work was performed with the help of the Large Scale Data Facility at the Karlsruhe Institute of Technology funded by the Ministry of Science, Research and the Arts Baden-Württemberg and by the Federal Ministry of Education and Research. The authors gratefully acknowledge the computing time provided on the high-performance computer HoreKa by the National High-Performance Computing Center at KIT (NHR@KIT). This center is jointly supported by the Federal Ministry of Education and Research and the Ministry of Science, Research and the Arts of Baden-Württemberg, as part of the National High-Performance Computing (NHR) joint funding program (https://www.nhr-verein.de/en/our-partners). HoreKa is partly funded by the German Research Foundation (DFG).
\end{acknowledgements}
\financialsupport{This project was funded by the NHR Call for Collaboration Project MICRO. In addition, MS gratefully acknowledges financial support by the Karlsruhe House of Young Scientists (KHYS). CH acknowledges funding through the Horizon Europe Cluster 5 project CleanCloud (Grant no. 101137639). TB acknowledges funding from the Swiss State Secretariat for Education, Research and Innovation (SERI) for the Horizon Europe project AI4PEX (Grant agreement ID: 101137682 and SERI no 23.00546).}

\appendix

\section{Additional results}\label{sec:appendix_results}
\subsection{Classification}
In Table~\ref{table:results_classification}, we present the $F_1$ score and the MCC for the classification models for 10 instantaneous and accumulated process rates for different output time steps, as shown in Fig.~\ref{fig:results_classification_bars}.
\begin{table*}[t]
    \centering
    \caption{MCC and $F_1$ score of the classification models for 10 instantaneous process rates with a 10-minute output time step and process rates accumulated over a one-minute, 10-minute and 30-minute output time step, computed on the test set.}
    \label{table:results_classification}
    \renewcommand\baselinestretch{1.0}\selectfont
    \begin{tabular}{l c c c c c c c c}
        \tophline
        \multirow{2}{*}{Process rate} & \multicolumn{2}{c}{Logistic regression} & \multicolumn{2}{c}{Random forest (RF)} & \multicolumn{2}{c}{Gradient boosting (XGB)} & \multicolumn{2}{c}{Neural network (NN)} \\
        \cline{2-9} & {$F_1$} & {MCC} & {$F_1$} & {MCC} & {$F_1$} & {MCC} & {$F_1$} & {MCC}  \\
        \middlehline
        \multicolumn{9}{l}{\textbf{Instantaneous process rates, 10-minute output time step}} \\
        \quad QR\_AC     & $0.86$ & $0.84$ & $0.96$ & $0.96$ & $0.96$ & $0.96$ & $0.96$ & $0.95$ \\
        \quad QR\_ACC    & $0.44$ & $0.40$ & $0.76$ & $0.72$ & $0.76$ & $0.73$ & $0.76$ & $0.72$ \\
        \quad QNC\_SC    & $0.88$ & $0.86$ & $0.95$ & $0.93$ & $0.95$ & $0.94$ & $0.95$ & $0.94$ \\
        \quad QNR\_SC    & $0.61$ & $0.55$ & $0.99$ & $0.99$ & $0.96$ & $0.98$ & $0.99$ & $0.99$ \\
        \quad QR\_EVAP   & $0.84$ & $0.54$ & $0.91$ & $0.75$ & $0.91$ & $0.75$ & $0.88$ & $0.66$ \\
        \quad QI\_HET    & $0.00$ & $0.02$ & $0.19$ & $0.23$ & $0.20$ & $0.24$ & $0.07$ & $0.13$ \\
        \quad QR\_RF     & $0.02$ & $0.05$ & $0.97$ & $0.97$ & $0.97$ & $0.97$ & $0.97$ & $0.97$ \\
        \quad QR\_MELT   & $0.26$ & $0.31$ & $0.94$ & $0.94$ & $0.94$ & $0.94$ & $0.81$ & $0.80$ \\
        \quad QX\_RIMING & $0.31$ & $0.32$ & $0.99$ & $0.99$ & $0.99$ & $0.99$ & $0.93$ & $0.91$ \\
        \quad QI\_DEP    & $0.46$ & $0.38$ & $0.90$ & $0.87$ & $0.93$ & $0.90$ & $0.82$ & $0.88$ \\
        \multicolumn{9}{l}{\textbf{Accumulated process rates, one-minute output time step}} \\
        \quad QR\_AC     & $0.87$ & $0.85$ & $0.96$ & $0.95$ & $0.96$ & $0.95$ & $0.96$ & $0.95$ \\
        \quad QR\_ACC    & $0.51$ & $0.46$ & $0.77$ & $0.72$ & $0.76$ & $0.72$ & $0.75$ & $0.70$ \\
        \quad QNC\_SC    & $0.86$ & $0.83$ & $0.93$ & $0.92$ & $0.93$ & $0.92$ & $0.93$ & $0.92$ \\
        \quad QNR\_SC    & $0.66$ & $0.74$ & $0.99$ & $0.99$ & $0.98$ & $0.99$ & $0.99$ & $0.99$ \\
        \quad QR\_EVAP   & $0.87$ & $0.58$ & $0.91$ & $0.74$ & $0.92$ & $0.74$ & $0.90$ & $0.70$ \\
        \quad QI\_HET    & $0.00$ & $0.00$ & $0.20$ & $0.25$ & $0.23$ & $0.27$ & $0.08$ & $0.13$ \\
        \quad QR\_RF     & $0.07$ & $0.13$ & $0.97$ & $0.97$ & $0.97$ & $0.97$ & $0.97$ & $0.97$ \\
        \quad QR\_MELT   & $0.27$ & $0.31$ & $0.95$ & $0.94$ & $0.94$ & $0.94$ & $0.83$ & $0.82$ \\
        \quad QX\_RIMING & $0.32$ & $0.33$ & $0.99$ & $0.98$ & $0.99$ & $0.99$ & $0.95$ & $0.95$ \\
        \quad QI\_DEP    & $0.47$ & $0.39$ & $0.90$ & $0.86$ & $0.93$ & $0.90$ & $0.79$ & $0.86$ \\    
        \multicolumn{9}{l}{\textbf{Accumulated process rates, 10-minute output time step}} \\
        \quad QR\_AC     & $0.83$ & $0.80$ & $0.90$ & $0.88$ & $0.90$ & $0.88$ & $0.90$ & $0.88$ \\
        \quad QR\_ACC    & $0.55$ & $0.47$ & $0.76$ & $0.68$ & $0.76$ & $0.68$ & $0.74$ & $0.65$ \\
        \quad QNC\_SC    & $0.76$ & $0.71$ & $0.82$ & $0.77$ & $0.82$ & $0.77$ & $0.81$ & $0.77$ \\
        \quad QNR\_SC    & $0.61$ & $0.70$ & $0.94$ & $0.93$ & $0.93$ & $0.93$ & $0.93$ & $0.93$ \\
        \quad QR\_EVAP   & $0.91$ & $0.54$ & $0.94$ & $0.73$ & $0.94$ & $0.73$ & $0.94$ & $0.70$ \\
        \quad QI\_HET    & $0.02$ & $0.07$ & $0.27$ & $0.30$ & $0.30$ & $0.32$ & $0.26$ & $0.26$ \\
        \quad QR\_RF     & $0.08$ & $0.13$ & $0.92$ & $0.92$ & $0.92$ & $0.92$ & $0.91$ & $0.91$ \\
        \quad QR\_MELT   & $0.32$ & $0.35$ & $0.93$ & $0.92$ & $0.93$ & $0.92$ & $0.81$ & $0.79$ \\
        \quad QX\_RIMING & $0.39$ & $0.38$ & $0.96$ & $0.95$ & $0.96$ & $0.95$ & $0.92$ & $0.90$ \\
        \quad QI\_DEP    & $0.46$ & $0.38$ & $0.87$ & $0.79$ & $0.90$ & $0.82$ & $0.89$ & $0.84$ \\
        \multicolumn{9}{l}{\textbf{Accumulated process rates, 30-minute output time step}} \\
        \quad QR\_AC     & $0.75$ & $0.72$ & $0.83$ & $0.79$ & $0.83$ & $0.79$ & $0.82$ & $0.77$ \\
        \quad QR\_ACC    & $0.59$ & $0.34$ & $0.73$ & $0.56$ & $0.73$ & $0.56$ & $0.70$ & $0.51$ \\
        \quad QNC\_SC    & $0.67$ & $0.56$ & $0.75$ & $0.62$ & $0.75$ & $0.61$ & $0.70$ & $0.56$ \\
        \quad QNR\_SC    & $0.58$ & $0.71$ & $0.79$ & $0.84$ & $0.79$ & $0.85$ & $0.78$ & $0.84$ \\
        \quad QR\_EVAP   & $0.94$ & $0.43$ & $0.96$ & $0.63$ & $0.96$ & $0.63$ & $0.95$ & $0.57$ \\
        \quad QI\_HET    & $0.06$ & $0.11$ & $0.35$ & $0.35$ & $0.39$ & $0.38$ & $0.38$ & $0.36$ \\
        \quad QR\_RF     & $0.07$ & $0.08$ & $0.82$ & $0.81$ & $0.82$ & $0.81$ & $0.81$ & $0.80$ \\
        \quad QR\_MELT   & $0.35$ & $0.38$ & $0.90$ & $0.89$ & $0.90$ & $0.89$ & $0.79$ & $0.77$ \\
        \quad QX\_RIMING & $0.43$ & $0.40$ & $0.91$ & $0.89$ & $0.91$ & $0.89$ & $0.86$ & $0.82$ \\
        \quad QI\_DEP    & $0.46$ & $0.37$ & $0.85$ & $0.73$ & $0.87$ & $0.76$ & $0.84$ & $0.76$ \\    
        \bottomhline
    \end{tabular}
\end{table*} 

\subsection{Regression}
In Table~\ref{table:results_regression}, we present the RMSE and the $R^2$ score for the regression models for 10 instantaneous and accumulated process rates for different output time steps, as shown in Fig.~\ref{fig:results_regression_bars}.
\begin{table*}[t]
    \centering
    \caption{RMSE and $R^2$ score of the regression models for 10 instantaneous process rates and a 10-minute output time step and process rates accumulated over a one-minute, 10-minute and 30-minute output time step, computed on the test set. For each time step and process rate, the best performing model, in terms of the $R^2$ score, is highlighted in bold. The time step \unit{t} is either $t^\text{out}$ or $t^\text{fast}$. Continued on the next page.}
    \label{table:results_regression}
    \renewcommand\baselinestretch{0.6}\selectfont
    \begin{tabular}{l c c c c}
        \tophline
        \multirow{2}{*}{Process rate} & \multicolumn{2}{c}{Linear regression} & \multicolumn{2}{c}{Random forest (RF)} \\
        \cline{2-3} \cline{4-5} & {RMSE / \unit{kg}\,\unit{kg}$^{-1}\,\unit{t}^{-1}[\text{s}]$} & {$R^2$} & {RMSE / \unit{kg}\,\unit{kg}$^{-1}\,\unit{t}^{-1}[\text{s}]$} & {$R^2$}  \\
        \middlehline
        \multicolumn{5}{l}{\textbf{Instantaneous process rates, 10-minute output time step}} \\
        \quad QR\_AC  & $7.58 \times 10^{-7}$ & $0.16$ & $3.60 \times 10^{-7}$ & $\mathbf{0.81}$ \\
        \quad QR\_ACC & $4.14 \times 10^{-6}$ & $0.45$ & $4.23 \times 10^{-7}$ & $0.99$ \\
        \quad QNC\_SC & $8.16 \times 10^{4}\,\unit{kg}^{-1}$  & $0.59$ & $5.40 \times 10^{3}\,\unit{kg}^{-1}$ & $1.00$ \\
        \quad QNR\_SC\_POS & $28.89\,\unit{kg}^{-1}$ & $0.06$ & $3.74\,\unit{kg}^{-1}$ & $0.98$ \\
        \quad QNR\_SC\_NEG & $292.09\,\unit{kg}^{-1}$ & $0.52$  & $39.66\,\unit{kg}^{-1}$ & $\mathbf{0.99}$ \\
        \quad QR\_EVAP & $4.06 \times 10^{-7}$ & $0.61$  & $9.75 \times 10^{-8}$ & $0.98$ \\
        \quad QI\_HET & $2.31 \times 10^{-6}$ & $0.00$  & $1.05 \times 10^{-6}$ & $\mathbf{0.79}$ \\
        \quad QR\_RF & $4.90 \times 10^{-6}$ & $0.00$  & $3.39 \times 10^{-6}$ & $0.52$ \\
        \quad QR\_MELT & $1.03 \times 10^{-5}$ & $0.00$  & $1.25 \times 10^{-6}$ & $0.99$ \\
        \quad QX\_RIMING & $5.19 \times 10^{-6}$ & $0.26$  & $2.00 \times 10^{-6}$ & $0.89$ \\
        \quad QI\_DEP\_POS & $1.05 \times 10^{-6}$ & $0.39$  & $5.57 \times 10^{-7}$ & $\mathbf{0.83}$ \\
        \quad QI\_DEP\_NEG & $2.50 \times 10^{-7}$ & $0.19$  & $1.67 \times 10^{-7}$ & $\mathbf{0.64}$ \\
        \multicolumn{5}{l}{\textbf{Accumulated process rates, one-minute output time step}} \\
        \quad QR\_AC & $1.41 \times 10^{-6}$ & $0.40$  & $7.29 \times 10^{-7}$ & $\mathbf{0.84}$ \\
        \quad QR\_ACC & $1.18 \times 10^{-5}$ & $0.45$  & $2.20 \times 10^{-6}$ & $0.98$ \\
        \quad QNC\_SC & $2.17 \times 10^{5}\,\unit{kg}^{-1}$ & $0.60$  & $1.89 \times 10^{4}\,\unit{kg}^{-1}$ & $1.00$ \\
        \quad QNR\_SC\_POS & $63.60\,\unit{kg}^{-1}$ & $0.07$ & $5.67\,\unit{kg}^{-1}$ & $\mathbf{0.99}$ \\
        \quad QNR\_SC\_NEG & $728.34\,\unit{kg}^{-1}$ & $0.52$ & $150.21\,\unit{kg}^{-1}$ & $0.98$ \\
        \quad QR\_EVAP & $1.22 \times 10^{-6}$ & $0.60$ & $3.38 \times 10^{-7}$ & $0.97$ \\
        \quad QI\_HET & $3.69 \times 10^{-6}$ & $0.03$ & $2.80 \times 10^{-6}$ & $\mathbf{0.44}$  \\
        \quad QR\_RF & $1.16 \times 10^{-5}$ & $0.02$ & $7.73 \times 10^{-6}$ & $\mathbf{0.57}$ \\
        \quad QR\_MELT & $1.89 \times 10^{-5}$ & $0.62$ & $3.57 \times 10^{-6}$ & $0.99$ \\
        \quad QX\_RIMING & $1.50 \times 10^{-5}$ & $0.28$ & $5.74 \times 10^{-6}$ & $0.90$ \\
        \quad QI\_DEP\_POS & $2.99 \times 10^{-6}$ & $0.40$ & $1.49 \times 10^{-6}$ & $\mathbf{0.85}$ \\
        \quad QI\_DEP\_NEG & $6.81 \times 10^{-7}$ & $0.20$ & $4.70 \times 10^{-7}$ & $0.62$ \\        
        \multicolumn{5}{l}{\textbf{Accumulated process rates, 10-minute output time step}} \\
        \quad QR\_AC & $7.40 \times 10^{-6}$ & $0.20$  & $6.94 \times 10^{-6}$ & $0.30$ \\
        \quad QR\_ACC & $8.58 \times 10^{-5}$ & $0.35$  & $5.00 \times 10^{-5}$ & $0.78$ \\
        \quad QNC\_SC & $1.89 \times 10^{6}\,\unit{kg}^{-1}$ & $0.50$  & $9.39 \times 10^{5}\,\unit{kg}^{-1}$ & $\mathbf{0.88}$ \\
        \quad QNR\_SC\_POS & $362.84\,\unit{kg}^{-1}$ & 0.00 & $176.02\,\unit{kg}^{-1}$ & $0.76$ \\
        \quad QNR\_SC\_NEG & $4.13 \times 10^{3}\,\unit{kg}^{-1}$ & $0.39$  & $3.45 \times 10^{3}\,\unit{kg}^{-1}$ & $\mathbf{0.57}$ \\        
        \quad QR\_EVAP & $1.18 \times 10^{-5}$ & $0.46$  & $6.49 \times 10^{-6}$ & $0.83$ \\
        \quad QI\_HET & $9.62 \times 10^{-6}$ & $0.00$  & $9.40 \times 10^{-6}$ & $0.05$ \\
        \quad QR\_RF & $6.96 \times 10^{-5}$ & $0.00$  & $6.16 \times 10^{-5}$ & $0.22$ \\
        \quad QR\_MELT & $1.73 \times 10^{-4}$ & $0.55$  & $7.42 \times 10^{-5}$ & $0.92$ \\
        \quad QX\_RIMING & $1.12 \times 10^{-4}$ & $0.29$  & $5.99 \times 10^{-5}$ & $0.80$ \\
        \quad QI\_DEP\_POS & $2.47 \times 10^{-5}$ & $0.37$  & $1.95 \times 10^{-5}$ & $0.61$ \\
        \quad QI\_DEP\_NEG & $5.03 \times 10^{-6}$ & $0.24$  & $3.73 \times 10^{-6}$ & $0.58$ \\ 
        \multicolumn{5}{l}{\textbf{Accumulated process rates, 30-minute output time step}} \\
        \quad QR\_AC & $1.34 \times 10^{-5}$ & $0.11$  & $1.28 \times 10^{-5}$ & $0.18$ \\
        \quad QR\_ACC & $1.78 \times 10^{-4}$ & $0.19$  & $1.59 \times 10^{-4}$ & $\mathbf{0.36}$ \\
        \quad QNC\_SC & $4.17 \times 10^{6}\,\unit{kg}^{-1}$ & $0.37$  & $3.69 \times 10^{6}\,\unit{kg}^{-1}$ & $0.51$ \\
        \quad QNR\_SC\_POS & $714.65\,\unit{kg}^{-1}$ & $-0.02$  & $591.46\,\unit{kg}^{-1}$ & $0.30$ \\
        \quad QNR\_SC\_NEG & $1.07 \times 10^{4}\,\unit{kg}^{-1}$ & $0.22$  & $1.05 \times 10^{4}\,\unit{kg}^{-1}$ & $\mathbf{0.25}$ \\
        \quad QR\_EVAP & $3.24 \times 10^{-5}$ & $0.26$  & $2.26 \times 10^{-5}$ & $0.64$ \\
        \quad QI\_HET & $1.45 \times 10^{-5}$ & $0.00$  & $1.44 \times 10^{-5}$ & $0.00$ \\
        \quad QR\_RF & $1.11 \times 10^{-4}$ & $0.00$  & $1.05 \times 10^{-4}$ & 0.11 \\
        \quad QR\_MELT & $4.62 \times 10^{-4}$ & $0.44$  & $3.05 \times 10^{-4}$ & $0.76$ \\
        \quad QX\_RIMING & $2.59 \times 10^{-4}$ & $0.19$  & $1.83 \times 10^{-4}$ & $0.60$ \\
        \quad QI\_DEP\_POS & $6.27 \times 10^{-5}$ & $0.35$  & $5.72 \times 10^{-5}$ & $\mathbf{0.46}$ \\
        \quad QI\_DEP\_NEG & $1.30 \times 10^{-5}$ & $0.22$  & $1.06 \times 10^{-5}$ & $0.49$ \\
    \bottomhline
    \end{tabular}
\end{table*}

\begin{table*}[t]
    \centering
    \caption{Continuation of Table~\ref{table:results_regression}. RMSE and $R^2$ score of the regression models for 12 instantaneous process rates and a 10-minute output time step and process rates accumulated over a one-minute, 10-minute and 30-minute output time step, computed on the test set. For each time step and process rate, the best performing model, in terms of the $R^2$ score, is highlighted in bold. The time step \unit{t} is either $t^\text{out}$ or $t^\text{fast}$.}
    \renewcommand\baselinestretch{0.6}\selectfont
    \begin{tabular}{l c c c c}
        \tophline
        \multirow{2}{*}{Process rate} & \multicolumn{2}{c}{Gradient boosting (XGB)} & \multicolumn{2}{c}{Neural network (NN)} \\
        \cline{2-3} \cline{4-5} & {RMSE / \unit{kg}\,\unit{kg}$^{-1}\,\unit{t}^{-1}[\text{s}]$} & {$R^2$} & {RMSE / \unit{kg}\,\unit{kg}$^{-1}\,\unit{t}^{-1}[\text{s}]$} & {$R^2$}  \\
        \middlehline
        \multicolumn{5}{l}{\textbf{Instantaneous process rates, 10-minute output time step}} \\
        \quad QR\_AC & $4.86 \times 10^{-7}$ &  $0.65$  & $4.13 \times 10^{-7}$ & $0.75$ \\
        \quad QR\_ACC & $2.84 \times 10^{-6}$ & $0.74$  & $3.11 \times 10^{-7}$ & $\mathbf{1.00}$ \\
        \quad QNC\_SC & $3.83 \times 10^{4}\,\unit{kg}^{-1}$ &  $0.91$  & $5.73 \times 10^{3}\,\unit{kg}^{-1}$ & $\mathbf{1.00}$ \\
        \quad QNR\_SC\_POS & $8.74\,\unit{kg}^{-1}$ & $0.91$  & $0.73\,\unit{kg}^{-1}$ & $\mathbf{1.00}$ \\
        \quad QNR\_SC\_NEG & $179.45\,\unit{kg}^{-1}$ & $0.82$  & $61.92\,\unit{kg}^{-1}$ & $0.98$ \\
        \quad QR\_EVAP & $2.59 \times 10^{-7}$ & $0.84$ & $7.13 \times 10^{-8}$ & $\mathbf{0.99}$ \\
        \quad QI\_HET & $1.80 \times 10^{-6}$ & $0.39$ & $1.34 \times 10^{-6}$ & $0.66$ \\
        \quad QR\_RF & $3.75 \times 10^{-6}$ & $0.41$ & $2.80 \times 10^{-6}$ & $\mathbf{0.67}$ \\
        \quad QR\_MELT & $1.68 \times 10^{-6}$ & $0.97$ & $9.78 \times 10^{-7}$ & $\mathbf{0.99}$ \\
        \quad QX\_RIMING & $1.99 \times 10^{-6}$ & $0.89$ & $1.91 \times 10^{-6}$ & $\mathbf{0.90}$ \\
        \quad QI\_DEP\_POS & $6.74 \times 10^{-7}$ & $0.75$ & $9.27 \times 10^{-7}$ & $0.53$ \\
        \quad QI\_DEP\_NEG & $3.86 \times 10^{-7}$ & $-0.92$ & $2.51 \times 10^{-7}$ & $0.19$ \\
        \multicolumn{5}{l}{\textbf{Accumulated process rates, one-minute output time step}} \\
        \quad QR\_AC & $7.64 \times 10^{-7}$ & $0.82$ & $8.24 \times 10^{-7}$ & $0.79$ \\
        \quad QR\_ACC & $8.37 \times 10^{-6}$ & $0.73$ & $2.15 \times 10^{-6}$ & $\mathbf{0.98}$ \\
        \quad QNC\_SC & $9.07 \times 10^{4}\,\unit{kg}^{-1}$ & $0.93$ & $1.49 \times 10^{4}\,\unit{kg}^{-1}$ & $\mathbf{1.00}$ \\
        \quad QNR\_SC\_POS & $17.40\,\unit{kg}^{-1}$ & $0.93$ & $6.69\,\unit{kg}^{-1}$ & $0.99$ \\
        \quad QNR\_SC\_NEG & $553.35\,\unit{kg}^{-1}$ & $0.72$ & $132.57\,\unit{kg}^{-1}$ & $\mathbf{0.98}$ \\
        \quad QR\_EVAP & $7.46 \times 10^{-7}$ & $0.85$ & $2.13 \times 10^{-7}$ & $\mathbf{0.99}$ \\
        \quad QI\_HET & $2.91 \times 10^{-6}$ & $0.40$ & $3.02 \times 10^{-6}$ & $0.35$ \\
        \quad QR\_RF & $9.07 \times 10^{-6}$ & $0.41$ & $8.82 \times 10^{-6}$ & $0.44$ \\
        \quad QR\_MELT & $5.01 \times 10^{-6}$ & $0.97$ & $2.85 \times 10^{-6}$ & $\mathbf{0.99}$ \\
        \quad QX\_RIMING & $5.01 \times 10^{-6}$ & $\mathbf{0.92}$ & $2.32 \times 10^{-4}$ & $-170.76$ \\
        \quad QI\_DEP\_POS & $1.84 \times 10^{-6}$ & $0.77$ & $2.52 \times 10^{-6}$ & $0.57$ \\
        \quad QI\_DEP\_NEG & $4.49 \times 10^{-7}$ & $0.65$ & $3.55 \times 10^{-7}$ & $\mathbf{0.78}$ \\        
        \multicolumn{5}{l}{\textbf{Accumulated process rates, 10-minute output time step}} \\
        \quad QR\_AC & $7.02 \times 10^{-6}$ & $0.28$ & $6.72 \times 10^{-6}$ & $\mathbf{0.34}$ \\
        \quad QR\_ACC & $6.26 \times 10^{-5}$ & $0.65$ & $4.66 \times 10^{-5}$ & $\mathbf{0.81}$ \\
        \quad QNC\_SC & $1.12 \times 10^{6}\,\unit{kg}^{-1}$ & $0.83$ & $1.26 \times 10^{6}\,\unit{kg}^{-1}$ & $0.78$ \\
        \quad QNR\_SC\_POS & $181.89\,\unit{kg}^{-1}$ & $0.75$ & $156.59\,\unit{kg}^{-1}$ & $\mathbf{0.81}$ \\
        \quad QNR\_SC\_NEG & $3.62 \times 10^{3}\,\unit{kg}^{-1}$ & $0.53$ & $3.96 \times 10^{3}\,\unit{kg}^{-1}$ & $0.44$ \\        
        \quad QR\_EVAP & $7.22 \times 10^{-6}$ & $0.79$ & $6.04 \times 10^{-6}$ & $\mathbf{0.86}$ \\
        \quad QI\_HET & $9.48 \times 10^{-6}$ & $0.03$ & $9.16 \times 10^{-6}$ & $\mathbf{0.10}$ \\
        \quad QR\_RF & $6.23 \times 10^{-5}$ & $0.20$ & $5.84 \times 10^{-5}$ & $\mathbf{0.30}$ \\
        \quad QR\_MELT & $7.00 \times 10^{-5}$ & $0.93$ & $6.80 \times 10^{-5}$ & $\mathbf{0.93}$ \\
        \quad QX\_RIMING & $5.49 \times 10^{-5}$ & $\mathbf{0.83}$  & $8.31 \times 10^{-5}$ & 0.61 \\
        \quad QI\_DEP\_POS & $2.35 \times 10^{-5}$ & $0.43$  & $1.93 \times 10^{-5}$ & $\mathbf{0.61}$ \\
        \quad QI\_DEP\_NEG & $3.57 \times 10^{-6}$ & $\mathbf{0.62}$  & $3.85 \times 10^{-6}$ & 0.56 \\ 
        \multicolumn{5}{l}{\textbf{Accumulated process rates, 30-minute output time step}} \\
        \quad QR\_AC & $1.29 \times 10^{-5}$ & $\mathbf{0.18}$ & $1.30 \times 10^{-5}$ & $0.17$ \\
        \quad QR\_ACC & $1.65 \times 10^{-4}$ & $0.30$ & $1.59 \times 10^{-4}$ & $0.35$ \\
        \quad QNC\_SC & $3.87 \times 10^{6}\,\unit{kg}^{-1}$ & $0.46$ & $3.66 \times 10^{6}\,\unit{kg}^{-1}$ & $\mathbf{0.52}$ \\
        \quad QNR\_SC\_POS & $589.06\,\unit{kg}^{-1}$ & $\mathbf{0.30}$ & $2.35 \times 10^{3}\,\unit{kg}^{-1}$ & $-10.10 $\\
        \quad QNR\_SC\_NEG & $1.08 \times 10^{4}\,\unit{kg}^{-1}$ & $0.21$ & $3.56 \times 10^{6}\,\unit{kg}^{-1}$ & $-8.57 \times 10^{4}$ \\
        \quad QR\_EVAP & $2.27 \times 10^{-5}$ & $0.64$ & $2.17 \times 10^{-5}$ & $\mathbf{0.67}$ \\
        \quad QI\_HET & $1.97 \times 10^{-5}$ &$-0.86$ & $1.46 \times 10^{-5}$ & $-0.02$ \\
        \quad QR\_RF & $1.10 \times 10^{-4}$ & $0.02$ & $1.04 \times 10^{-4}$ & $\mathbf{0.13}$ \\
        \quad QR\_MELT & $2.98 \times 10^{-4}$ & $\mathbf{0.77}$  & $3.00 \times 10^{-4}$ & $0.77$ \\
        \quad QX\_RIMING & $1.83 \times 10^{-4}$ & $\mathbf{0.60}$  & $1.84 \times 10^{-4}$ & $0.59$ \\
        \quad QI\_DEP\_POS & $6.54 \times 10^{-5}$ & $0.30$ & $5.83 \times 10^{-5}$ & $0.44$ \\
        \quad QI\_DEP\_NEG & $1.03 \times 10^{-5}$ & $0.52$ & $1.03 \times 10^{-5}$ & $\mathbf{0.52}$ \\
    \bottomhline
    \end{tabular}
\end{table*}
\clearpage

\subsection{Dependence of the model performance on the output time step}
\begin{figure*}[t]
    \centering
    \includegraphics[width=12cm]{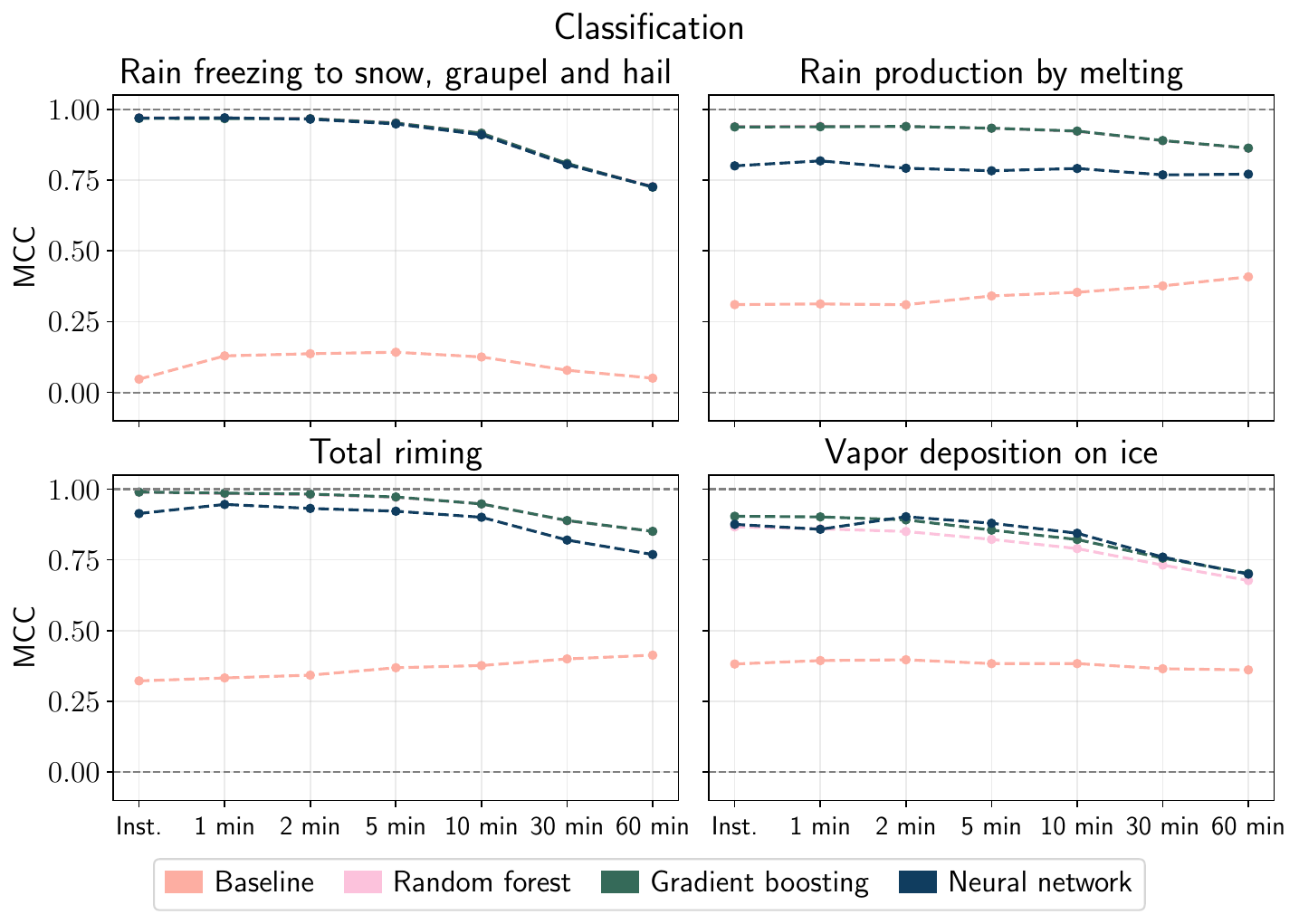}
    \caption{Same as Fig.~\ref{fig:time_steps_classification} for the process rates QR\_RF (upper left), QR\_MELT (upper right), QX\_RIMING (lower left) and QI\_DEP (lower right).}
    \label{fig:time_steps_classification_app}
\end{figure*}

\begin{figure*}[t]
    \centering
    \includegraphics[width=12cm]{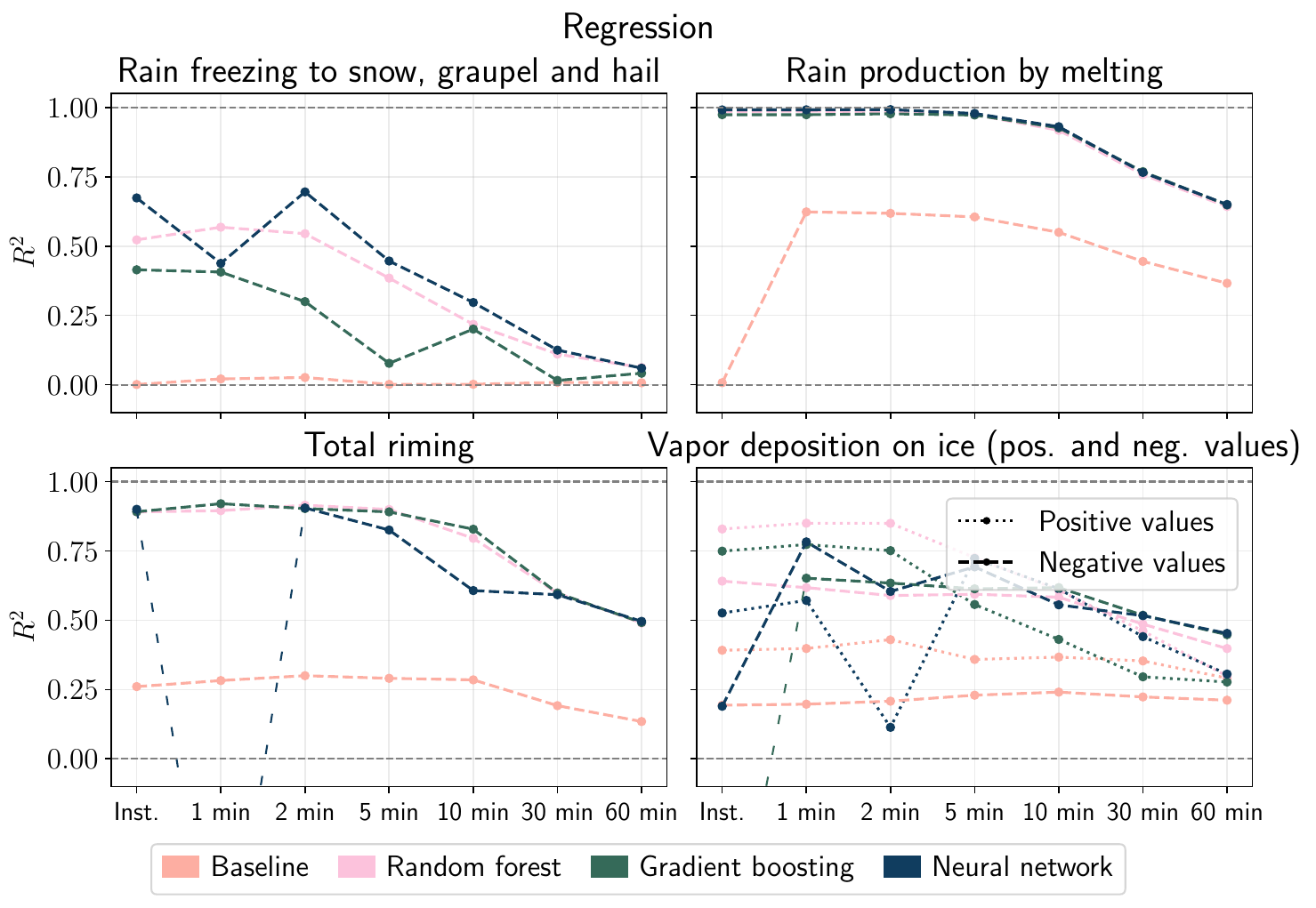}
    \caption{Same as Fig.~\ref{fig:time_steps_classification} but for the regression models with the $R^2$ score as the evaluation metric and for the process rates QR\_RF (upper left), QR\_MELT (upper right), QX\_RIMING (lower left) and QI\_DEP (lower right). Note that the sub-figure for vapor deposition on ice includes both the results for the models for the positive and negative contribution to the total process rate.}
    \label{fig:time_steps_regression_app}
\end{figure*}

Corresponding to the results presented in Sect.~\ref{sec:time_step_comparison}, in Fig.~\ref{fig:time_steps_classification_app} and Fig.~\ref{fig:time_steps_regression_app}, we show the output time step dependence of the model performance for QR\_RF, QR\_MELT, QX\_RIMING and QI\_DEP for the classification and regression models, respectively.

\subsection{Impact of wind fields and vertical velocity as additional input features on the model performance}\label{sec:wind_fields}
\begin{figure*}[t]
    \centering
    \includegraphics[width=12cm]{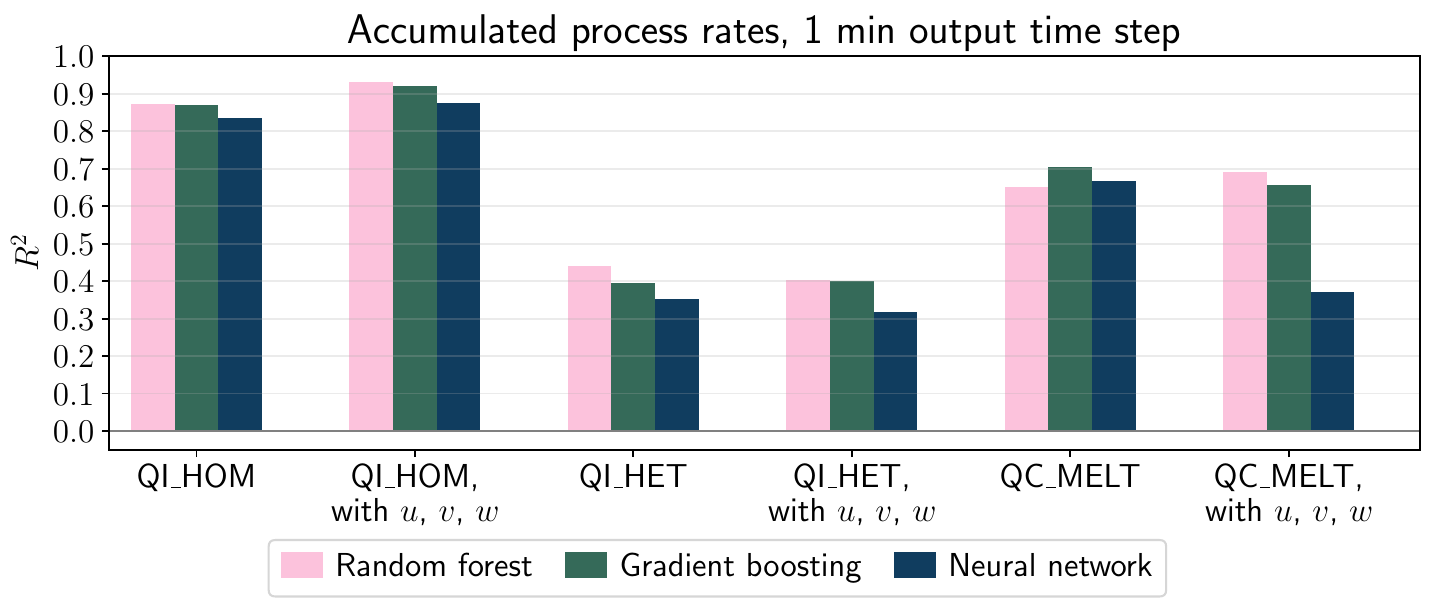}
    \caption{Impact of the zonal and meridional wind components $u$ and $v$ and the vertical velocity $w$ as additional input features on the performance of the regression models for the prediction of the accumulated process rates QI\_HOM, QI\_HET and QC\_MELT for the one-minute output time step. Shown is the $R^2$ score for the RF (pink), gradient boosting model (green) and NN (blue).}
    \label{fig:results_wind_fields}
\end{figure*}

\begin{figure*}[t]
    \centering
    \includegraphics[width=12cm]{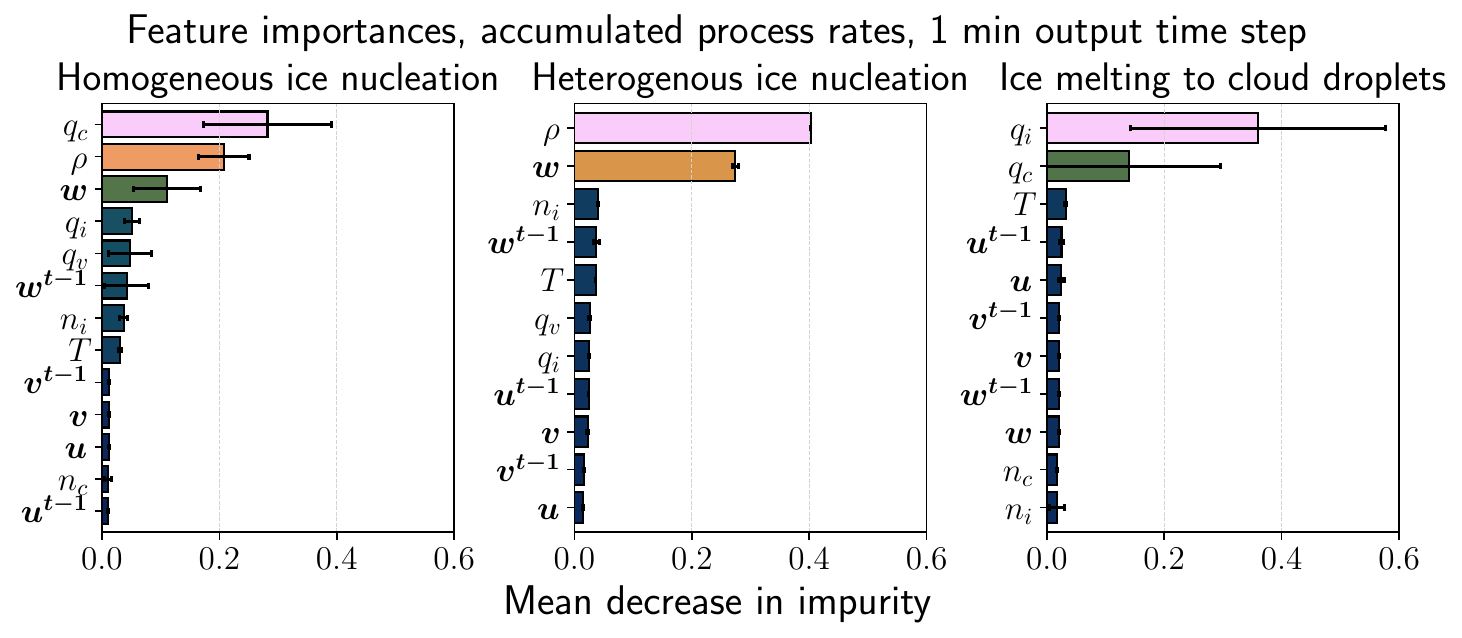}
    \caption{Feature importance of the RF regression models for QI\_HOM, QI\_HET and QC\_MELT accumulated over a one-minute output time step. For simplicity, we show only the 15 most important features for each process rate. Except for $u$, $v$ and $w$, we also do not show the importance of the variables at the previous time step, i.e. $q_k^{t-1}$ and $n_k^{t-1}$.}
    \label{fig:feature_importance_wind_fields}
\end{figure*}

As described in Sect.~\ref{sec:sampling}, our approach is fully local. Furthermore, the choice of input features is aligned with the implementation of the microphysics scheme in ICON, which is purely local as well. 
Nevertheless, it is interesting to investigate the influence of supplementary non-local information on the performance of the ML models.
In a related study, \citet{wangNonLocalParameterizationAtmospheric2022} found that including non-local features and the local vertical velocity or horizontal wind divergence in the set of model input features can substantially improve the offline performance of a parameterization.
Although including non-local features, e.g. from the neighboring grid cells, is not straightforward, we can include the wind fields $u$ and $v$ and the vertical velocity $w$ in the set of input features. In Fig~\ref{fig:results_wind_fields}, we show this for the regression models for QI\_HOM, QI\_HET and QC\_MELT, which are particularly hard to predict, 
for the one-minute output time step. 
Note that the vertical velocity $w$ is not available at the same model level as the other input features and the process rates, but at so-called half-levels above and below. To capture the effect of updrafts on the microphysical processes, we choose to use the values from the half-level below.
To that end, we repeat the training workflow, including the hyperparameter tuning, with the extended set of input features
\begin{equation*}
    \left[q_k, \,q_k^{t-1}, \,n_k, \,n_k^{t-1}, \,q_v, \,T, \,p, \,\rho \right] + \left[u,\,v,\,w\right], \qquad k \in \left\{c, \,r, \,i, \,s, \,g, \,h \right\} \, .
\end{equation*}
We find that including the wind fields as additional input features is beneficial for the models for predicting QI\_HOM, while there is no substantial improvement for QI\_HET and QC\_MELT. To further analyze these dependencies, we compute the feature importance for the RF models (Fig.~\ref{fig:feature_importance_wind_fields}), which reveals a strong dependence on the vertical velocity $w$, which is the second most important feature for QI\_HET and the third most important feature for QI\_HOM. However, for QC\_MELT, we find that $u$, $v$ and $w$ are less important features.

\subsection{Predicting averaged process rates instead of accumulated process rates}
\begin{figure*}[t]
    \centering
    \includegraphics[width=12cm]{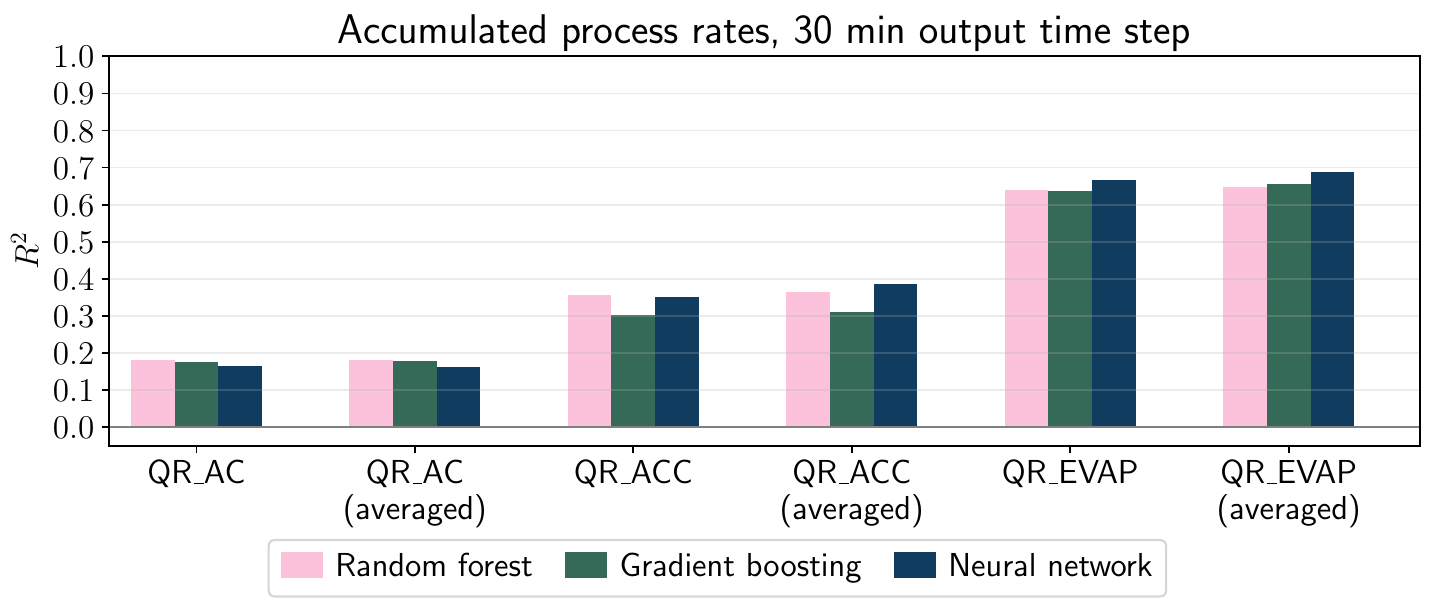}
    \caption{Comparison of the performance of the regression models designed to predict the accumulated process rates and the regression models designed to predict the averaged process rates for QR\_AC, QR\_ACC and QR\_EVAP accumulated over a 30-minute output time step.}
    \label{fig:compare_average}
\end{figure*}

In Fig.~\ref{fig:compare_average}, we present a comparison of the performance of regression models that are trained to predict accumulated process rates versus averaged process rates for an output time step of 30 minutes for the three exemplary process rates QR\_AC, QR\_ACC and QR\_EVAP. In this context, \textit{averaged} means that the accumulated process rate values are divided by the number of internal fast physics time steps in the output time step interval. It is apparent that the differences between the two prediction tasks are negligible.

\subsection{Using an adjusted threshold}\label{sec:adjusted_threshold}
\begin{figure*}[t]
    \centering
    \includegraphics[width=12cm]{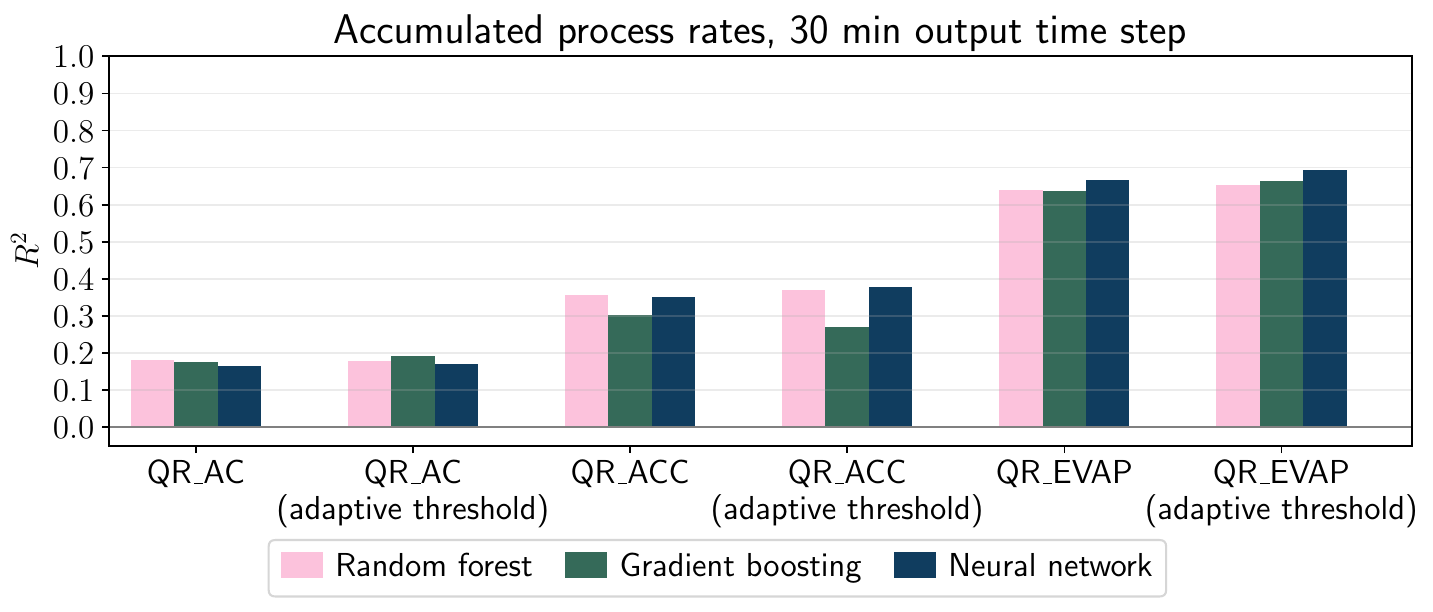}
    \caption{Comparison of the performance of the regression models trained on data with an adjusted threshold versus the default threshold for QR\_AC, QR\_ACC and QR\_EVAP accumulated over a 30-minute output time step.}
    \label{fig:compare_adjusted_threshold}
\end{figure*}

As mentioned in Sect.~\ref{sec:time_step_comparison}, the threshold used during data pre-processing could slightly skew the comparison between different output time steps. To investigate this in more detail, in Fig.~\ref{fig:compare_adjusted_threshold}, we show a comparison of the regression models trained on data with a single threshold for all output time steps versus a threshold that is adjusted for each output time step to ensure a fair comparison. It is apparent that the effect of the exact choice of threshold is negligible.

\subsection{Output distributions}\label{sec:output_distributions}
\begin{figure*}[t]
    \centering
    \includegraphics[width=12cm]{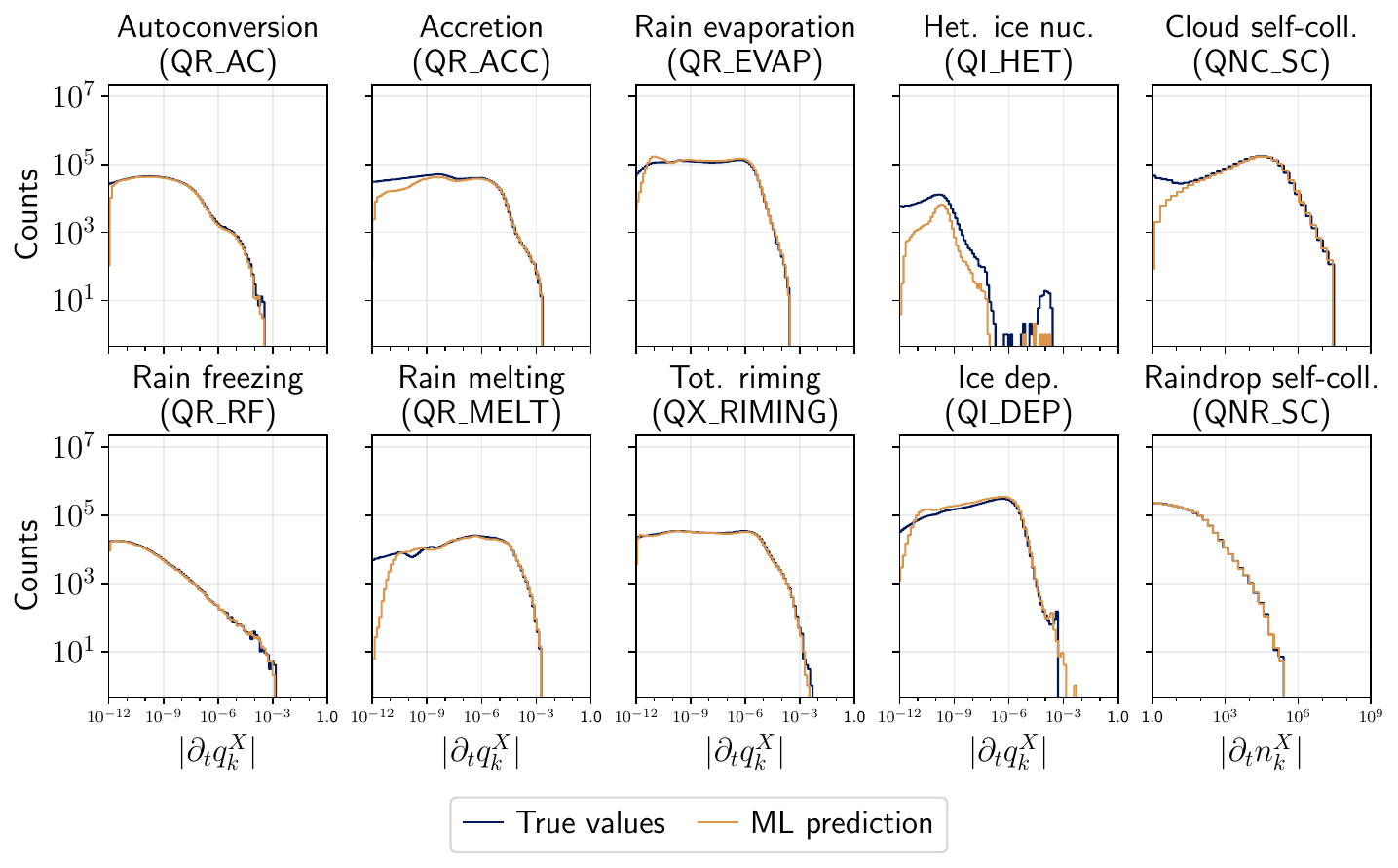}
    \caption{Distribution of the true values (dark blue) and predicted values (orange) in the test set of 10 microphysical process rates integrated over a one-minute output time step. For better visualization, the distributions of the absolute values are shown.}
    \label{fig:distributions_combined_evaluation_acc_1min}
\end{figure*}

\begin{figure*}[t]
    \centering
    \includegraphics[width=12cm]{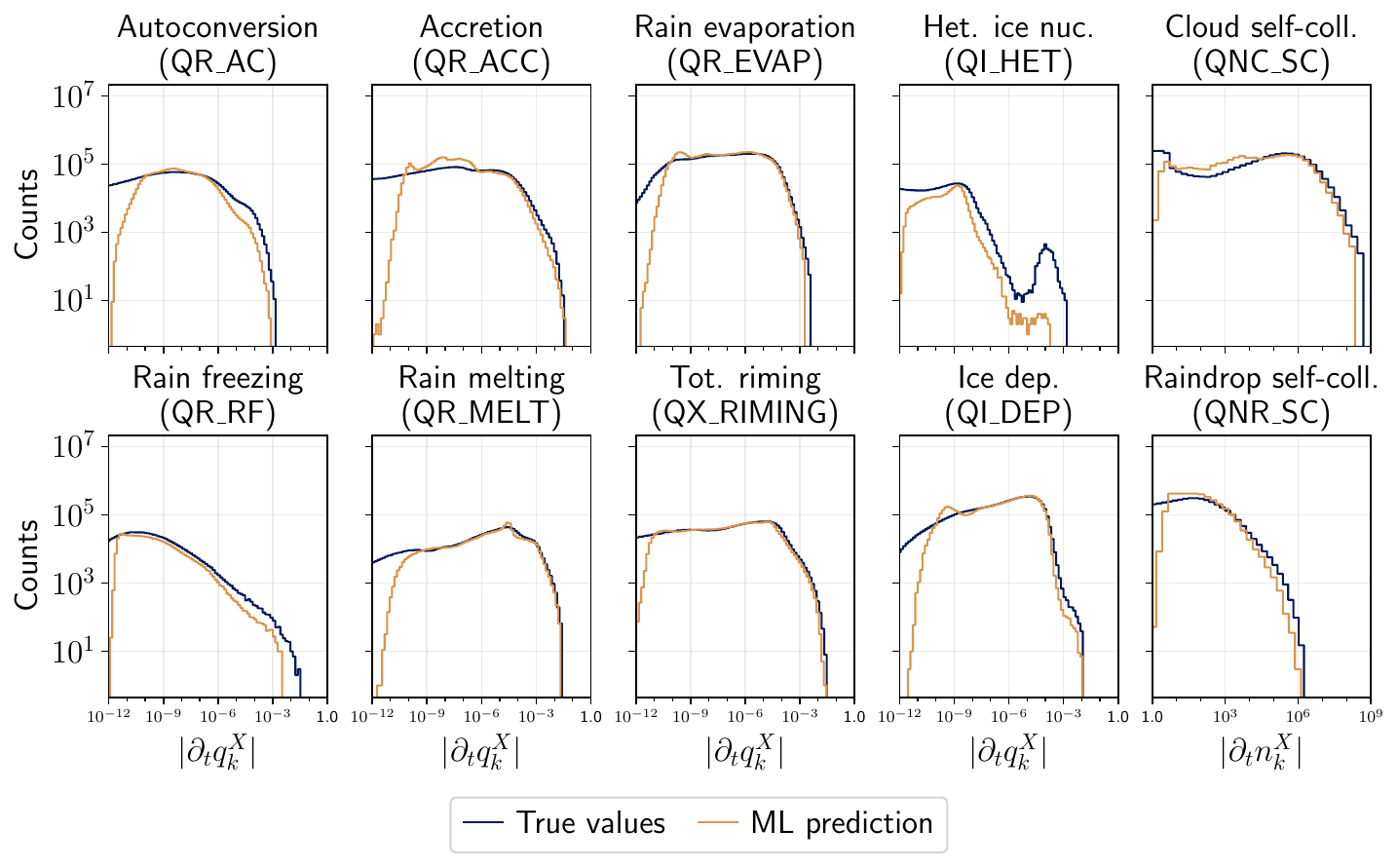}
    \caption{Same as Fig.~\ref{fig:distributions_combined_evaluation_acc_10min} but for process rates integrated over a 30-minute output time step.}
    \label{fig:distributions_combined_evaluation_acc_30min}
\end{figure*}

Related to Fig.~\ref{fig:distributions_combined_evaluation_inst_10min} and Fig.~\ref{fig:distributions_combined_evaluation_acc_10min} in Sect.~\ref{sec:combined_results}, in Fig.~\ref{fig:distributions_combined_evaluation_acc_1min} and Fig.~\ref{fig:distributions_combined_evaluation_acc_30min} we show the distribution of the true and predicted values for 10 microphysical process rates for a one-minute and 30-minute output time step, respectively.

\subsection{Combined evaluation for remaining process rates}
Table~\ref{table:results_combined_evaluation} presents the RMSE and the $R^2$ score for 10 selected process rates, with instantaneous output and a 10-minute output time step and accumulated output with a one-minute, 10-minute and 30-minute output time step. Similarly, in Table~\ref{table:results_combined_evaluation_remaining_appendix}, we list the RMSE and the $R^2$ score for the 15 other process rates not included in Table~\ref{table:results_combined_evaluation}, as shown in Fig.~\ref{fig:results_additional_process_rates_combined_evaluation}. Additionally, we show the $R^2_\mathrm{log}$ score computed with the logarithmically transformed true and predicted values.

\begin{table*}[t]
    \centering
    \caption{RMSE and $R^2$ score of the combined classification-regression models for 10 selected process rates, computed on a separate test set. Additionally, we show the $R^2_\mathrm{log}$ score computed with the logarithmically transformed true and predicted values. In some cases, the $R^2$ score is not a reliable metric for the overall model performance, the respective values are colored in light gray (see main text, Sect.~\ref{sec:combined_results}, for details).
    }    \label{table:results_combined_evaluation}
    \begin{tabular}{l c c c c c c}
        \tophline
        \multirow{2}{*}{Process rate} & \multicolumn{3}{c}{Inst., \unit{10\,min}} & \multicolumn{3}{c}{Acc., \unit{1\,min}} \\
        \cline{2-4} \cline{5-7} & {RMSE / \unit{kg}\,\unit{kg}$^{-1}\,(20\,\text{s})^{-1}$} & {$R^2$} & {$R^2_\mathrm{log}$} & {RMSE / \unit{kg}\,\unit{kg}$^{-1}\,(600\,\text{s})^{-1}$} & {$R^2$} & {$R^2_\mathrm{log}$} \\
        \middlehline
QR\_AC & $1.29 \times 10^{-7}$ & $0.79$ & $0.94$ & $3.12 \times 10^{-7}$ & $0.82$ & $0.94$\\
QR\_ACC & $1.24 \times 10^{-7}$ & $1.00$ & $0.73$ & $5.49 \times 10^{-7}$ & $0.99$ & $0.72$\\
QNC\_SC & $2.83 \times 10^{3}\,\unit{kg}^{-1}$ & $1.00$ & $0.97$ & $7.71 \times 10^{3}\,\unit{kg}^{-1}$ & $1.00$ & $0.97$\\
QNR\_SC & $25.19\,\unit{kg}^{-1}$ & $0.97$ & $0.98$ & $55.72\,\unit{kg}^{-1}$ & $0.98$ & $0.99$\\
QR\_EVAP & $1.99 \times 10^{-7}$ & $0.86$ & $0.54$ & $5.79 \times 10^{-7}$ & $0.87$ & $0.55$\\
QI\_HET & $3.81 \times 10^{-7}$ & $0.00$ & $0.00$ & $4.54 \times 10^{-7}$ & $0.06$ & $0.02$\\
QR\_RF & $3.27 \times 10^{-7}$ & $0.59$ & $0.92$ & $1.07 \times 10^{-6}$ & $0.57$ & $0.94$\\
QR\_MELT & $2.87 \times 10^{-7}$ & $0.99$ & $0.97$ & $8.21 \times 10^{-7}$ & $0.99$ & $0.96$\\
QX\_RIMING & $7.04 \times 10^{-7}$ & $0.92$ & $0.98$ & $3.53 \times 10^{-6}$ & $0.82$ & $0.98$\\
QI\_DEP & $8.70 \times 10^{-7}$ & $0.06$ & $0.87$ & $6.51 \times 10^{-5}$ & \textcolor{lightgray}{$-569.77$} & $0.90$\\
\middlehline
        Mean score & {} & {0.72} & {0.79} & {} & {0.62} & {0.80} \\
        \bottomhline \\
        \tophline
        \multirow{2}{*}{Process rate} & \multicolumn{3}{c}{Acc., \unit{10\,min}} & \multicolumn{3}{c}{Acc., \unit{30\,min}} \\
        \cline{2-4} \cline{5-7} & {RMSE / \unit{kg}\,\unit{kg}$^{-1}\,(600\,\text{s})^{-1}$} & {$R^2$} & {$R^2_\mathrm{log}$} & {RMSE / \unit{kg}\,\unit{kg}$^{-1}\,(1800\,\text{s})^{-1}$} & {$R^2$} & {$R^2_\mathrm{log}$} \\
        \middlehline
QR\_AC & $3.06 \times 10^{-6}$ & $0.31$ & $0.85$ & $6.62 \times 10^{-6}$ & $0.19$ & $0.70$\\
QR\_ACC & $2.81 \times 10^{-5}$ & $0.78$ & $0.64$ & $1.17 \times 10^{-4}$ & $0.33$ & $0.39$\\
QNC\_SC & $8.52 \times 10^{5}\,\unit{kg}^{-1}$ & $0.66$ & $0.87$ & $2.72 \times 10^{6}\,\unit{kg}^{-1}$ & $0.53$ & $0.68$\\
QNR\_SC & $2.51 \times 10^{4}\,\unit{kg}^{-1}$ & \textcolor{lightgray}{$-69.82$} & $0.95$ & $1.26 \times 10^{6}\,\unit{kg}^{-1}$ & \textcolor{lightgray}{$-3.52 \times 10^{4}$} & $0.82$\\
QR\_EVAP & $5.66 \times 10^{-6}$ & $0.84$ & $0.59$ & $2.19 \times 10^{-5}$ & $0.66$ & $0.54$\\
QI\_HET & $1.95 \times 10^{-6}$ & $0.00$ & $0.04$ & $3.75 \times 10^{-6}$ & $0.00$ & $0.04$\\
QR\_RF & $1.62 \times 10^{-5}$ & $0.26$ & $0.90$ & $3.52 \times 10^{-5}$ & $0.11$ & $0.71$\\
QR\_MELT & $2.17 \times 10^{-5}$ & $0.93$ & $0.92$ & $1.06 \times 10^{-4}$ & $0.77$ & $0.86$\\
QX\_RIMING & $2.77 \times 10^{-5}$ & $0.81$ & $0.93$ & $9.04 \times 10^{-5}$ & $0.61$ & $0.84$\\
QI\_DEP & $1.54 \times 10^{-5}$ & $0.55$ & $0.89$ & $4.11 \times 10^{-5}$ & $0.41$ & $0.84$\\
        \middlehline
        Mean score & {} & {0.41} & {0.76} & {} & {0.26} & {0.64} \\
        \bottomhline
    \end{tabular}
    \belowtable{The $R^2_\mathrm{log}$ score is computed as $R^2(\log(\lvert y_\mathrm{true}\rvert + \varepsilon),\, \log(\lvert y_\mathrm{pred}\rvert + \varepsilon))$ where $\varepsilon = 1$ for the process rates related to a change in the number concentration (i.e., QNC\_SC, QNR\_SC) and $\varepsilon = 10^{-13}$ for all other process rates.}
    \belowtable{To avoid distortion towards negative values, the scores $R^2 < -1$ are set to $R^2 = -1$ in the computation of the mean score. The computation of the mean RMSE is omitted due to different units of QNC\_SC and QNR\_SC.}
\end{table*}

\begin{table*}[t]
    \centering
    \caption{Same as Table~\ref{table:results_combined_evaluation} for the 15 remaining process rates.}    \label{table:results_combined_evaluation_remaining_appendix}
    \begin{tabular}{l c c c c c c}
        \tophline
        \multirow{2}{*}{Process rate} & \multicolumn{3}{c}{Inst., \unit{10\,min}} & \multicolumn{3}{c}{Acc., \unit{10\,min}} \\
        \cline{2-4} \cline{5-7} & {RMSE / \unit{kg}\,\unit{kg}$^{-1}\,(20\,\text{s})^{-1}$} & {$R^2$} & {$R^2_\mathrm{log}$} & {RMSE / \unit{kg}\,\unit{kg}$^{-1}\,(600\,\text{s})^{-1}$} & {$R^2$} & {$R^2_\mathrm{log}$} \\
        \middlehline
QSGH\_EVAP & $2.81 \times 10^{-8}$ & $0.97$ & $0.94$ & $1.89 \times 10^{-6}$ & $0.82$ & $0.91$\\
QI\_HOM & $2.58 \times 10^{-6}$ & $0.94$ & $0.65$ & $1.02 \times 10^{-4}$ & $0.44$ & $0.54$\\
QC\_MELT & $3.91 \times 10^{-11}$ & $0.67$ & $0.52$ & $1.15 \times 10^{-9}$ & $0.38$ & $0.48$\\
QI\_RF & $2.77 \times 10^{-8}$ & $0.91$ & $0.91$ & $3.46 \times 10^{-6}$ & $0.15$ & $0.90$\\
QG\_RF & $1.86 \times 10^{-7}$ & $0.52$ & $0.92$ & $5.77 \times 10^{-6}$ & $0.29$ & $0.88$\\
QH\_RF & $2.26 \times 10^{-7}$ & $0.53$ & $0.93$ & $7.03 \times 10^{-6}$ & $0.29$ & $0.77$\\
QS\_DEP & $1.43 \times 10^{-7}$ & $0.93$ & $0.87$ & $7.49 \times 10^{-6}$ & $0.74$ & $0.88$\\
QG\_DEP & $1.41 \times 10^{-7}$ & $0.98$ & $0.97$ & $1.35 \times 10^{-5}$ & $0.71$ & $0.94$\\
QH\_DEP & $1.00 \times 10^{-8}$ & $0.96$ & $0.52$ & $8.31 \times 10^{-7}$ & $0.54$ & $0.65$\\
QC\_RIME\_I & $1.29 \times 10^{-7}$ & $0.43$ & $0.94$ & $1.97 \times 10^{-6}$ & $0.38$ & $0.76$\\
QC\_RIME\_S & $2.74 \times 10^{-7}$ & $0.18$ & $0.93$ & $4.93 \times 10^{-6}$ & $0.24$ & $0.75$\\
QC\_RIME\_GH & $7.09 \times 10^{-8}$ & $1.00$ & $0.97$ & $1.34 \times 10^{-5}$ & $0.83$ & $0.86$\\
QR\_RIME\_I & $8.80 \times 10^{-7}$ & $-2.11$ & $0.97$ & $1.47 \times 10^{-3}$ & \textcolor{lightgray}{$-5.20 \times 10^{4}$} & $0.80$\\
QR\_RIME\_S & $3.39 \times 10^{-8}$ & $0.61$ & $0.97$ & $1.88 \times 10^{-6}$ & $0.14$ & $0.82$\\
QR\_RIME\_GH & $1.01 \times 10^{-6}$ & $0.79$ & $0.97$ & $2.32 \times 10^{-5}$ & $0.83$ & $0.93$\\
\middlehline
        Mean score & {} & {0.63} & {0.87} & {} & {0.39} & {0.79} \\
        \bottomhline
    \end{tabular}
    \belowtable{The $R^2_\mathrm{log}$ score is computed as $R^2(\log(\lvert y_\mathrm{true}\rvert + \varepsilon),\, \log(\lvert y_\mathrm{pred}\rvert + \varepsilon))$ where $\varepsilon = 10^{-13}$ for all other process rates.}
\end{table*}

\subsection{Prediction intervals with conformalized quantile regression}
In Table~\ref{table:results_combined_evaluation_cqr}, Table~\ref{table:results_combined_evaluation_cqr_rf} and Table~\ref{table:results_combined_evaluation_cqr_xgb} we list the values of the prediction interval coverage probability (PICP) for 10 selected process rates and different output time steps for the quantile NN, quantile RF and quantile XGB models, respectively. The results listed in Table~\ref{table:results_combined_evaluation_cqr} are shown in Fig.~\ref{fig:results_combined_evaluation_cqr}. In Table~\ref{table:results_combined_evaluation_cqr_nmpiw}, we show the normalized mean prediction interval width (NMPIW) for the prediction intervals obtained with the quantile NN.

 \begin{table*}[t]
     \centering
     \caption{Prediction interval coverage probability (PICP) and $\text{PICP}_\text{rounded}$ for the prediction intervals rounded to the nearest order of magnitude (see text for details) for 10 process rates obtained with the combined classification-regression model and the quantile NN. Cases where $\text{PICP} \geq 90\%$ are marked in green.}
     \label{table:results_combined_evaluation_cqr}
     \begin{tabular}{l c c c c c c c c}
         \tophline
             \multirow{2}{*}{Process rate} & \multicolumn{2}{c}{Inst., \unit{10\,min}} & \multicolumn{2}{c}{Acc., \unit{1\,min}} & \multicolumn{2}{c}{Acc., \unit{10\,min}} & \multicolumn{2}{c}{Acc., \unit{30\,min}} \\
             \cline{2-9} & {PICP /\,\%} & {$\text{PICP}_\text{rounded}$ /\,\%} & {PICP /\,\%} & {$\text{PICP}_\text{rounded}$ /\,\%} & {PICP /\,\%} & {$\text{PICP}_\text{rounded}$ /\,\%} & {PICP /\,\%} & {$\text{PICP}_\text{rounded}$ /\,\%}\\
\middlehline
QR\_AC & \textcolor{ForestGreen}{$91.48$} & \textcolor{ForestGreen}{$98.35$} & $72.74$ & \textcolor{ForestGreen}{$96.90$} & $84.95$ & \textcolor{ForestGreen}{$93.39$} & $78.91$ & $86.42$\\
QR\_ACC & $79.13$ & $89.07$ & $88.27$ & \textcolor{ForestGreen}{$99.34$} & $89.20$ & \textcolor{ForestGreen}{$96.65$} & $69.34$ & $73.63$\\
QNC\_SC & \textcolor{ForestGreen}{$90.81$} & \textcolor{ForestGreen}{$99.14$} & \textcolor{ForestGreen}{$90.06$} & \textcolor{ForestGreen}{$98.52$} & $88.80$ & \textcolor{ForestGreen}{$95.59$} & \textcolor{ForestGreen}{$92.23$} & \textcolor{ForestGreen}{$97.90$}\\
QNR\_SC & \textcolor{ForestGreen}{$99.80$} & \textcolor{ForestGreen}{$99.99$} & \textcolor{ForestGreen}{$91.10$} & \textcolor{ForestGreen}{$99.65$} & $89.00$ & \textcolor{ForestGreen}{$97.36$} & \textcolor{ForestGreen}{$91.14$} & \textcolor{ForestGreen}{$96.02$}\\
QR\_EVAP & $79.08$ & $87.20$ & $78.97$ & $87.40$ & \textcolor{ForestGreen}{$90.05$} & \textcolor{ForestGreen}{$95.62$} & $84.58$ & $89.25$\\
QI\_HET & $43.15$ & $48.37$ & $46.32$ & $50.82$ & $45.19$ & $51.65$ & $48.27$ & $53.31$\\
QI\_HET (SCP)   & $83.43$ & \textcolor{ForestGreen}{$91.27$} & $88.00$ & \textcolor{ForestGreen}{$92.57$} & \textcolor{ForestGreen}{$92.26$} & \textcolor{ForestGreen}{$96.39$} & \textcolor{ForestGreen}{$94.52$} & \textcolor{ForestGreen}{$97.13$}\\
QR\_RF & $88.36$ & \textcolor{ForestGreen}{$98.34$} & \textcolor{ForestGreen}{$90.67$} & \textcolor{ForestGreen}{$98.57$} & $87.90$ & \textcolor{ForestGreen}{$96.49$} & $84.65$ & \textcolor{ForestGreen}{$90.74$}\\
QR\_MELT & \textcolor{ForestGreen}{$91.97$} & \textcolor{ForestGreen}{$98.93$} & \textcolor{ForestGreen}{$91.28$} & \textcolor{ForestGreen}{$99.24$} & \textcolor{ForestGreen}{$96.18$} & \textcolor{ForestGreen}{$99.28$} & $86.42$ & \textcolor{ForestGreen}{$92.13$}\\
QX\_RIMING & \textcolor{ForestGreen}{$92.76$} & \textcolor{ForestGreen}{$98.71$} & \textcolor{ForestGreen}{$95.25$} & \textcolor{ForestGreen}{$99.52$} & $86.37$ & \textcolor{ForestGreen}{$94.52$} & $84.73$ & \textcolor{ForestGreen}{$91.36$}\\
QI\_DEP & $86.24$ & \textcolor{ForestGreen}{$93.13$} & $85.71$ & \textcolor{ForestGreen}{$93.81$} & $82.30$ & $86.71$ & $78.45$ & $84.20$\\
    \bottomhline \\
    \end{tabular}
\end{table*}

  \begin{table*}[t]
     \centering
     \caption{Prediction interval coverage probability (PICP) and $\text{PICP}_\text{rounded}$ for the prediction intervals rounded to the nearest order of magnitude (see text for details) for 10 process rates obtained with the combined classification-regression model and the quantile RF model. Cases where $\text{PICP} \geq 90\%$ are marked in green.}
     \label{table:results_combined_evaluation_cqr_rf}
     \begin{tabular}{l c c c c c c c c}
         \tophline
             \multirow{2}{*}{Process rate} & \multicolumn{2}{c}{Inst., \unit{10\,min}} & \multicolumn{2}{c}{Acc., \unit{1\,min}} & \multicolumn{2}{c}{Acc., \unit{10\,min}} & \multicolumn{2}{c}{Acc., \unit{30\,min}} \\
             \cline{2-9} & {PICP /\,\%} & {$\text{PICP}_\text{rounded}$ /\,\%} & {PICP /\,\%} & {$\text{PICP}_\text{rounded}$ /\,\%} & {PICP /\,\%} & {$\text{PICP}_\text{rounded}$ /\,\%} & {PICP /\,\%} & {$\text{PICP}_\text{rounded}$ /\,\%}\\
    \middlehline
QR\_AC & \textcolor{ForestGreen}{$90.22$} & \textcolor{ForestGreen}{$98.15$} & \textcolor{ForestGreen}{$91.13$} & \textcolor{ForestGreen}{$97.63$} & $84.39$ & \textcolor{ForestGreen}{$93.90$} & $86.92$ & \textcolor{ForestGreen}{$92.85$}\\
QR\_ACC & $75.73$ & $87.75$ & \textcolor{ForestGreen}{$94.74$} & \textcolor{ForestGreen}{$99.04$} & $74.70$ & $81.59$ & $57.47$ & $61.63$\\
QNC\_SC & $89.54$ & \textcolor{ForestGreen}{$98.59$} & \textcolor{ForestGreen}{$90.79$} & \textcolor{ForestGreen}{$98.21$} & $89.79$ & \textcolor{ForestGreen}{$96.91$} & \textcolor{ForestGreen}{$91.09$} & \textcolor{ForestGreen}{$97.41$}\\
QNR\_SC & \textcolor{ForestGreen}{$99.86$} & \textcolor{ForestGreen}{$99.99$} & $89.01$ & \textcolor{ForestGreen}{$99.38$} & \textcolor{ForestGreen}{$90.16$} & \textcolor{ForestGreen}{$97.64$} & $89.91$ & \textcolor{ForestGreen}{$95.77$}\\
QR\_EVAP & $79.19$ & $85.92$ & $79.12$ & $86.66$ & $81.21$ & $89.32$ & $83.95$ & \textcolor{ForestGreen}{$90.02$}\\
QI\_HET & $38.91$ & $47.44$ & $44.17$ & $50.73$ & $45.29$ & $51.84$ & $46.01$ & $52.56$\\
QR\_RF & $87.52$ & \textcolor{ForestGreen}{$98.44$} & $89.19$ & \textcolor{ForestGreen}{$98.15$} & $83.03$ & \textcolor{ForestGreen}{$92.83$} & $82.29$ & \textcolor{ForestGreen}{$92.48$}\\
QR\_MELT & \textcolor{ForestGreen}{$92.62$} & \textcolor{ForestGreen}{$95.77$} & $89.70$ & \textcolor{ForestGreen}{$94.73$} & \textcolor{ForestGreen}{$93.73$} & \textcolor{ForestGreen}{$98.53$} & $84.78$ & \textcolor{ForestGreen}{$90.90$}\\
QX\_RIMING & \textcolor{ForestGreen}{$96.91$} & \textcolor{ForestGreen}{$98.94$} & $88.72$ & \textcolor{ForestGreen}{$96.15$} & $83.97$ & \textcolor{ForestGreen}{$94.56$} & $83.56$ & \textcolor{ForestGreen}{$92.00$}\\
QI\_DEP & $83.69$ & \textcolor{ForestGreen}{$91.80$} & $84.57$ & \textcolor{ForestGreen}{$92.10$} & $81.08$ & $87.94$ & $76.68$ & $83.88$\\
    \bottomhline \\
     \end{tabular}
 \end{table*}

  \begin{table*}[t]
     \centering
     \caption{Prediction interval coverage probability (PICP) and $\text{PICP}_\text{rounded}$ for the prediction intervals rounded to the nearest order of magnitude (see text for details) for 10 process rates obtained with the combined classification-regression model and the quantile XGB model. Cases where $\text{PICP} \geq 90\%$ are marked in green.}
     \label{table:results_combined_evaluation_cqr_xgb}
     \begin{tabular}{l c c c c c c c c}
         \tophline
             \multirow{2}{*}{Process rate} & \multicolumn{2}{c}{Inst., \unit{10\,min}} & \multicolumn{2}{c}{Acc., \unit{1\,min}} & \multicolumn{2}{c}{Acc., \unit{10\,min}} & \multicolumn{2}{c}{Acc., \unit{30\,min}} \\
             \cline{2-9} & {PICP /\,\%} & {$\text{PICP}_\text{rounded}$ /\,\%} & {PICP /\,\%} & {$\text{PICP}_\text{rounded}$ /\,\%} & {PICP /\,\%} & {$\text{PICP}_\text{rounded}$ /\,\%} & {PICP /\,\%} & {$\text{PICP}_\text{rounded}$ /\,\%}\\
    \middlehline
QR\_AC & \textcolor{ForestGreen}{$94.97$} & \textcolor{ForestGreen}{$99.41$} & \textcolor{ForestGreen}{$93.94$} & \textcolor{ForestGreen}{$99.27$} & $85.86$ & \textcolor{ForestGreen}{$94.14$} & $78.34$ & $85.93$\\
QR\_ACC & $77.52$ & $88.95$ & \textcolor{ForestGreen}{$95.04$} & \textcolor{ForestGreen}{$99.53$} & $78.21$ & $85.77$ & $57.45$ & $61.78$\\
QNC\_SC & $88.93$ & \textcolor{ForestGreen}{$98.98$} & $89.39$ & \textcolor{ForestGreen}{$99.05$} & $88.08$ & \textcolor{ForestGreen}{$97.48$} & $64.60$ & $71.61$\\
QNR\_SC & \textcolor{ForestGreen}{$99.73$} & \textcolor{ForestGreen}{$99.98$} & \textcolor{ForestGreen}{$90.78$} & \textcolor{ForestGreen}{$99.71$} & $88.39$ & \textcolor{ForestGreen}{$97.45$} & \textcolor{ForestGreen}{$91.17$} & \textcolor{ForestGreen}{$96.28$}\\
QR\_EVAP & $76.56$ & $88.31$ & $80.48$ & \textcolor{ForestGreen}{$90.64$} & $87.71$ & \textcolor{ForestGreen}{$94.14$} & $84.84$ & \textcolor{ForestGreen}{$90.92$}\\
QI\_HET & $40.30$ & $48.24$ & $40.93$ & $49.92$ & $56.26$ & $65.92$ & $46.01$ & $52.46$\\
QR\_RF & $87.73$ & \textcolor{ForestGreen}{$98.46$} & $88.55$ & \textcolor{ForestGreen}{$98.58$} & $86.33$ & \textcolor{ForestGreen}{$94.91$} & $84.26$ & \textcolor{ForestGreen}{$92.45$}\\
QR\_MELT & \textcolor{ForestGreen}{$95.74$} & \textcolor{ForestGreen}{$99.66$} & $88.18$ & \textcolor{ForestGreen}{$98.82$} & \textcolor{ForestGreen}{$91.93$} & \textcolor{ForestGreen}{$98.80$} & $88.21$ & \textcolor{ForestGreen}{$96.15$}\\
QX\_RIMING & \textcolor{ForestGreen}{$94.52$} & \textcolor{ForestGreen}{$99.37$} & \textcolor{ForestGreen}{$92.61$} & \textcolor{ForestGreen}{$99.02$} & $87.50$ & \textcolor{ForestGreen}{$96.22$} & $84.17$ & \textcolor{ForestGreen}{$93.16$}\\
QI\_DEP & $81.78$ & \textcolor{ForestGreen}{$94.12$} & $81.90$ & \textcolor{ForestGreen}{$94.11$} & $80.88$ & $88.73$ & $76.82$ & $84.19$\\
    \bottomhline \\
     \end{tabular}
 \end{table*}

\begin{sidewaystable*}[t]
     \centering
     \caption{Normalized mean prediction interval width (NMPIW) and $\text{NMPIW}_\text{rounded}$ for the prediction intervals rounded to the nearest order of magnitude (see Sect.~\ref{sec:results_conformal_prediction} for details) for 10 process rates obtained with the combined classification-regression model and the quantile NN.}
     \label{table:results_combined_evaluation_cqr_nmpiw}
     \begin{tabular}{l c c c c c c c c}
         \tophline
        \multirow{2}{*}{Process rate} & \multicolumn{2}{c}{Inst., \unit{10\,min}} & \multicolumn{2}{c}{Acc., \unit{1\,min}} & \multicolumn{2}{c}{Acc., \unit{10\,min}} & \multicolumn{2}{c}{Acc., \unit{30\,min}} \\
        \cline{2-9} & {NMPIW} & {$\text{NMPIW}_\text{rounded}$} & {NMPIW} & {$\text{NMPIW}_\text{rounded}$} & {NMPIW} & {$\text{NMPIW}_\text{rounded}$} & {NMPIW} & {$\text{NMPIW}_\text{rounded}$}\\
        \middlehline
QR\_AC & $1.93 \times 10^{-4}$ & $8.48 \times 10^{-4}$ & $2.98 \times 10^{-4}$ & $1.34 \times 10^{-3}$ & $8.97 \times 10^{-4}$ & $3.56 \times 10^{-3}$ & $3.21 \times 10^{-3}$ & 0.01\\
QR\_ACC & $3.31 \times 10^{-4}$ & $3.31 \times 10^{-3}$ & $2.06 \times 10^{-4}$ & $2.59 \times 10^{-3}$ & $6.81 \times 10^{-4}$ & $3.17 \times 10^{-3}$ & $9.04 \times 10^{-4}$ & $3.74 \times 10^{-3}$\\
QNC\_SC & $5.60 \times 10^{-4}$ & $6.21 \times 10^{-3}$ & $4.80 \times 10^{-4}$ & $6.76 \times 10^{-3}$ & $6.27 \times 10^{-3}$ & 0.03 & $2.71 \times 10^{-3}$ & 0.01\\
QNR\_SC & $2.38 \times 10^{-4}$ & $1.21 \times 10^{-3}$ & $5.66 \times 10^{-5}$ & $9.65 \times 10^{-4}$ & $5.50 \times 10^{-4}$ & $2.46 \times 10^{-3}$ & $1.04 \times 10^{-3}$ & $4.15 \times 10^{-3}$\\
QR\_EVAP & $5.66 \times 10^{-4}$ & $4.85 \times 10^{-3}$ & $1.24 \times 10^{-3}$ & $6.76 \times 10^{-3}$ & $1.09 \times 10^{-3}$ & $5.02 \times 10^{-3}$ & $4.06 \times 10^{-3}$ & 0.02\\
QI\_HET & 0.01 & 0.05 & $1.79 \times 10^{-3}$ & $7.72 \times 10^{-3}$ & 0.45 & 0.95 & $3.48 \times 10^{-3}$ & 0.01\\
QI\_HET (SCP) & $4.74 \times 10^{-3}$ & $2.16 \times 10^{-2}$ & $2.50 \times 10^{-6}$ & $1.24 \times 10^{-4}$ & $5.47 \times 10^{-6}$ & $4.16 \times 10^{-5}$ & $6.05 \times 10^{-6}$ & $3.63 \times 10^{-5}$ \\

QR\_RF & $1.64 \times 10^{-4}$ & $8.62 \times 10^{-4}$ & $3.64 \times 10^{-3}$ & $5.04 \times 10^{-3}$ & $1.71 \times 10^{-4}$ & $6.79 \times 10^{-4}$ & $1.13 \times 10^{-4}$ & $4.35 \times 10^{-4}$\\ 
QR\_MELT & $1.32 \times 10^{-3}$ & 0.01 & $1.93 \times 10^{-3}$ & 0.02 & $7.04 \times 10^{-3}$ & 0.03 & $9.56 \times 10^{-3}$ & 0.04\\
QX\_RIMING & $3.40 \times 10^{-3}$ & 0.01 & $2.87 \times 10^{-4}$ & $1.40 \times 10^{-3}$ & $1.61 \times 10^{-3}$ & $6.88 \times 10^{-3}$ & $3.04 \times 10^{-3}$ & 0.01\\
QI\_DEP & $9.09 \times 10^{-4}$ & $4.15 \times 10^{-3}$ & $1.60 \times 10^{-3}$ & $7.63 \times 10^{-3}$ & $1.81 \times 10^{-3}$ & $5.52 \times 10^{-3}$ & $1.58 \times 10^{-3}$ & $7.48 \times 10^{-3}$\\
         \bottomhline \\
     \end{tabular}
 \end{sidewaystable*}

\subsection{Spatial distribution}\label{sec:spatial_distribution}
\begin{figure*}[t]
    \includegraphics[width=12cm]{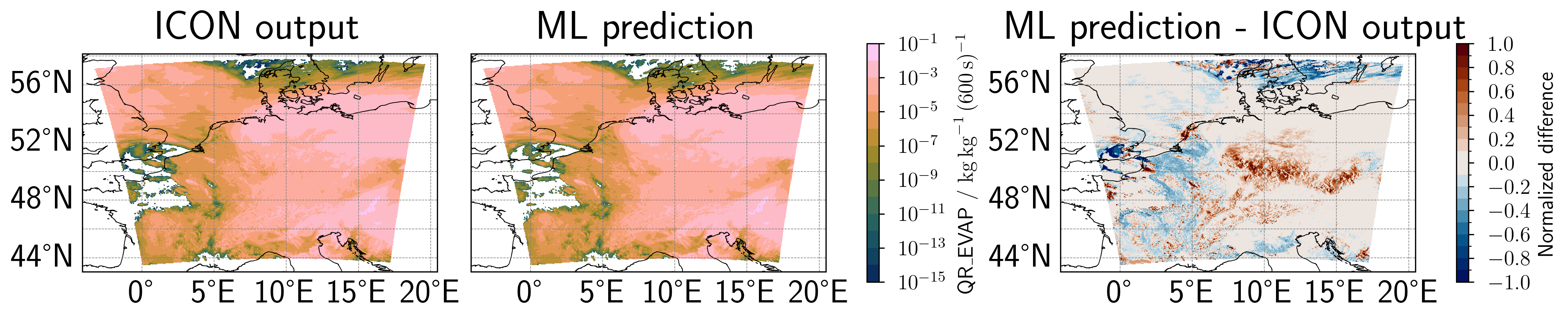}
    \includegraphics[width=12cm]{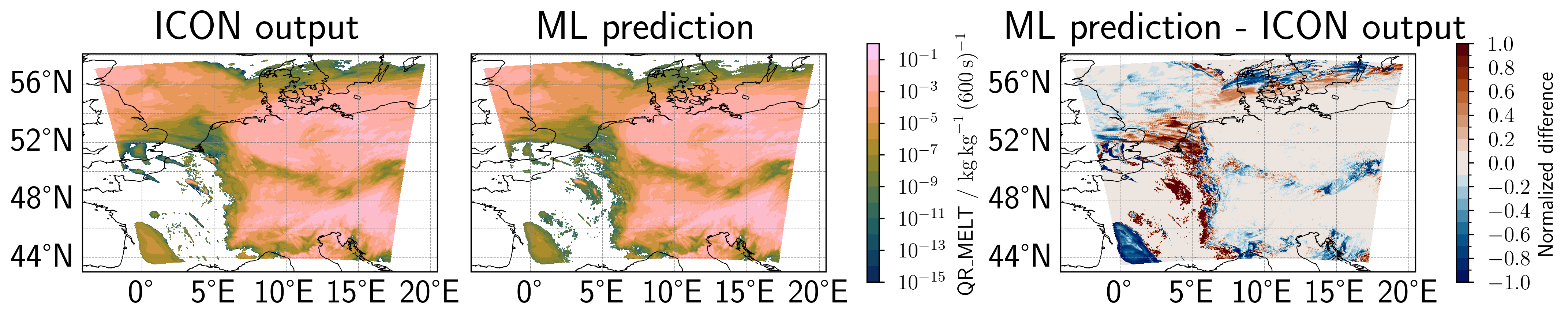}\\
    \includegraphics[width=12cm]{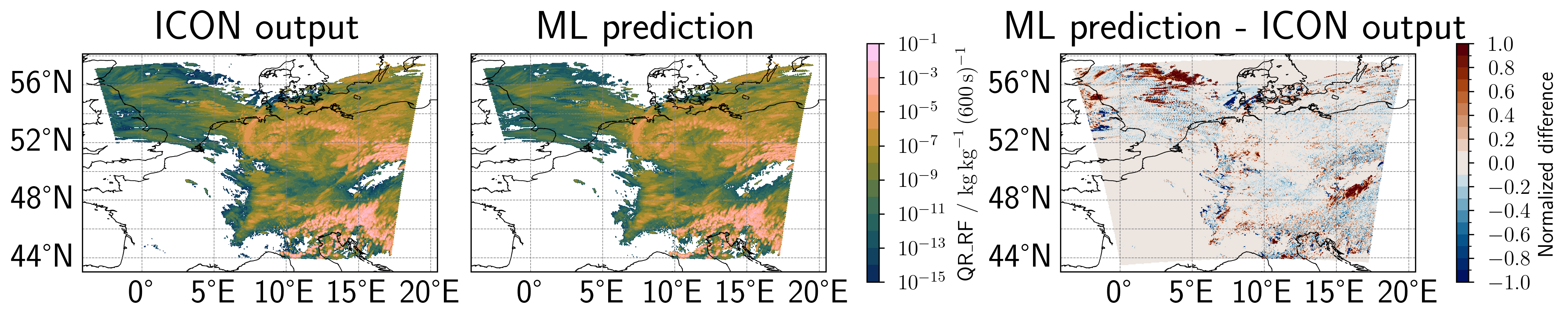}
    \includegraphics[width=12cm]{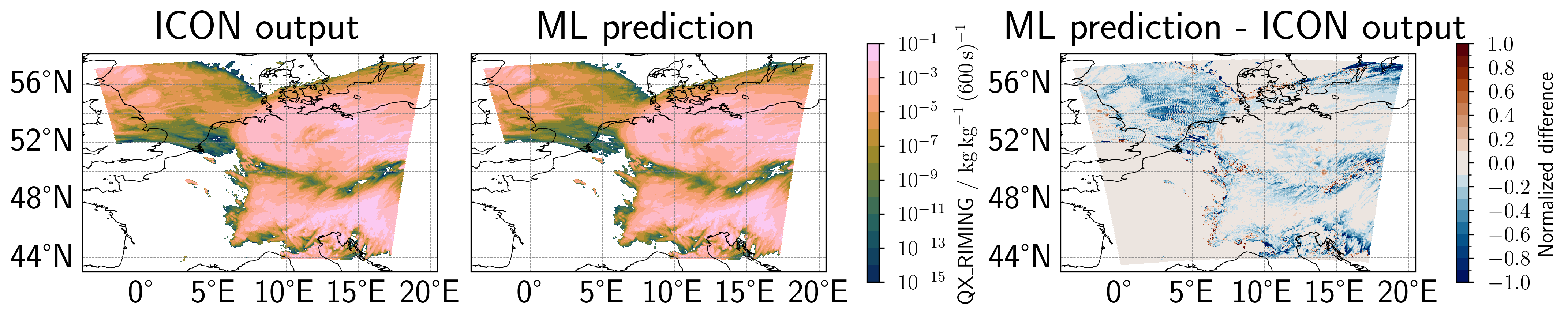}\\
    \includegraphics[width=12cm]{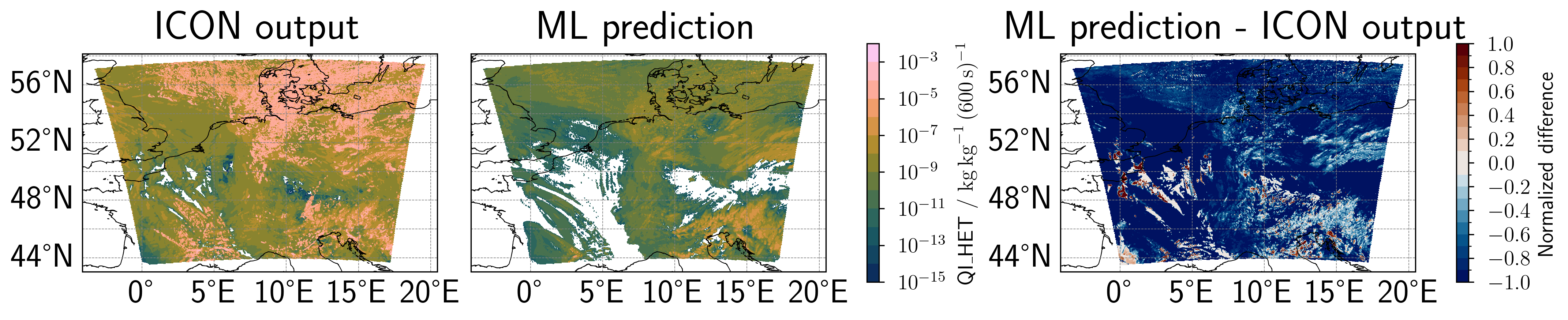}
    \includegraphics[width=12cm]{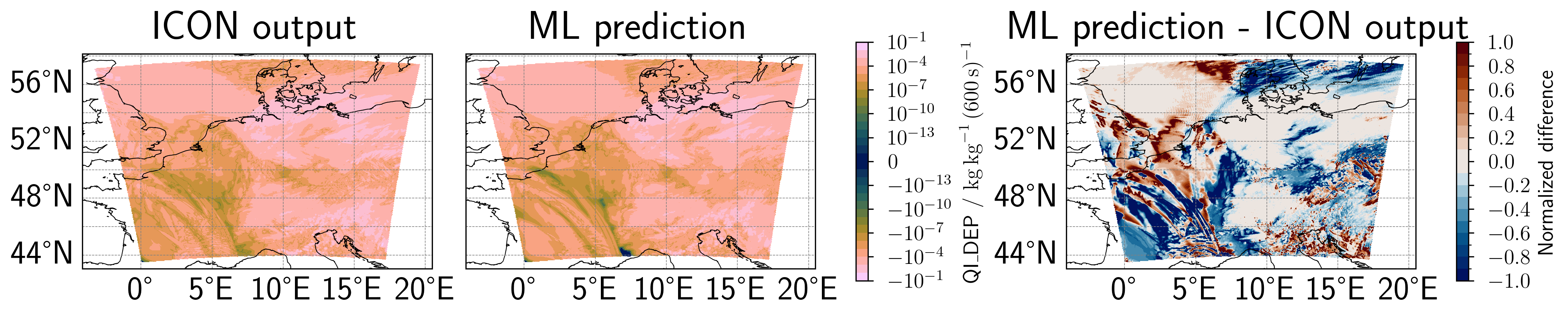}\\
    \caption{Average spatial distribution of QR\_EVAP, QR\_MELT, QR\_RF, QX\_RIMING, QI\_HET and QI\_DEP, accumulated over a 10-minute output time step for 23 June 2023 with the ICON-D2 domain, for the ICON model output (left) and the ML prediction (center). The average is computed over all height levels and the \unit{24\,hour}-simulation interval. In the right column, we show the difference between the simulation output and the prediction, normalized to the respective range of the process rates.}
    \label{fig:2d_plots_20230623_additional_rates}
\end{figure*}

\begin{figure*}[t]
    \includegraphics[width=11cm]{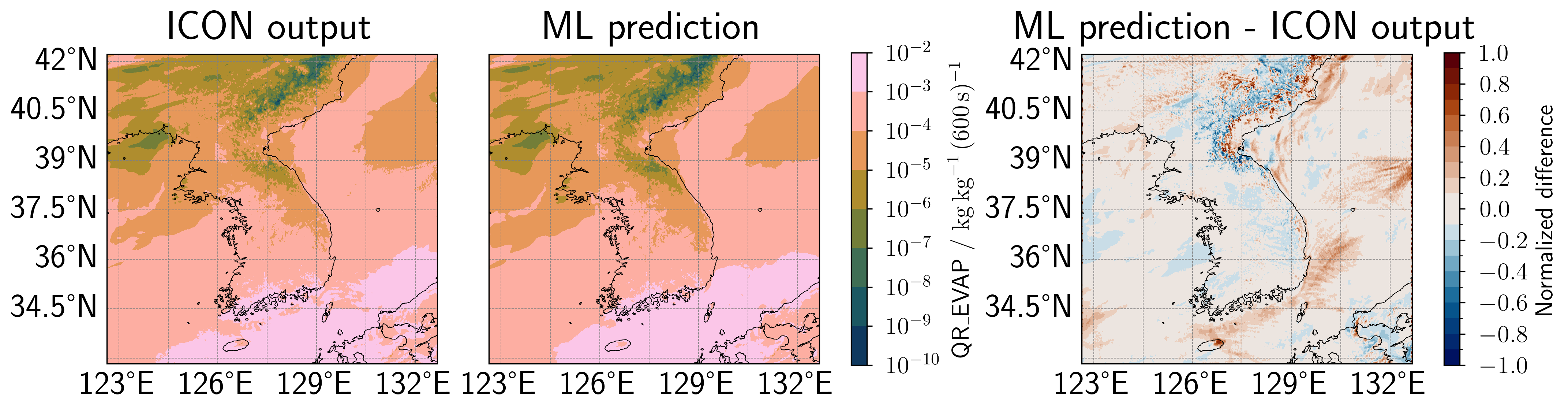}
    \includegraphics[width=11cm]{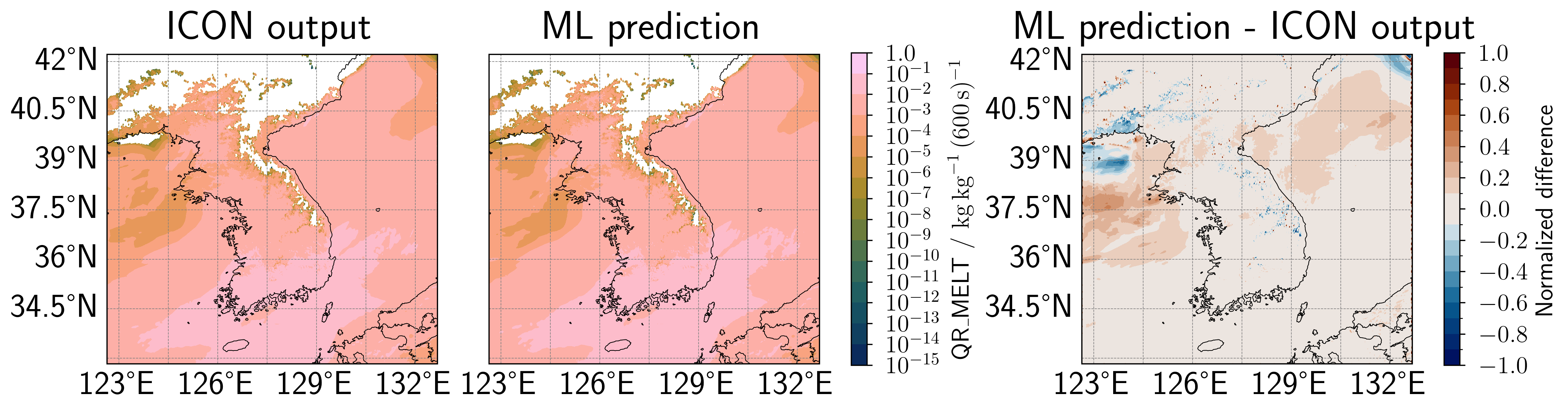}\\
    \includegraphics[width=11cm]{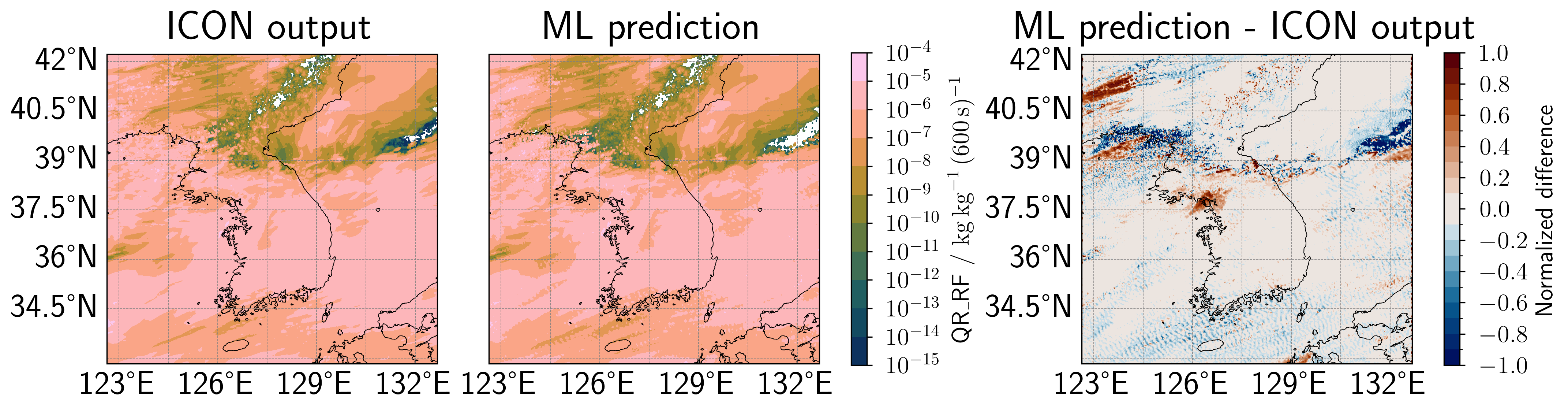}
    \includegraphics[width=11cm]{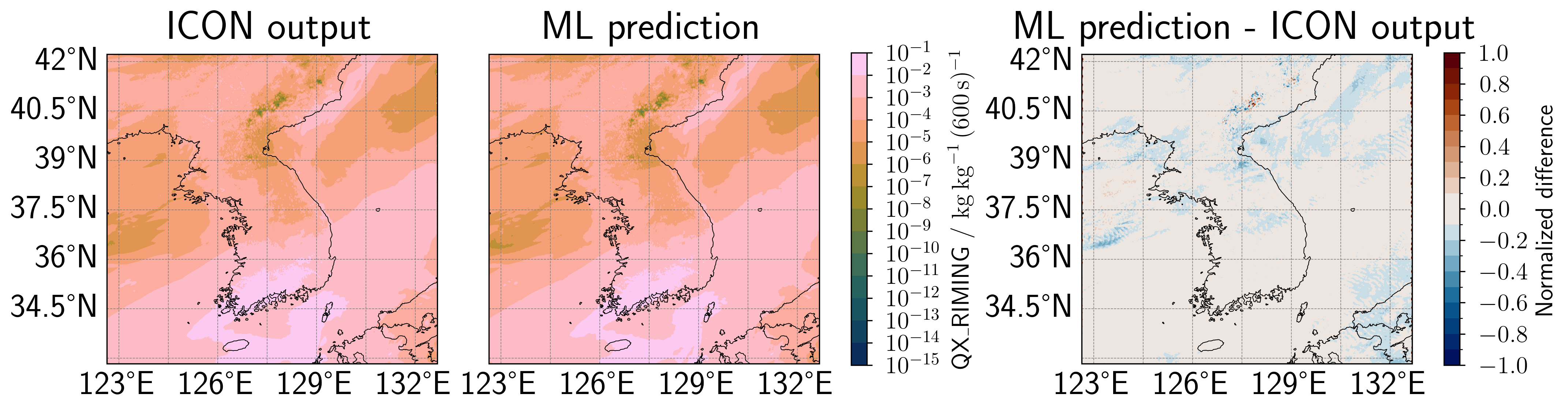}\\
    \includegraphics[width=11cm]{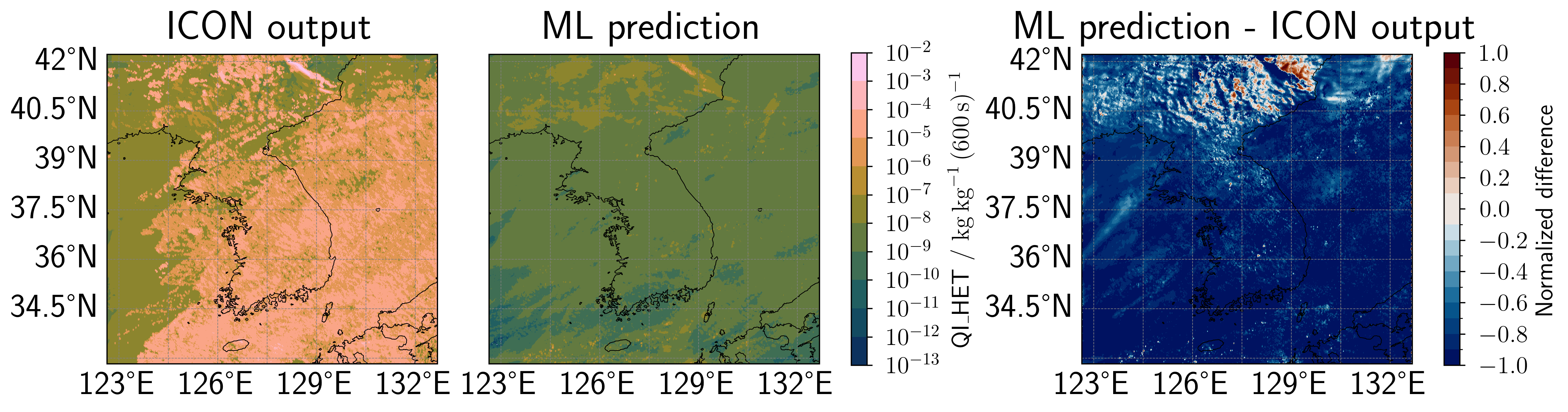}
    \includegraphics[width=11cm]{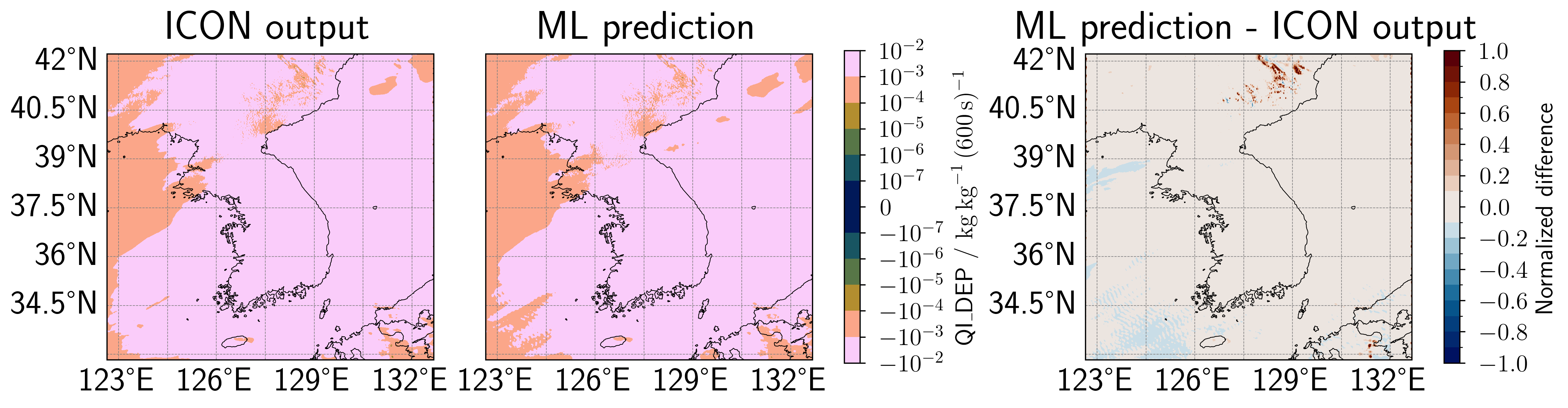}\\
    \caption{Average spatial distribution of QR\_EVAP, QR\_MELT, QR\_RF, QX\_RIMING, QI\_HET and QI\_DEP, accumulated over a 10-minute output time step for 7 March 2018 with the ICE-POP domain (Fig.~\ref{fig:icepop_domain}), for the ICON model output (left) and the ML prediction (center). The average is computed over all height levels and the \unit{24\,hour}-simulation interval. In the right column, we show the difference between the simulation output and the prediction, normalized to the respective range of the process rates.}
    \label{fig:2d_plots_icepop_additional_rates}
\end{figure*}

\begin{figure*}[t]
    \includegraphics[width=10cm]{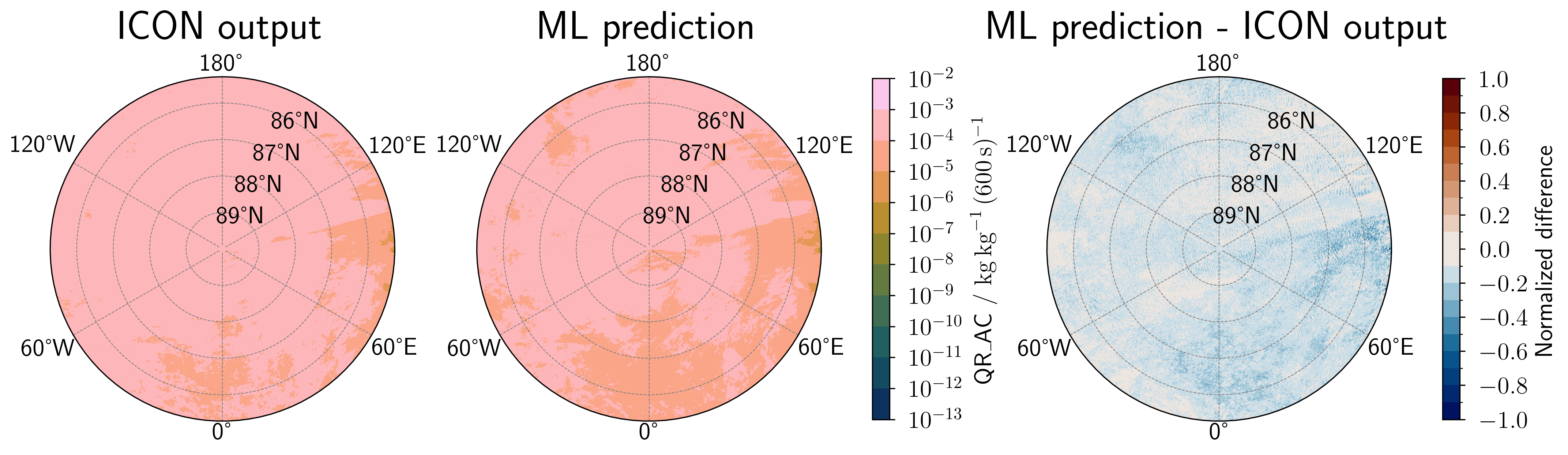}
    \includegraphics[width=10cm]{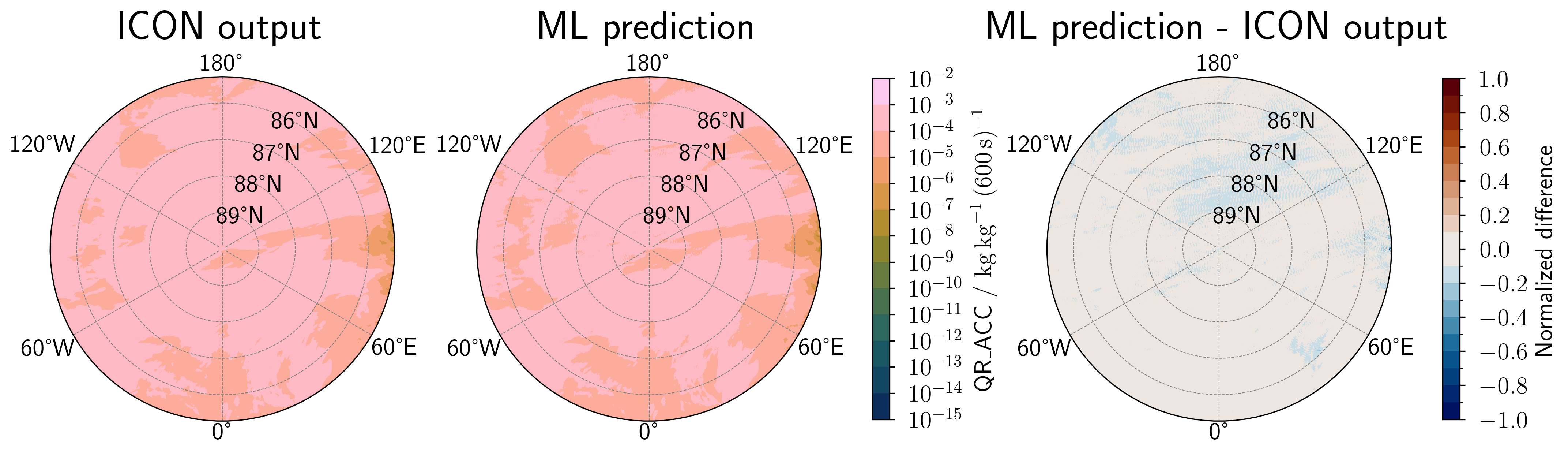}\\
    \includegraphics[width=10cm]{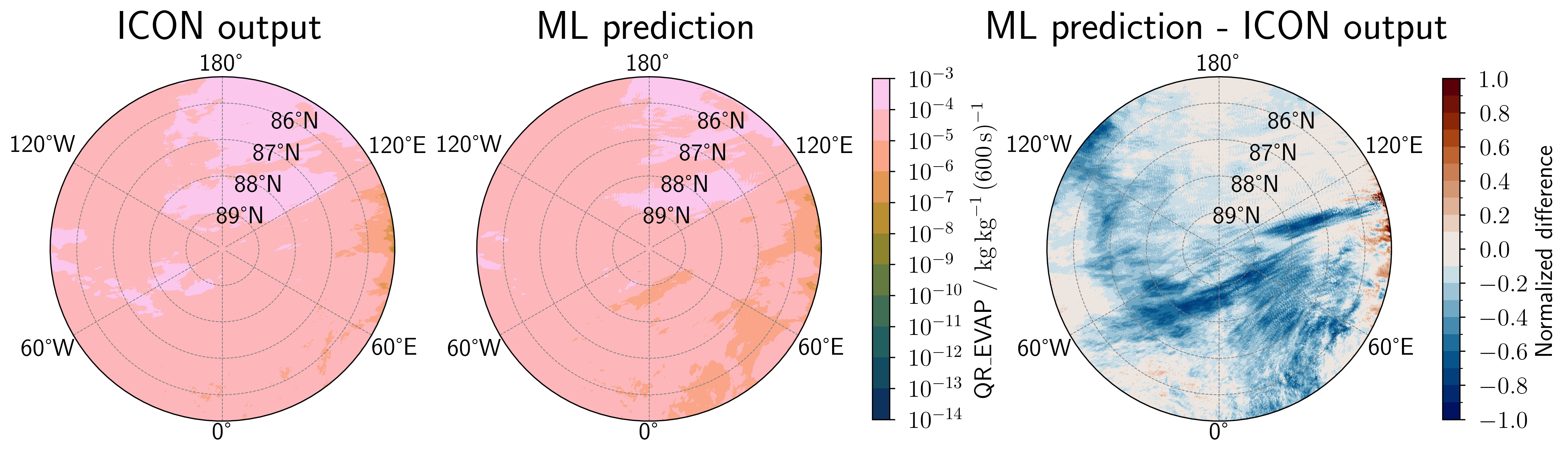}
    \includegraphics[width=10cm]{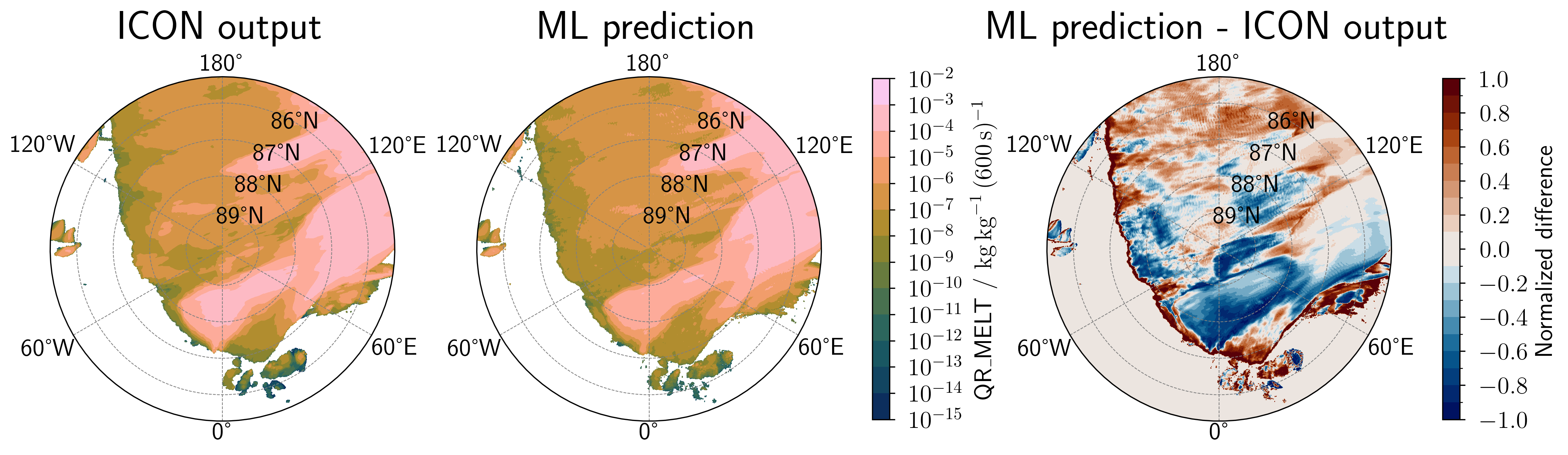}\\
    \includegraphics[width=10cm]{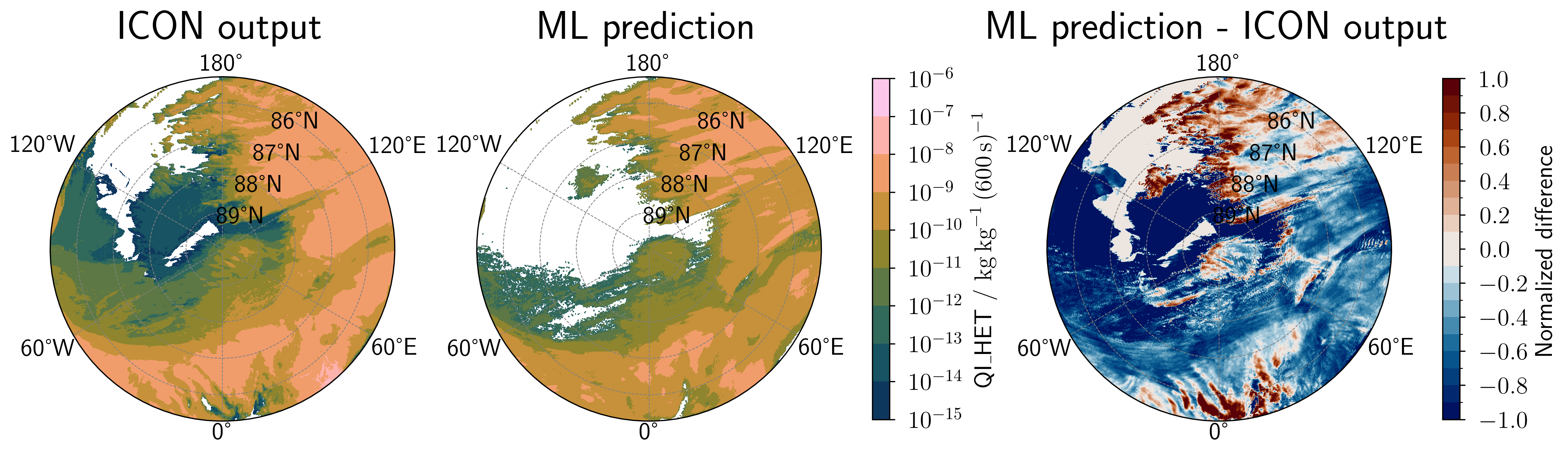}
    \includegraphics[width=10cm]{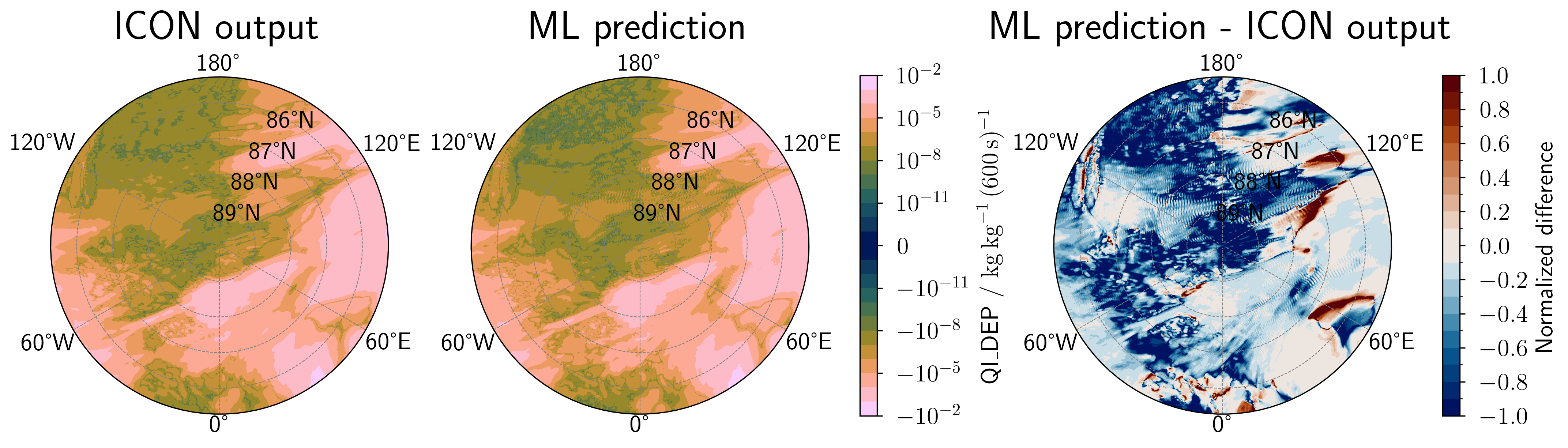}\\
    \caption{Average spatial distribution of QR\_AC, QR\_ACC, QR\_EVAP, QR\_MELT, QI\_HET and QI\_DEP, accumulated over a 10-minute output time step for 3 September 2020 with the MOSAiC domain (Fig.~\ref{fig:mosaic_domain}), for the ICON model output (left) and the ML prediction (center). The average is computed over all height levels and the 24 hour-simulation interval. In the right column, we show the difference between the simulation output and the prediction, normalized to the respective range of the process rates.}
    \label{fig:2d_plots_mosaic_additional_rates}
\end{figure*}

Corresponding to Sect.~\ref{sec:model_output}, we show the average spatial distribution for additional process rates for 23 June 2023 on the ICON-D2 domain (Fig.~\ref{fig:2d_plots_20230623_additional_rates}), for 7 March 2018 on the ICE-POP domain (Fig.~\ref{fig:2d_plots_icepop_additional_rates}) and for 3 September 2020 on the MOSAiC domain (Fig.~\ref{fig:2d_plots_mosaic_additional_rates}).

\section{Modeling aspects}
\label{sec:appendix_model_aspects}
\subsection{Selection of input data}
\label{sec:appendix_data_selection}
\begin{figure*}[t]
    \centering
    \includegraphics[width=\textwidth]{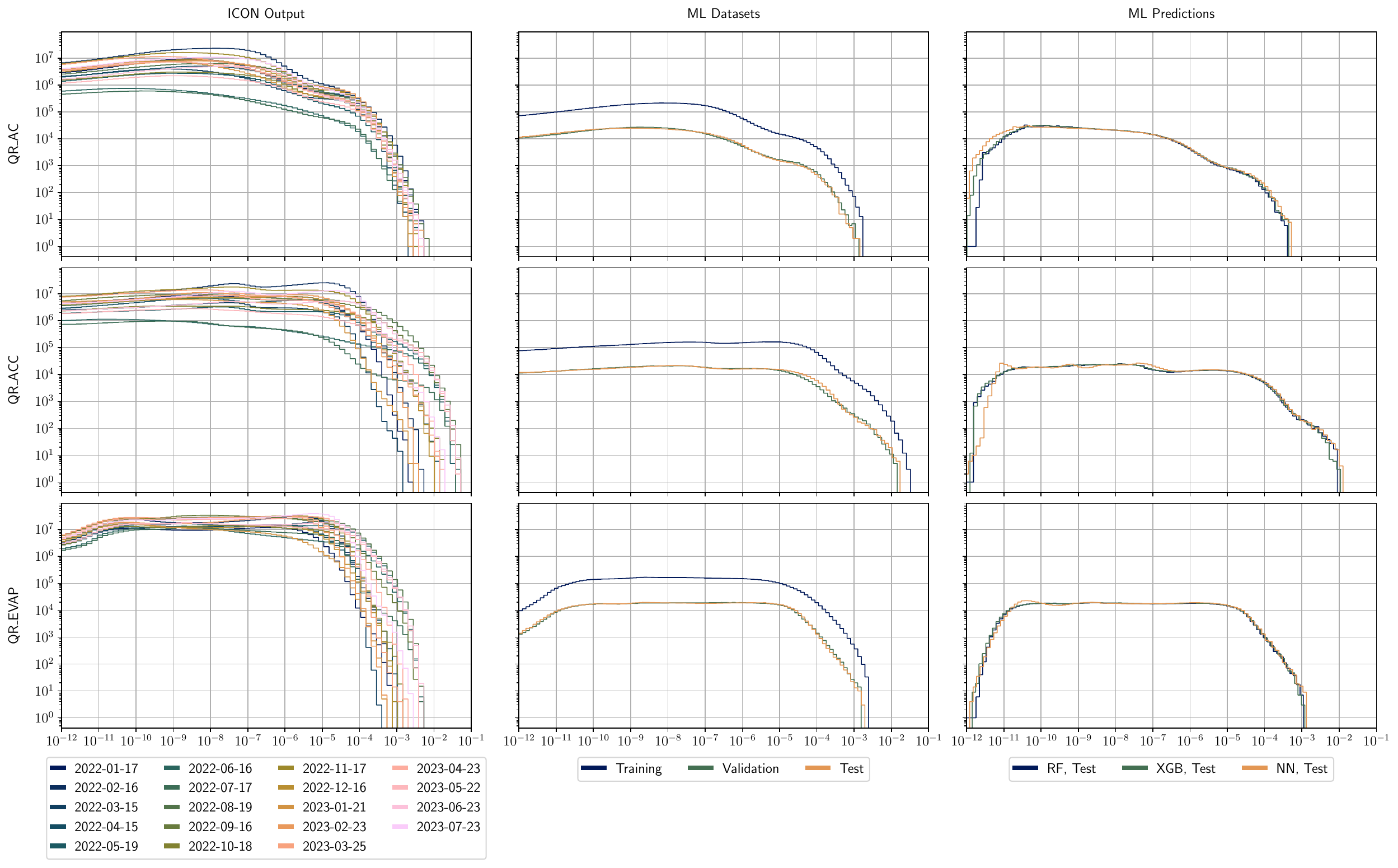} \\
    \caption{Histograms of the process rates accumulated over a one-minute output time step for the ICON model simulations of the different days (left, Table~\ref{table:full_training_data_overview}) and the sampled training, validation and test datasets (center). On the right, the distribution of the predicted values with the RF, gradient boosting and NN model is shown.}
    \label{fig:histograms_data_distribution_1}
\end{figure*}

\begin{figure*}[t]
    \centering
    \includegraphics[width=\textwidth]{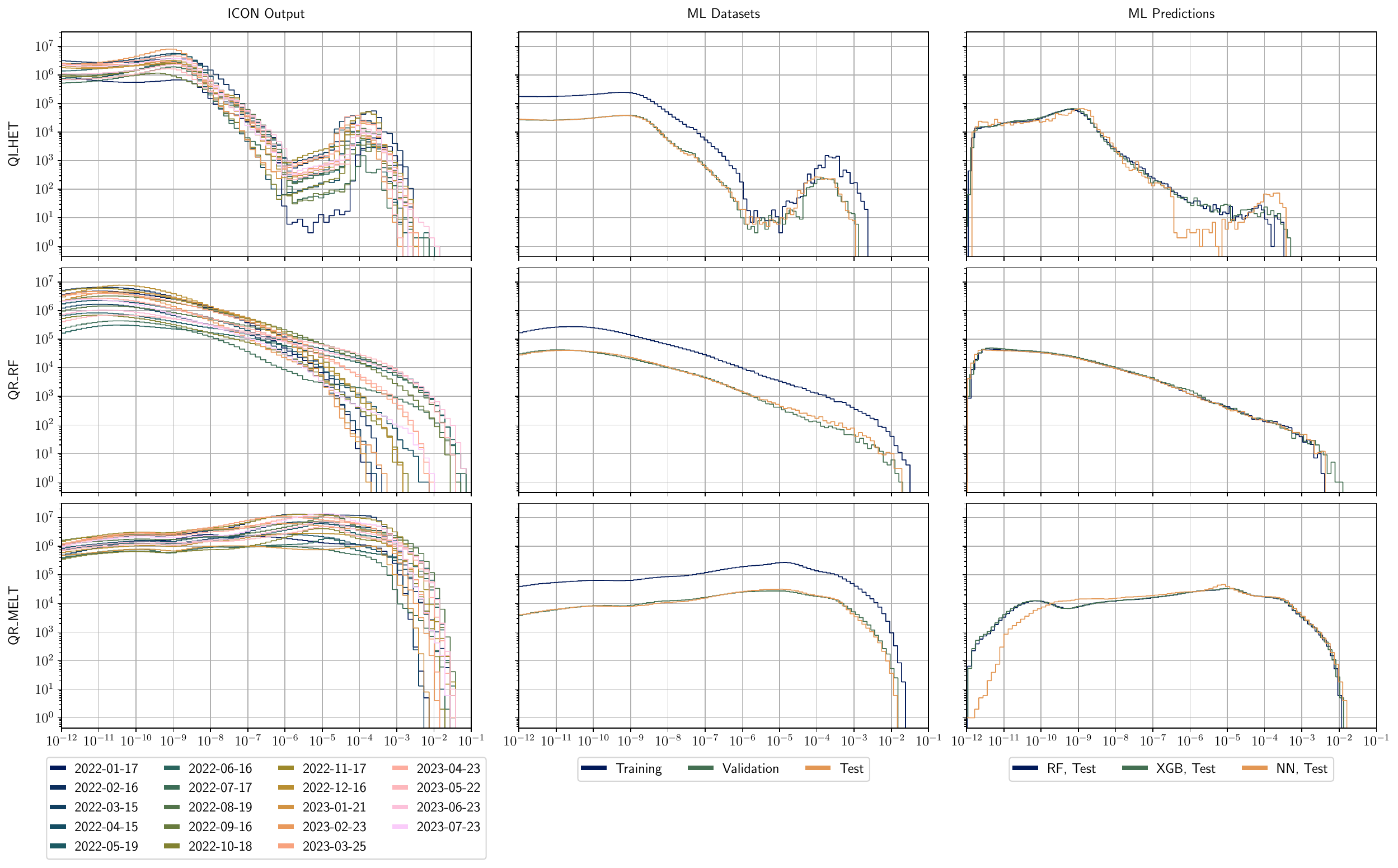} \\
    \caption{Histograms of the process rates accumulated over a one-minute output time step for the ICON model simulations of the different days (left, Table~\ref{table:full_training_data_overview}) and the sampled training, validation and test datasets (center). On the right, the distribution of the predicted values with the RF, gradient boosting and NN model is shown.}
    \label{fig:histograms_data_distribution_2}
\end{figure*}

\begin{figure*}[t]
    \centering
    \includegraphics[width=\textwidth]{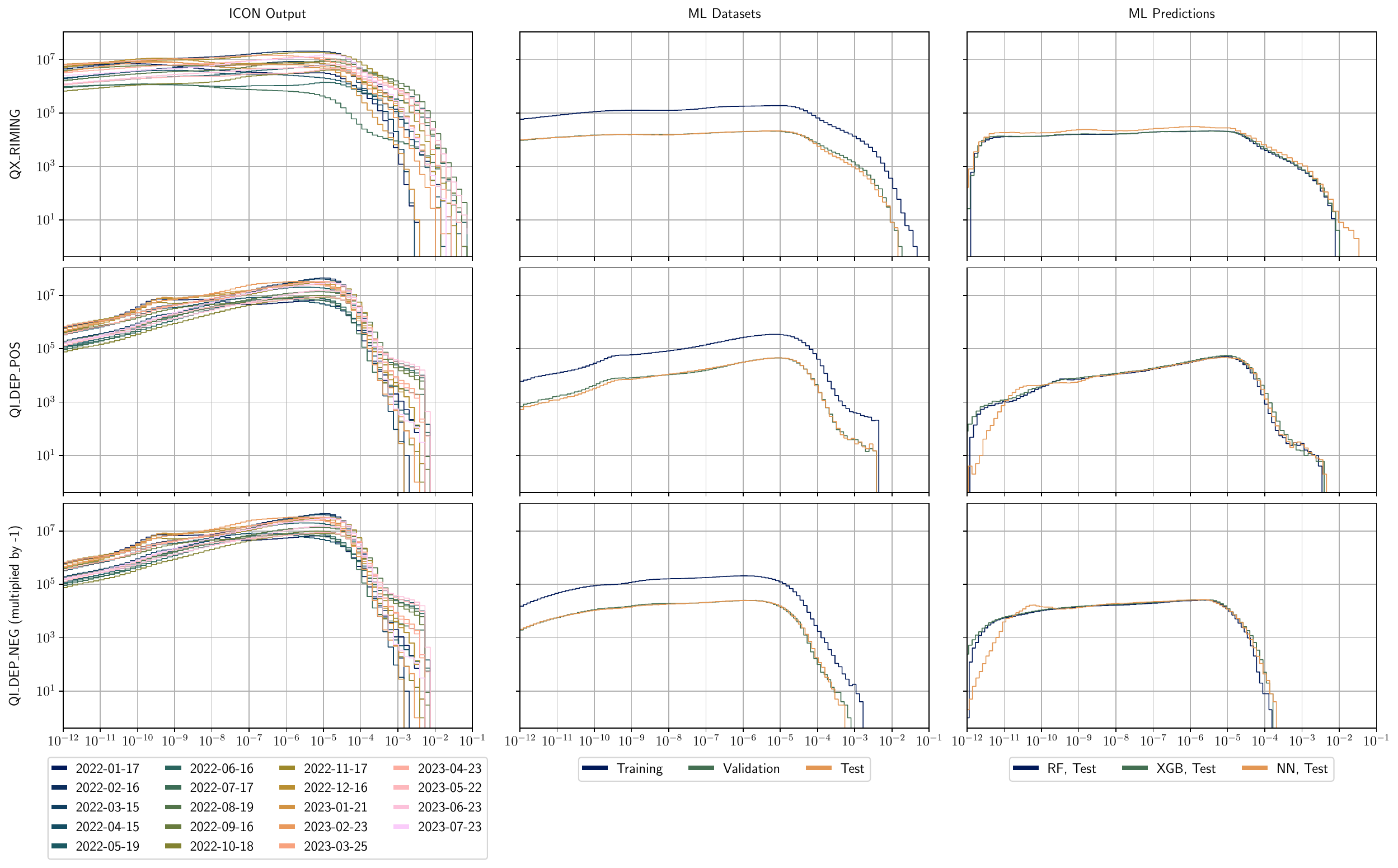} \\
    \caption{Histograms of the process rates accumulated over a one-minute output time step for the ICON model simulations of the different days (left, Table~\ref{table:full_training_data_overview}) and the sampled training, validation and test datasets (center). On the right, the distribution of the predicted values with the RF, gradient boosting and NN model is shown.}
    \label{fig:histograms_data_distribution_3}
\end{figure*}

The two-hourly intervals selected for the training, validation and test datasets (see Sect.~\ref{sec:sampling}) are given in Table~\ref{table:full_training_data_overview}. Furthermore, in Fig.~\ref{fig:histograms_data_distribution_1}, Fig.~\ref{fig:histograms_data_distribution_2} and Fig.~\ref{fig:histograms_data_distribution_3}, we show histograms of the process rate values of the simulation output for the simulated days, the training, validation and test dataset as well as the ML model prediction.
\begin{table*}[t]
    \centering
    \caption{Overview of the simulated days and the 2-hour time intervals used for the training, validation and test datasets.}
    \label{table:full_training_data_overview}
    \begin{tabular}[t]{l l l}
        \tophline 
            {Dataset} & {Date} & {Time interval} \\
        \middlehline
            {\textbf{Training}} & {2022-02-16} & {20:00 UTC - 21:59 UTC, 22:00 UTC - 23:59 UTC} \\
            {} & {2022-04-15} & {14:00 UTC - 15:59 UTC, 18:00 UTC - 19:59 UTC, 22:00 UTC - 23:59 UTC} \\
            {} & {2022-06-16} & {10:00 UTC - 11:59 UTC, 12:00 UTC - 13:59 UTC, 14:00 UTC - 15:59 UTC} \\
            {} & {2022-08-19} & {12:00 UTC - 13:59 UTC, 16:00 UTC - 17:59 UTC, 18:00 UTC - 19:59 UTC} \\
            {} & {2022-10-18} & {14:00 UTC - 15:59 UTC, 16:00 UTC - 17:59 UTC, 18:00 UTC - 19:59 UTC}  \\
            {} & {2022-12-16} & {00:00 UTC - 01:59 UTC, 02:00 UTC - 03:59 UTC, 08:00 UTC - 09:59 UTC}  \\
            \middlehline
            {\textbf{Validation}} & {2022-01-17} & {00:00 UTC - 01:59 UTC, 02:00 UTC - 03:59 UTC, 06:00 UTC - 07:59 UTC} \\
            {} & {} & {18:00 UTC - 19:59 UTC} \\
            {} & {2022-03-15} & {00:00 UTC - 01:59 UTC, 02:00 UTC - 03:59 UTC, 10:00 UTC - 11:59 UTC} \\
            {} & {} & {16:00 UTC - 17:59 UTC} \\
            {} & {2022-05-19} & {12:00 UTC - 13:59 UTC, 16:00 UTC - 17:59 UTC} \\
            {} & {2022-07-17} & {02:00 UTC - 03:59 UTC, 10:00 UTC - 11:59 UTC} \\
            {} & {2022-09-16} & {06:00 UTC - 07:59 UTC, 14:00 UTC - 15:59 UTC, 22:00 UTC - 23:59 UTC}  \\
            {} & {2022-11-17} & {20:00 UTC - 21:59 UTC, 22:00 UTC - 23:59 UTC}  \\
            \middlehline
            {\textbf{Testing}} & {2023-01-21} & {04:00 UTC - 05:59 UTC, 10:00 UTC - 11:59 UTC, 16:00 UTC - 17:59 UTC} \\
            {} & {} & {20:00 UTC - 21:59 UTC, 22:00 UTC - 23:59 UTC} \\
            {} & {2023-02-23} & {16:00 UTC - 17:59 UTC, 22:00 UTC - 23:59 UTC} \\
            {} & {2023-03-25} & {00:00 UTC - 01:59 UTC, 12:00 UTC - 13:59 UTC, 14:00 UTC - 15:59 UTC} \\
            {} & {2023-04-23} & {12:00 UTC - 13:59 UTC, 20:00 UTC - 21:59 UTC} \\
            {} & {2023-05-22} & {14:00 UTC - 15:59 UTC} \\
            {} & {2023-06-23} & {10:00 UTC - 11:59 UTC, 12:00 UTC - 13:59 UTC, 14:00 UTC - 15:59 UTC} \\
            {} & {} & {22:00 UTC - 23:59 UTC}  \\
            {} & {2023-07-23} & {10:00 UTC - 11:59 UTC, 12:00 UTC - 13:59 UTC, 14:00 UTC - 15:59 UTC} \\
            {} & {} & {20:00 UTC - 21:59 UTC}  \\
    \bottomhline
    \end{tabular}
\end{table*}

\subsection{Selection of the model architecture for each process rate for the combined classification-regression model}
In Table~\ref{table:classification_validation_scores} and Table~\ref{table:regression_validation_scores}, we show the selection of the model architecture for the classification and regression model in the two-step framework for each process rate. The selection is based on the validation performance, measured by the $F_1$ (classification) and the $R^2$ score (regression).

\begin{table*}[t]
    \centering
    \caption{Selection of the classification model architecture for each process rate based on validation performance in terms of the $F_1$ score for the instantaneous process rates with a 10-minute output time step. Highest (best) scores are highlighted in bold.}
    \label{table:classification_validation_scores}
    \begin{tabular}{l c c c c}
    \tophline
    \multirow{2}{*}{Process rate (target)} & \multicolumn{3}{c}{Model performance ($F_1$)} & \multirow{2}{*}{Selected classifier} \\
    \cline{2-4}
    & RF & XGB & NN & \\
    \middlehline
QR\_AC       & $\mathbf{0.96}$ & 0.96          & 0.96 & {RF}\\
QR\_ACC      & $0.77$ & $\mathbf{0.77}$ & $0.77$ & {XGB}\\
QNC\_SC      & $0.94$ & $\mathbf{0.94}$ & $0.94$ & {XGB}\\
QNR\_SC      & $0.99$ & $0.96$ & $\mathbf{0.99}$ & {NN}\\
QR\_EVAP     & $0.91$ & $\mathbf{0.91}$ & $0.88$ & {XGB}\\
QSGH\_EVAP   & $\mathbf{0.96}$ & $0.95$ & $0.89$ & {RF}\\
QI\_HOM      & $\mathbf{0.60}$ & $0.45$ & $0.56$ & {RF}\\
QI\_HET      & $0.22$ & $\mathbf{0.24}$ & $0.04$ & {XGB}\\
QC\_MELT     & $\mathbf{0.53}$ & $0.51$ & $0.00$ & {RF}\\
QR\_MELT     & $\mathbf{0.94}$ & $0.94$ & $0.81$ & {RF}\\
QR\_RF       & $\mathbf{0.98}$ & $0.98$ & $0.98$ & {RF}\\
QI\_RF       & $\mathbf{0.98}$ & $0.98$ & $0.98$ & {RF}\\
QG\_RF       & $0.98$ & $0.97$ & $\mathbf{0.98}$ & {NN}\\
QH\_RF       & $0.96$ & $\mathbf{0.97}$ & $0.95$ & {XGB}\\
QI\_DEP      & $0.94$ & $\mathbf{0.95}$ & $0.83$ & {XGB}\\
QS\_DEP      & $0.93$ & $\mathbf{0.94}$ & $0.88$ & {XGB}\\
QG\_DEP      & $\mathbf{0.96}$ & $0.96$ & $0.93$ & {RF}\\
QH\_DEP      & $0.92$ & $\mathbf{0.96}$ & $0.47$ & {XGB}\\
QC\_RIME\_I  & $\mathbf{0.97}$ & $0.97$ & $0.97$ & {RF}\\
QC\_RIME\_S  & $\mathbf{0.97}$ & $0.96$ & $0.97$ & {RF}\\
QC\_RIME\_GH & $\mathbf{0.98}$ & $0.97$ & $0.97$ & {RF}\\
QR\_RIME\_I  & $\mathbf{0.98}$ & $0.97$ & $0.90$ & {RF}\\
QR\_RIME\_S  & $\mathbf{0.98}$ & $0.97$ & $0.69$ & {RF}\\
QR\_RIME\_GH & $0.99$ & $\mathbf{0.99}$ & $0.93$ & {XGB}\\
QX\_RIMING   & $0.99$ & $\mathbf{0.99}$ & $0.93$ & {XGB}\\
        \bottomhline
    \end{tabular}
\end{table*}

\begin{table*}[t]
    \centering
    \caption{Selection of the regression model architecture for each process rate based on validation performance in terms of the $R^2$ score for the instantaneous process rates with a 10-minute output time step. Highest (best) scores are highlighted in bold.}
    \label{table:regression_validation_scores}
    \begin{tabular}{l c c c c}
    \tophline
    \multirow{2}{*}{Process rate (target)} & \multicolumn{3}{c}{Model performance ($R^2$)} & \multirow{2}{*}{Selected regressor} \\
    \cline{2-4}
    & RF & XGB & NN & \\
    \middlehline
QR\_AC       & $\mathbf{0.87}$ & $0.79$ & $0.77$ & {RF}\\
QR\_ACC      & $0.99$ & $0.74$ & $\mathbf{1.00}$ & {NN}\\
QNC\_SC      & $1.00$ & $0.93$ & $\mathbf{1.00}$ & {NN} \\
QNR\_SC\_POS & $0.99$ & $0.93$ & $\mathbf{1.00}$ & {NN}\\
QNR\_SC\_NEG & $\mathbf{0.97}$ & $0.88$ & $0.96$ & {RF}\\
QR\_EVAP     & $0.97$ & $0.86$ & $\mathbf{0.99}$ & {NN}\\
QSGH\_EVAP   & $0.95$ & $0.94$ & $\mathbf{1.00}$ & {NN}\\
QI\_HOM      & $\mathbf{0.87}$ & $0.86$ & $0.76$ & {RF}\\
QI\_HET      & $\mathbf{0.81}$ & $0.32$ & $0.78$ & {RF}\\
QC\_MELT     & $0.87$ & $\mathbf{0.87}$ & $0.71$ & {XGB}\\
QR\_MELT     & $0.99$ & $0.98$ & $\mathbf{0.99}$ & {NN}\\
QR\_RF       & $0.49$ & $0.36$ & $\mathbf{0.67}$ & {NN}\\
QI\_RF       & $0.79$ & $\mathbf{0.87}$ & $0.83$ & {XGB}\\
QG\_RF       & $0.69$ & $0.64$ & $\mathbf{0.86}$ & {NN}\\
QH\_RF       & $0.61$ & $\mathbf{0.63}$ & $0.43$ & {XGB}\\
QC\_RIME\_I  & $0.66$ & $0.53$ & $\mathbf{0.66}$ & {NN}\\
QC\_RIME\_S  & $\mathbf{0.90}$ & $0.80$ & $0.89$ & {RF}\\
QC\_RIME\_GH & $0.99$ & $0.93$ & $\mathbf{1.00}$ & {NN}\\
QR\_RIME\_I  & $0.84$ & $0.70$ & $\mathbf{0.90}$ & {NN}\\
QR\_RIME\_S  & $\mathbf{0.86}$ & $0.48$ & $0.76$ & {RF}\\
QR\_RIME\_GH & $0.93$ & $0.91$ & $\mathbf{0.94}$ & {NN}\\
QX\_RIMING   & $0.91$ & $\mathbf{0.91}$ & $0.90$ & {XGB}\\
QI\_DEP\_POS & $\mathbf{0.83}$ & $0.75$ & $-1.90$ & {RF}\\
QI\_DEP\_NEG & $0.59$ & $\mathbf{0.72}$ & $0.45$ & {XGB}\\
QS\_DEP\_POS & $0.89$ & $0.95$ & $\mathbf{0.99}$ & {NN}\\
QS\_DEP\_NEG & $0.86$ & $0.92$ & $\mathbf{0.98}$ & {NN}\\
QG\_DEP\_POS & $0.94$ & $0.94$ & $\mathbf{0.99}$ & {NN}\\
QG\_DEP\_NEG & $0.93$ & $0.95$ & $\mathbf{1.00}$ & {NN}\\
QH\_DEP\_POS & $\mathbf{0.97}$ & $0.82$ & $0.96$ & {RF}\\
QH\_DEP\_NEG & $0.85$ & $0.58$ & $\mathbf{0.92}$ & {NN}\\
        \bottomhline
    \end{tabular}
\end{table*}

\subsection{Definition of process rates and ICON output variables}
\label{sec:ICON_output_variables} 
In Table~\ref{table:ICON_output_variables}, we list the output variables of the ICON model used as inputs to the ML models for the prediction of the microphysical process rates together with their respective unit. In addition, in Table~\ref{table:microphysical_process_rates}, we list the microphysical process rates together with their respective unit. Table~\ref{table:features_models} lists the process rates (target variables) and the input features used for the prediction of each process rate, as well as the selected model architecture for the classification and regression step.
\begin{table*}[t]
    \centering
    \caption{ICON output variables.}
    \label{table:ICON_output_variables}
    \begin{tabular}{c c l}
        \tophline
            {Variable} & {Unit} & {Description} \\
        \middlehline
            {$q_c$} & {\unit{kg}\,\unit{kg}$^{-1}$} & {Specific cloud water content} \\
            {$q_r$} & {\unit{kg}\,\unit{kg}$^{-1}$} & {Specific rain content} \\
            {$q_i$} & {\unit{kg}\,\unit{kg}$^{-1}$} & {Specific cloud ice content} \\
            {$q_s$} & {\unit{kg}\,\unit{kg}$^{-1}$} & {Specific snow content} \\
            {$q_g$} & {\unit{kg}\,\unit{kg}$^{-1}$} & {Specific graupel content} \\
            {$q_h$} & {\unit{kg}\,\unit{kg}$^{-1}$} & {Specific hail content} \\
            {$n_c$} & {\quad\;\unit{kg}$^{-1}$} & {Cloud droplet number concentration} \\
            {$n_r$} & {\quad\;\unit{kg}$^{-1}$} & {Rain drop number concentration} \\
            {$n_i$} & {\quad\;\unit{kg}$^{-1}$} & {Cloud ice number concentration} \\
            {$n_s$} & {\quad\;\unit{kg}$^{-1}$} & {Snow number concentration} \\
            {$n_g$} & {\quad\;\unit{kg}$^{-1}$} & {Graupel number concentration} \\
            {$n_h$} & {\quad\;\unit{kg}$^{-1}$} & {Hail number concentration} \\
            {$q_v$} & {\unit{kg}\,\unit{kg}$^{-1}$} & {Specific humidity} \\
            {$\rho$} & {\unit{kg}\,\unit{m}$^{-3}$\,} & {Density} \\
            {$p$} & {\unit{Pa}} & {Pressure} \\
            {$T$} & {\unit{K}} & {Temperature} \\
            {$u$} & {\unit{m}\,\unit{s}$^{-1}$} & {Zonal wind} \\
            {$v$} & {\unit{m}\,\unit{s}$^{-1}$} & {Meridional wind} \\
            {$w$} & {\unit{m}\,\unit{s}$^{-1}$} & {Vertical veloctiy} \\
        \bottomhline
    \end{tabular}
\end{table*}

\begin{table*}[t]
    \centering
    \caption{Microphysical process rates with corresponding unit, value range and the hydrometeor category that is updated with the process rate. The time step $\unit{t}$ is the output time step $t^\text{out}$ for the accumulated process rates and the fast-physics time step $t^\text{fast}$ for the instantaneous process rates.}
    \label{table:microphysical_process_rates}
    \begin{tabular}{l c c c l}
        \tophline
            {Process rate} & {Unit} & {Hydrometeor category} & {Value range} & {Process description}\\
        \middlehline
            {QR\_AC}       & {\unit{kg}\,\unit{kg}$^{-1}\,\unit{t}^{-1}$} & {$q_r$}           & {$\;\;\;\;\;\;[0, \infty)$} & {Autoconversion}\\
            {QR\_ACC}      & {\unit{kg}\,\unit{kg}$^{-1}\,\unit{t}^{-1}$} & {$q_r$}           & {$\;\;\;\;\;\;[0, \infty)$} & {Accretion}\\
            {QNC\_SC}      & {\quad\;\unit{kg}$^{-1}\,\unit{t}^{-1}$}            & {$n_c$}           & {$(-\infty, 0]\;\;\;$} & {Cloud droplet self-collection}\\
            {QNR\_SC}      & {\quad\;\unit{kg}$^{-1}\,\unit{t}^{-1}$}            & {$n_r$}           & {$(-\infty, \infty)$} & {Raindrop self-collection}\\
            {QR\_EVAP}     & {\unit{kg}\,\unit{kg}$^{-1}\,\unit{t}^{-1}$} & {$q_r$}           & {$\;\;\;\;\;\;[0, \infty)$} & {Rain evaporation}\\
            {QSGH\_EVAP}   & {\unit{kg}\,\unit{kg}$^{-1}\,\unit{t}^{-1}$} & {$q_s, q_g, q_h$} & {$\;\;\;\;\;\;[0, \infty)$} & {Evaporation of melting snow, graupel, hail}\\
            {QI\_HOM}      & {\unit{kg}\,\unit{kg}$^{-1}\,\unit{t}^{-1}$} & {$q_i$}           & {$\;\;\;\;\;\;[0, \infty)$} & {Homogeneous ice nucleation}\\
            {QI\_HET}      & {\unit{kg}\,\unit{kg}$^{-1}\,\unit{t}^{-1}$} & {$q_i$}           & {$\;\;\;\;\;\;[0, \infty)$} & {Heterogeneous ice nucleation}\\
            {QC\_MELT}     & {\unit{kg}\,\unit{kg}$^{-1}\,\unit{t}^{-1}$} & {$q_c$}           & {$\;\;\;\;\;\;[0, \infty)$} & {Ice melting to cloud droplets}\\
            {QR\_MELT}     & {\unit{kg}\,\unit{kg}$^{-1}\,\unit{t}^{-1}$} & {$q_r$}           & {$\;\;\;\;\;\;[0, \infty)$} & {Rain production by melting}\\
            {QR\_RF}       & {\unit{kg}\,\unit{kg}$^{-1}\,\unit{t}^{-1}$} & {$q_r$}           & {$\;\;\;\;\;\;[0, \infty)$} & {Rain freezing to snow, graupel and hail}\\
            {QI\_RF}       & {\unit{kg}\,\unit{kg}$^{-1}\,\unit{t}^{-1}$} & {$q_i$}           & {$\;\;\;\;\;\;[0,\infty)$} & {Rain freezing to ice}\\
            {QG\_RF}       & {\unit{kg}\,\unit{kg}$^{-1}\,\unit{t}^{-1}$} & {$q_g$}           & {$\;\;\;\;\;\;[0,\infty)$} & {Rain freezing to graupel}\\
            {QH\_RF}       & {\unit{kg}\,\unit{kg}$^{-1}\,\unit{t}^{-1}$} & {$q_h$}           & {$\;\;\;\;\;\;[0,\infty)$} & {Rain freezing to hail}\\
            {QX\_RIMING}   & {\unit{kg}\,\unit{kg}$^{-1}\,\unit{t}^{-1}$} & {all $q_k$}       & {$\;\;\;\;\;\;[0,\infty)$} & {Total riming}\\
            {QC\_RIME\_I}  & {\unit{kg}\,\unit{kg}$^{-1}\,\unit{t}^{-1}$} & {$q_c$}           & {$\;\;\;\;\;\;[0,\infty)$} & {Riming of ice with cloud droplets}\\
            {QR\_RIME\_I}  & {\unit{kg}\,\unit{kg}$^{-1}\,\unit{t}^{-1}$} & {$q_r$}           & {$\;\;\;\;\;\;[0,\infty)$} & {Riming of ice with raindrops}\\
            {QC\_RIME\_S}  & {\unit{kg}\,\unit{kg}$^{-1}\,\unit{t}^{-1}$} & {$q_c$}           & {$\;\;\;\;\;\;[0,\infty)$} & {Riming of snow with cloud droplets}\\
            {QR\_RIME\_S}  & {\unit{kg}\,\unit{kg}$^{-1}\,\unit{t}^{-1}$} & {$q_r$}           & {$\;\;\;\;\;\;[0,\infty)$} & {Riming of snow with raindrops}\\
            {QC\_RIME\_GH} & {\unit{kg}\,\unit{kg}$^{-1}\,\unit{t}^{-1}$} & {$q_c$}           & {$\;\;\;\;\;\;[0,\infty)$} & {Riming of graupel or hail with cloud droplets}\\
            {QR\_RIME\_GH} & {\unit{kg}\,\unit{kg}$^{-1}\,\unit{t}^{-1}$} & {$q_r$}           & {$\;\;\;\;\;\;[0, \infty)$} & {Riming of graupel or hail with raindrops}\\
            {QI\_DEP}      & {\unit{kg}\,\unit{kg}$^{-1}\,\unit{t}^{-1}$} & {$q_i$}           & {$(-\infty, \infty)$} & {Vapor deposition on ice}\\
            {QS\_DEP}      & {\unit{kg}\,\unit{kg}$^{-1}\,\unit{t}^{-1}$} & {$q_s$}           & {$(-\infty, \infty)$} & {Vapor deposition on snow}\\
            {QG\_DEP}      & {\unit{kg}\,\unit{kg}$^{-1}\,\unit{t}^{-1}$} & {$q_g$}           & {$(-\infty, \infty)$} & {Vapor deposition on graupel}\\
            {QH\_DEP}      & {\unit{kg}\,\unit{kg}$^{-1}\,\unit{t}^{-1}$} & {$q_h$}           & {$(-\infty, \infty)$} & {Vapor deposition on hail}\\
        \bottomhline
    \end{tabular}
\end{table*}

\begin{table*}[t]
    \centering
    \caption{Input features and the choice of architecture for the corresponding classification and regression model. For the process rates that can take positive and negative values (QNR\_SC, QI\_DEP, QS\_DEP, QG\_DEP, QH\_DEP), we use the same model architecture for the models that predict the positive values and those that predict the negative values.}
    \label{table:features_models}
    \begin{tabular}{l l l l}
        \tophline
            {Process rate (target)} & {ICON output variable (features)} & {Classifier} & {Regressor}\\
        \middlehline
            {QR\_AC}       & {$q_c,\,q_r,\,n_c,\,n_r,\,q_v,\,\rho$} & {RF} & {RF}\\
            {QR\_ACC}      & {$q_c,\,q_r,\,n_c,\,n_r,\,q_v,\,\rho$} & {XGB} & {NN}\\
            {QNC\_SC}      & {$q_c,\,q_r,\,n_c,\,n_r,\,q_v,\,\rho$} & {XGB} & {NN}\\
            {QNR\_SC}      & {$q_c,\,q_r,\,n_c,\,n_r,\,q_v,\,\rho$} & {NN} & {NN} \\
            {QR\_EVAP}     & {$q_c,\,q_r,\,n_r,\,q_v,\,\rho,\,T,\,p$} & {XGB} & {NN}\\
            {QSGH\_EVAP}   & {$q_s,\,q_g,\,q_h,\,n_s,\,n_g,\,n_h,\,q_v,\,T,\,\rho$} & {RF} & {NN} \\
            {QI\_HOM}      & {$q_c,\,q_i,\,n_c,\,n_i,\,q_v,\,\rho,\,T$} & {RF} & {RF} \\
            {QI\_HET}      & {$q_i,\,n_i,\,q_v,\,\rho,\,T$} & {XGB} & {RF} \\
            {QC\_MELT}     & {$q_c,\,q_i,\,n_c,\,n_i,\,q_v,\,T$} & {RF} & {XGB} \\
            {QR\_MELT}     & {$q_c,\,q_r,\,q_s,\,q_g,\,q_h,\,n_c,\,n_r,\,n_s,\,n_g,\,n_h,\,q_v,\,\rho,\,T$} & {RF} & {NN} \\
            {QR\_RF}       & {$q_r,\,n_r,\,T$} & {RF} & {NN}\\
            {QI\_RF}       & {$q_r,\,q_i,\,n_r,\,n_i,\,T$} & {RF} & {XGB} \\
            {QG\_RF}       & {$q_r,\,q_g,\,n_r,\,n_g,\,T$} & {NN} & {NN} \\
            {QH\_RF}       & {$q_r,\,q_h,\,n_r,\,n_h,\,T$} & {XGB} & {XGB} \\
            {QX\_RIMING}   & {$q_c,\,q_r,\,q_i,\,q_s,\,q_g,\,q_h,\,n_c,\,n_r,\,n_i,\,n_s,\,n_g,\,n_h,\,q_v,\,\rho,\,T$} & {XGB} & {XGB} \\
            {QC\_RIME\_I}  & {$q_c,\,q_i,\,q_g,\,n_c,\,n_i,\,n_g,\,\rho,\,T$} & {RF} & {NN} \\
            {QR\_RIME\_I}  & {$q_r,\,q_i,\,q_g,\,n_r,\,n_i,\,n_g,\,\rho,\,T$} & {RF} & {NN} \\
            {QC\_RIME\_S}  & {$q_c,\,q_i,\,q_g,\,q_s,\,n_c,\,n_i,\,n_s,\,n_g,\,\rho,\,T$} & {RF} & {RF} \\
            {QR\_RIME\_S}  & {$q_c,\,q_r,\,q_i,\,q_s,\,q_g,\,q_h,\,n_c,\,n_r,\,n_i,\,n_s,\,n_g,\,n_h,\,q_v,\,\rho,\,T,\,p$} & {RF} & {RF} \\
            {QC\_RIME\_GH} & {$q_c,\,q_g,\,q_h,\,n_c,\,n_g,\,n_h,\,\rho,\,T$} & {RF} & {NN} \\
            {QR\_RIME\_GH} & {$q_r,\,q_g,\,q_h,\,n_r,\,n_g,\,n_h,\,\rho,\,T$} & {XGB} & {NN} \\
            {QI\_DEP}      & {$q_i,\,n_i,\,q_v,\,\rho,\,T,\,p$} & {XGB} & {NN} \\
            {QS\_DEP}      & {$q_s,\,n_s,\,q_v,\,\rho,\,T,\,p$} & {XGB} & {NN} \\
            {QG\_DEP}      & {$q_g,\,n_g,\,q_v,\,\rho,\,T,\,p$} & {RF} & {NN} \\
            {QH\_DEP}      & {$q_h,\,n_h,\,q_v,\,\rho,\,T,\,p$} & {XGB} & {NN} \\
        \bottomhline
    \end{tabular}
\end{table*}

\subsection{Hyperparameter space}\label{sec:hyperparameter_space}
In Table~\ref{table:hyperparameter_space}, we list the ranges of possible hyperparameter values tested during hyperparameter tuning with a random search for the different model architectures.
\begin{table*}[t]
    \centering
    \caption{Hyperparameter ranges tested during hyperparameter tuning for different ML algorithms. The loss function for the NN models is a tuned only for the regression models, for the classification models, we use categorical cross-entropy.}
    \label{table:hyperparameter_space}
    \begin{tabular}{l l l}
        \tophline
            {Model architecture} & {Hyperparameter} & {Range of values} \\
        \middlehline
            {\textbf{Random forest}} & Num. estimators & {50, 100, 150, 200, 250, 300, 400, 500} \\
            {} & Max. depth & {10, 20, 30, 50, 70, 100} \\
            {} & Min. samples leaf & {1, 2, 5, 7, 10} \\
            {} & Min. samples split & {2, 3, 4, 5, 8, 10, 12, 15, 20, 30, 50} \\
            {} & Max. features & {sqrt, log2, 0.25, 0.5, 0.75, 1.0} \\
            {} & Max. samples & {0.5, 0.75, 1.0} \\
            {\textbf{Gradient boosting}} & Num. estimators & {150, 200, 250, 300, 400, 500} \\
            {} & Max. depth & {3, 4, 5, 6, 7, 8, 9, 10, 11, 12, 15, 20} \\
            {} & Learning rate & {0.01, 0.02, 0.05, 0.1, 0.2, 0.3, 0.4, 0.5} \\
            {} & Min. child weight & {1, 2, 3, 4, 5} \\
            {} & Subsample ratio & {0.5, 0.6, 0.6, 0.7, 0.8, 0.9, 1.0} \\
            {} & Gamma & {0.0, 0.1, 0.2, 0.3, 0.4, 0.5} \\
            {} & L1 regularization & {0.0, 0.01, 0.1, 0.5, 1.0, 5.0, 10.0, 20.0} \\
            {} & L2 regularization & {0.1, 0.5, 1.0, 2.0, 3.0, 4.0, 5.0, 7.0, 10.0, 20.0} \\
            {} & Column subsample ratio & {0.5, 0.6, 0.7, 0.8, 0.9, 1.0} \\
            {\textbf{Neural network}} & Num. nodes & {32, 64, 96, 128, 160, 192, 224, 256, 512} \\
            {} & Num. layers & {2, 3, 4, 5} \\
            {} & Learning rate & {$10^{-3}$, $10^{-4}$} \\
            {} & Weight decay & {$10^{-4}$, $10^{-5}$, $10^{-6}$} \\
            {} & Batch size & {32, 64, 96, 128, 160, 192, 224, 256} \\
            {} & Activation function & ReLU, Tanh \\
            {} & Loss & {MSE, MAE, Huber ($\delta = \{0.5,\, 1.0,\,2.0\}$)} \\
        \bottomhline
    \end{tabular}
\end{table*}

\noappendix
\clearpage
\bibliographystyle{copernicus}
\bibliography{references}
\end{document}